%% file: static.tex
\documentclass[11pt,a4paper]{article}

\pdfoutput=1

\usepackage{jheppub}

\include{prolog}

\newcommand{\rev}[1]{#1}

\allowdisplaybreaks[2]

\title{Two-current correlations in the pion on the lattice}

\abstract{We perform a systematic study of the correlation functions of two quark currents in a pion using lattice QCD.  We obtain good signals for all but one of the relevant Wick contractions of quark fields.  We investigate the quark mass dependence of our results and test the importance of correlations between the quark and the antiquark in the pion.  Our lattice data are compared with predictions from chiral perturbation theory.}

\preprint{\vbox{
\hbox{DESY 18-105, CERN-TH-2018-147}}}

\author[a,b]{Gunnar S. Bali}
\author[c]{Peter C. Bruns}
\author[a]{Luca Castagnini}
\author[d,e]{Markus Diehl}
\author[f]{Jonathan R. Gaunt}
\author[g]{Benjamin Gl{\"a}{\ss}le}
\author[a]{Andreas Sch{\"a}fer}
\author[h]{Andr{\'e} Sternbeck}
\author[a]{and Christian Zimmermann}

\affiliation[a]{Institute for Theoretical Physics, University of Regensburg, 93040 Regensburg, Germany}
\affiliation[b]{Department of Theoretical Physics, Tata Institute of Fundamental Research, Homi Bhabha Road, Mumbai 400005, India}
\affiliation[c]{Nuclear Physics Institute, 25069 \v{R}e\v{z}, Czech Republic}
\affiliation[d]{Fachbereich Physik, University of Hamburg, 22761 Hamburg, Germany}
\affiliation[e]{Deutsches Elektronen-Synchroton DESY, 22607 Hamburg, Germany}
\affiliation[f]{CERN Theory Division, 1211 Geneva 23, Switzerland}
\affiliation[g]{Zentrum f\"ur Datenverarbeitung, Universit\"at T\"ubingen, W\"achterstr.\ 76, 72074 T\"ubingen, Germany}
\affiliation[h]{Theoretisch-Physikalisches Institut, Friedrich-Schiller-Universit\"at Jena, 07743 Jena, Germany}

\begin{document}

\maketitle

\input{intro}
\input{theory}
\input{lattice}
\input{data}
\input{graphs}
\input{isospin}
\input{summary}


\section*{Acknowledgements}

We gratefully acknowledge input from Sara Collins and Philipp Wein and discussions with Michael Engelhardt.  \rev{We used a modified version of the Chroma~\cite{Edwards:2004sx} software package, along with the locally deflated domain decomposition solver implementation of openQCD~\cite{Luscher:2012av}. The gauge ensembles have been generated by the QCDSF and RQCD collaborations on the QPACE computer using BQCD~\cite{Nakamura:2010qh,Nakamura:2011cd}. The discrete Fourier transforms in section~\ref{sec:rms} were computed using the package FFTW3 \cite{Frigo:2005}, and the graphs in figures~\ref{fig:chpt} and \ref{fig:contractions} were produced with JaxoDraw~\cite{Binosi:2003yf,Binosi:2008ig}.}
The simulations used for this work were performed with resources provided by the North-German Supercomputing Alliance (HLRN).  This work was supported by the Deutsche Forschungsgemeinschaft SFB/TRR~55.


\addcontentsline{toc}{section}{References}

\bibliographystyle{JHEP}
\bibliography{static}

\end{document}

%% file: prolog.tex
\usepackage[english]{babel}

\usepackage{subfigure}
\usepackage{bbold}    
\usepackage{braket}   

\usepackage{tikz}
\usetikzlibrary{arrows}
\usetikzlibrary{decorations.pathmorphing}
\usetikzlibrary{matrix}
\tikzstyle{snakeline} = [decorate, decoration={snake}]
\tikzstyle{block} = [draw,rectangle,thick]

\newcommand{\half}{{\textstyle\frac{1}{2}}}

\newcommand{\re}{\operatorname{Re}}
\newcommand{\im}{\operatorname{Im}}

\newcommand{\gev}{\operatorname{GeV}}
\newcommand{\mev}{\operatorname{MeV}}

\newcommand{\fm}{\operatorname{fm}}

\newcommand{\ms}{\mskip 1.5mu}
\newcommand{\bs}{\mskip -1.5mu}

\newcommand{\msbar}{\scalebox{0.6}{$\overline{\mathrm{MS}}$}}

\newcommand{\mvec}[1]{\vec{\mskip 0.5mu #1}\mskip 1.5mu}

\newcommand{\unit}{\mathbb{1}}

\newcommand{\mpi}{m_\pi}

\newcommand{\mat}[4]{\bra{\pi^{#1}} \mathcal{O}_i^{#3}(y)\ms \mathcal{O}_j^{#4}(0) \ket{\pi^{#2}}}

\newcommand{\os}{\langle S S \rangle}
\newcommand{\op}{\langle P P \rangle}
\newcommand{\ov}{\langle V^0 V^0 \rangle}
\newcommand{\oa}{\langle A^0 A^0 \rangle}

%% file: intro.tex
\section{Introduction}
\label{sec:intro}

The study of hadronic form factors has a long history in QCD and remains an active field of study.  Defined from the matrix elements of single currents, form factors contain information about the spatial distribution of charge inside a hadron.  Going beyond this, the matrix elements of \emph{two} currents at different points in space are sensitive to charge \emph{correlations} and thus yield qualitatively new information about QCD bound states \rev{and their many-body structure.}

Hadronic matrix elements of one or two currents can be calculated in lattice QCD.  \rev{In fact, it was realised long ago that two-current correlators on the lattice can be regarded as gauge invariant probes of hadron wave functions.  Following earlier work
\cite{Barad:1984px,Barad:1985qd,Wilcox:1986ge,Wilcox:1986dk,Wilcox:1990zc,Chu:1990ps,Lissia:1991gv,Burkardt:1994pw},} detailed computations for the vector, scalar and pseudoscalar currents were presented in \cite{Alexandrou:2002nn,Alexandrou:2003qt} and extended studies for the vector current in~\cite{Alexandrou:2008ru}.  \rev{These studies focused on  a broad range of physics and observables, such as confinement \cite{Barad:1984px,Barad:1985qd}, the size of hadrons \cite{Wilcox:1986ge,Wilcox:1986dk,Wilcox:1990zc,Burkardt:1994pw}, comparison with quark models \cite{Lissia:1991gv}, or the non-spherical shape of hadrons with spin $1$ or larger \cite{Alexandrou:2002nn,Alexandrou:2003qt,Alexandrou:2008ru}.}

The computation of two-current correlators on the lattice involves many different Wick contractions of the quark fields, including disconnected graphs and graphs with all-to-all propagators.  This presents major challenges, for the sheer amount of calculations and for obtaining statistically significant results.  Whilst the work in \cite{Barad:1984px,Barad:1985qd,Wilcox:1986ge,Wilcox:1986dk,Wilcox:1990zc,Chu:1990ps,Lissia:1991gv,Burkardt:1994pw,Alexandrou:2002nn,Alexandrou:2003qt,Alexandrou:2008ru} focused on the graph denoted by $C_1$ in figure~\ref{fig:contractions}, we study all relevant contractions for a meson here (leaving the case of baryons for future work).  We also extend the set of currents mentioned above by the axial current.  Last but not least, increased computing power has allowed us to use larger lattices and a smaller pion mass than previous studies.  Indeed, we will see that finite volume effects can be appreciable for the quantities we are interested in.

At large distances between the two currents, the correlation functions we consider can be evaluated in chiral perturbation theory, \rev{which describes the low-energy limit of QCD in terms of mesons and their interactions.}  A leading-order calculation, supplemented by an estimate of higher orders using resonance exchange graphs, has been performed in \cite{Bruns:2015yto}, and we will compare our lattice results with these predictions.

An extension of the work presented here allows us to compute matrix elements that, as explained in \cite{Diehl:2011yj}, can be connected with \rev{the Mellin moments of double parton distributions.  These distributions} are necessary to compute coherent hard scattering on two spatially separated partons inside a hadron.  \rev{An additional challenge in this case is that one must compute correlation functions for all vector or tensor components of the inserted currents and then extract their twist-two part.}  Corresponding results for the pion will be presented in a forthcoming paper.  The ultimate goal is to extend such studies to double parton distributions in the nucleon, which is of acute interest for the precise description of high-multiplicity final states at the LHC and at possible future hadron colliders.

This paper is organised as follows.  In section \ref{sec:theory}, we derive a number of general properties of two-current matrix elements of the pion and recall the predictions of chiral perturbation theory relevant to our study.  Properties of the different Wick contractions and their computation on the lattice are described in section~\ref{sec:lattice}.  The general quality of our data and the presence of lattice artefacts is investigated in section~\ref{sec:data}.  Results for individual lattice contractions are shown in section~\ref{sec:graphs}, including the quark mass dependence and the relevance of correlations between the two currents.  Physical matrix elements are presented in section~\ref{sec:matrix-elms} and compared with chiral perturbation theory.  A summary of our work is given in section~\ref{sec:summary}.

%% file: theory.tex
\section{Correlation functions of two currents}
\label{sec:theory}

The object of our study are correlation functions of two currents in a pion,
\begin{align}
\label{basic-matel}
\bra{\pi^k(p)} \mathcal{O}^{q_1 q_2}_i(y)\ms
  \mathcal{O}^{q_3 q_4}_j(0) \ket{\pi^{k'}(p)} \,,
\end{align}
where the pion charges $k, k' = +,-,0$ may differ between the bra and ket state, whereas the four-momentum $p$ is the same for both.  The matrix elements \eqref{basic-matel} are understood to be fully connected, with disconnected contributions like $\braket{\pi | \pi} \cdot \bra{0} \mathcal{O}_i(y)\ms \mathcal{O}_j(0) \ket{0}$ or $\bra{\pi} \mathcal{O}_i(y) \ket{0} \cdot \bra{0}
  \mathcal{O}_j(0) \ket{\pi}$ removed.   The currents we consider are
\begin{align}
\label{op-def}
\mathcal{O}^{q q'}_i(y) &= \bar{q}(y)\ms \Gamma_i \ms q'(y) \,,
\end{align}
where $q$ and $q'$ are $u$ or $d$ quark fields.  The full set of Dirac matrices $\Gamma_i$ corresponds to the scalar, pseudoscalar, vector, axial and tensor currents:
\begin{align}
\label{curr-def}
S_{q q'} &= \bar{q} \ms q' \,,
&
P_{q q'} &= i \ms \bar{q} \ms \gamma_5 \ms q' \,,
&
V_{q q'}^\mu &= \bar{q} \gamma^\mu\ms q' \,,
&
A_{q q'}^\mu &= \bar{q} \gamma^\mu \gamma_5\ms q' \,,
&
T_{q q'}^{\mu\nu}   &= \bar{q} \sigma^{\mu\nu} \ms q' \,.
\end{align}
Note that for equal quark flavours all currents are hermitian.

In the present work, we investigate lattice data for correlation functions of two equal currents $V^0$, $A^0$, $S$ or $P$.  In a pion at rest, these currents are associated with the vector, axial, scalar or pseudoscalar charges.  We abbreviate the corresponding matrix elements \eqref{basic-matel} as
\begin{align}
\label{stat-curr}
\ov, \oa, \os, \op,
\end{align}
respectively.  For later use, we shall however consider the full set of currents \eqref{op-def} in the theoretical exposition below.

The matrix elements $\os$ and $\op$ are boost invariant and thus depend only on the products  $y^2$ and $y p$ of four-vectors, given that
$p^2 = m_\pi^2$ is fixed.  On the lattice, we can compute the matrix elements for $y^0 = 0$ and for different pion momenta.  The rotation to Euclidean time and the method to extract hadronic matrix elements from a lattice calculation single out a particular reference frame.  It is hence a valuable check for lattice artefacts to verify that for a given $\mvec{y}^2$, the matrix elements with scalar and pseudoscalar currents are independent of the pion momentum when $\mvec{y} \cdot \mvec{p} = 0$, a condition that can be achieved for both zero and nonzero $\mvec{p}$.

The operators \eqref{op-def} transform under charge conjugation ($C$) and under the
combination of parity and time reversal ($PT$) as
\begin{align}
\mathcal{O}_{i}^{q q'}(y)
 & \, \underset{C}{\to} \, \eta_{C}^i\, \mathcal{O}_{i}^{q'\bs q}(y) \,,
&
\mathcal{O}_{i}^{q q'}(y)
 & \, \underset{PT}{\to} \, \eta_{PT}^i\, \mathcal{O}_{i}^{q'\bs q}(-y) \,,
\end{align}
with sign factors
\begin{align}
  \label{C-parity}
\eta_{C}^{i}
  &= +1 & ~\text{for}~ i &= S, P, A,
&
\eta_{C}^{i}
  &= -1 & ~\text{for}~ i &= V, T,
\intertext{and}
  \label{PT-parity}
\eta_{PT}^{i}
  &= +1 & ~\text{for}~ i &= S, P, V,
&
\eta_{PT}^{i}
  &= -1 & ~\text{for}~ i &= A, T.
\end{align}
We also define the products $\eta_C^{ij} = \eta_C^{i \phantom{j}\!\!}\ms \eta_C^j$ and $\eta_{PT}^{ij} = \eta_{PT}^{i \phantom{j}\!\!}\ms \eta_{PT}^{j}$.


\subsection{Isospin decomposition and constraints}
\label{sec:isospin-dec}

The matrix elements \eqref{basic-matel} are not all independent due to constraints from isospin symmetry and from the discrete symmetries just discussed.  To exploit isospin symmetry, it is convenient to use linear combinations with isospin $0$ and $1$,
\begin{align}
\mathcal{O}^{\text{s}}_i
  &= \mathcal{O}^{uu}_i + \mathcal{O}^{dd}_i \,,
&
\mathcal{O}^{\text{ns}}_i
  &= \mathcal{O}^{uu}_i - \mathcal{O}^{dd}_i \,,
\end{align}
as well as the isotriplet current
\begin{align}
\label{iso-current}
\mathcal{O}^a_i &= \bar{Q}\ms \tau^a \Gamma_i \ms Q \,, & a &= 1,2,3
\end{align}
with the Pauli matrices $\tau^a$ and the isodoublet $Q = (u, d)$ of quark fields.  One then has
\begin{align}
  \label{iso-operators}
\mathcal{O}^{\text{ns}}_i &= \mathcal{O}^3_i \,,
&
\mathcal{O}^{ud}_i
   &= \bigl( \mathcal{O}^1_i + i \mathcal{O}^2_i \bigr) /{2} \,,
&
\mathcal{O}^{du}_i
   &= \bigl( \mathcal{O}^1_i - i \mathcal{O}^2_i \bigr) /{2} \,.
\end{align}
Expressing the pion states in the isospin basis, we have
\begin{align}
  \label{iso-states}
|\pi^+\rangle &= \bigl( |\pi^1\rangle + i |\pi^2\rangle \bigr) /\sqrt{2} \,,
&
|\pi^-\rangle &= \bigl( |\pi^1\rangle - i |\pi^2\rangle \bigr) /\sqrt{2} \,,
&
|\pi^0\rangle &= |\pi^3\rangle \,.
\end{align}
We can now decompose
\begin{align}
  \label{iso-decomp}
\mat{d}{c}{\text{s}}{\text{s}} &= \delta^{cd}\ms F_0(y) \,,
\nonumber \\
\mat{d}{c}{a}{b} &= \delta^{ab} \delta^{cd} \ms F_1(y)
+ \bigl( \delta^{ac} \delta^{bd} + \delta^{ad} \delta^{bc} \bigr) \ms
         F_2(y)
+ \bigl( \delta^{ac} \delta^{bd} - \delta^{ad} \delta^{bc} \bigr) \ms
         i F_3(y) \,,
\nonumber \\
\mat{d}{c}{b}{\text{s}} &= i\epsilon^{bcd}\ms G_1(y) \,,
\nonumber \\
\mat{d}{c}{\text{s}}{b} &= i\epsilon^{bcd}\ms G_2(y) \,,
\end{align}
where for brevity we have suppressed the dependence of $F_n$, $G_n$ on the pion momentum $p$ and on the indices $i,j$ that specify the operators.  Taking the complex conjugate of \eqref{iso-decomp}, one readily finds that the isospin amplitudes $F_n$ and $G_n$ are real valued.

For the matrix elements in the quark flavor basis, we obtain
\begin{align}
\mat{+}{+}{\text{s}}{\text{s}}
  = \mat{0}{0}{\text{s}}{\text{s}} &= F_0 \,,
\nonumber \\
\label{mixed-res}
\mat{+}{+}{\text{ns}}{\text{s}}
  = - \sqrt{2}\ms \mat{0}{+}{du}{\text{s}} &= G_1 \,,
\nonumber \\
\mat{0}{0}{\text{ns}}{\text{s}} &= 0
\end{align}
and analogous relations for the matrix elements parameterised by $G_2$, now also suppressing the $y$-dependence of $F_n$ and $G_n$.
For matrix elements with two isotriplet operators, we have
\begin{align}
\mat{+}{+}{\text{ns}}{\text{ns}}
 &= 2\ms \mat{0}{0}{ud}{du} = 2\ms \mat{0}{0}{du}{ud} = F_1 \,,
\nonumber \\
\mat{0}{0}{\text{ns}}{\text{ns}} &= F_1 + 2 F_2 \,,
\nonumber \\
\mat{-}{+}{du}{du} &= F_2 \,,
\nonumber \\
\mat{+}{+}{ud}{du} &= ( F_1 + F_2 - i F_3 ) /2 \,,
\nonumber \\
\mat{+}{+}{du}{ud} &= ( F_1 + F_2 + i F_3 ) /2  \,,
\nonumber \\
\mat{0}{+}{\text{ns}}{du} &= ( F_2 - i F_3 ) /\sqrt{2} \,,
\nonumber \\
\mat{0}{+}{du}{\text{ns}} &= ( F_2 + i F_3 ) /\sqrt{2} \,.
\end{align}

From \eqref{iso-decomp} one also finds that the matrix elements with two isosinglet or with two isotriplet operators are even under the simultaneous exchange\footnote{According to
  \eqref{iso-operators} and \eqref{iso-states} this corresponds to
  changing the sign of $|\pi^a\rangle$ and $\mathcal{O}^a_i$ for $a=2$ in
  the isospin basis.}
\begin{align}
  \label{c-replacements}
|\pi^+\rangle & \leftrightarrow |\pi^-\rangle \,,
&
\mathcal{O}^{du}_i & \leftrightarrow \mathcal{O}^{ud}_i \,,
\end{align}
whereas the matrix elements with one isosinglet and one isotriplet operator are odd under that exchange.  The charge conjugate of a matrix element is obtained by applying \eqref{c-replacements} and multiplying with the product $\eta_C^{ij}$ of intrinsic $C$ parities of the two operators.  One readily sees that $G_n = 0$ if $\eta_C^{ij} = +1$ and that $F_n = 0$ if $\eta_C^{ij} = -1$.
The charge conjugation constraints give relations between matrix elements in the quark flavor basis.  For $\eta_C^{ij} = +1$ the vanishing of $G_1$ and $G_2$ implies in particular
\begin{align}
\mat{k}{k}{uu}{uu} &= \mat{k}{k}{dd}{dd} \,,
\nonumber \\
\mat{k}{k}{uu}{dd} &= \mat{k}{k}{dd}{uu}
\intertext{and thus}
\half\ms \mat{k}{k}{\text{s}}{\text{s}}
  &= \mat{k}{k}{uu}{uu} + \mat{k}{k}{uu}{dd} \,,
\nonumber \\
\half\ms \mat{k}{k}{\text{ns}}{\text{ns}}
  &= \mat{k}{k}{uu}{uu} - \mat{k}{k}{uu}{dd}
\end{align}
for $k= +,-,0$.


\subsection{Predictions from chiral perturbation theory}
\label{sec:chpt-pred}

\begin{figure*}
\begin{center}
\includegraphics[width=0.95\textwidth]{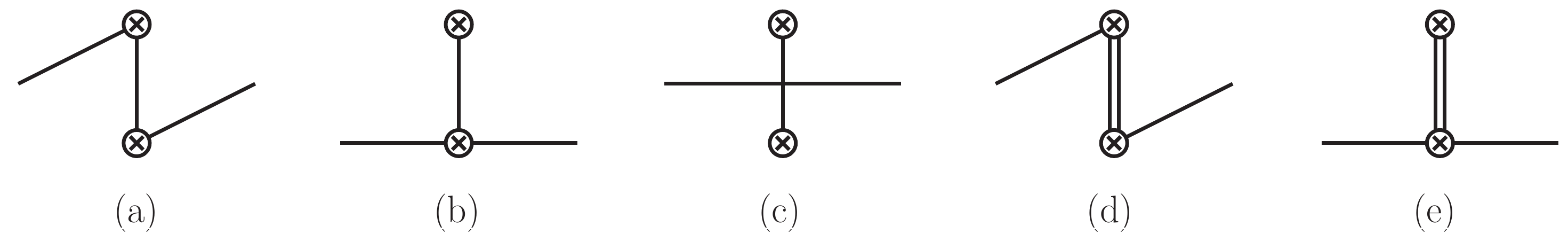}
\end{center}
\caption{\label{fig:chpt} Graphs for the matrix elements $\os, \op, \ov, \oa$ in chiral perturbation theory.  Single lines denote pions, double lines resonances, and crossed circles indicate current insertions.  The leading-order chiral Lagrangian gives rise to graph (a) for $\os, \ov$ and to graphs (b) and (c) for $\op, \oa$.  The resonance exchange graphs (d) and (e) can contribute to any of the four matrix elements, depending on the quantum numbers of the resonance.}
\end{figure*}

Consider pions at rest and take the matrix elements $\os, \op, \ov$ and $\oa$ at $y^0 = 0$ and large $|\mvec{y}|$.  These can then be computed in chiral perturbation theory.  The chiral expansion, on which this theory is based, requires the pion mass and momenta $p$ to be much smaller than $4 \pi F \sim 1 \gev$, where $F$ is the pion decay constant.  In position space, one should hence require $|\mvec{y}| \gg 0.2 \fm$.  At leading order in the chiral expansion, the matrix elements can be computed from the tree-level graphs in figure~\ref{fig:chpt}(a), (b) and (c).  The corresponding calculation is detailed in  \cite{Bruns:2015yto}.  As an estimate for higher-order contributions, the same work evaluated the resonance exchange graphs (d) and (e) from the appropriate leading-order Lagrangian in the approximation of vanishing pion four-momentum (thus setting $\mpi$ to zero).  Resonance exchange graphs with the topology of figure~\ref{fig:chpt}(c) would involve a vertex between pions and resonances, which vanishes in this approximation.  Considered were the lowest-lying resonances $\rho$, $a_1$, $a_0$, $\eta$ and $\sigma$.

The corresponding results are given in sections~II.A and~III.B of \cite{Bruns:2015yto}.  For convenience, we recast them into the notation used here, making use of the simplifications that arise when setting $y^0 =0$.  The isospin amplitudes $F_i$ defined here are related to those in \cite{Bruns:2015yto} as $F_0 = C^{00}$, $F_1 = C_1$, $F_2 = C_1 + C_2$ and $i F_3 = C_3$.  We also note that the isospin currents $V^a_\mu$ and $A^a_\mu$ in \cite{Bruns:2015yto} have an extra factor of $1/2$ compared with our currents~\eqref{iso-current}.

Vector and axial currents were only considered for the isovector case in \cite{Bruns:2015yto}, so that no predictions are available for $F_0$ in the  $\ov$ and $\oa$ channels.  Under the conditions stated above, the isotriplet amplitude $F_3$ is found to vanish for all channels $\os$, $\op$, $\ov$, $\oa$.

For the remaining matrix elements, we obtain the following results from \cite{Bruns:2015yto}, writing $y = |\mvec{y}|$ for simplicity.  The leading order chiral Lagrangian gives rise to the amplitudes
\begin{align}
F_{0\ms (\mathrm{LO})}^{{SS}} &= \frac{2 B^2 \mpi}{\pi^2 y}\, K_1(\mpi\ms y) \,,
&
F_{0\ms (\mathrm{LO})}^{{PP}} &= 0 \,,
\nonumber \\
F_{1\ms (\mathrm{LO})}^{{SS}} &= 0 \,,
&
F_{1\ms (\mathrm{LO})}^{{PP}}
  &= - \frac{B^2 \mpi^2}{2\pi^2}\, K_0(\mpi\ms y) \,,
\nonumber \\
F_{2\ms (\mathrm{LO})}^{{SS}} &= 0 \,, \phantom{\frac{1}{1}}
&
F_{2\ms (\mathrm{LO})}^{{PP}}
  &= \frac{B^2 \mpi^2}{2 \pi^2}
     \left[ K_0(\mpi y) - \frac{2 K_1(\mpi\ms y)}{\mpi\ms y} \right]
\end{align}
and
\begin{align}
F_{1\ms (\mathrm{LO})}^{{VV}}
  &= \frac{2 \mpi^3}{\pi^2 y}
     \left[ K_1(\mpi\ms y) - \frac{K_2(\mpi\ms y)}{\mpi\ms y} \right] \,,
\nonumber \\
F_{1\ms (\mathrm{LO})}^{{AA}}
  &= \frac{\mpi^3}{2 \pi^2 y}
     \left[ K_1(\mpi\ms y) + \frac{4 K_2(\mpi\ms y)}{\mpi\ms y} \right] \,,
\nonumber \\
F_{2\ms (\mathrm{LO})}^{{VV}}
  &= -\frac{1}{2} F_{1\ms (\mathrm{LO})}^{\mathrm{VV}} \,,
  \phantom{ \left[ \frac{0}{0} \right] }
\nonumber \\
F_{2\ms (\mathrm{LO})}^{{AA}}
  &= - \frac{\mpi^3}{2 \pi^2 y}
     \left[ K_1(\mpi\ms y) + \frac{2 K_2(\mpi\ms y)}{\mpi\ms y} \right] \,,
\end{align}
where $K_n$ denotes the Macdonald functions (modified Bessel functions of the second kind).  The pion decay constant $F$ and the chiral symmetry breaking parameter $B$ are defined as usual in chiral perturbation theory, see section~I.A in \cite{Bruns:2015yto}.  The resonance exchange contributions read
\begin{align}
F_{0\ms (\mathrm{R})}^{{SS}}
  &= - \frac{32 B^2 (c_m^\sigma)^2\, m_\sigma^{}}{\pi^2 F^2\ms y}\,
     K_1(m_\sigma\ms y) \,,
\nonumber \\[0.2em]
F_{0\ms (\mathrm{R})}^{{PP}}
  &= \frac{16 B^2}{\pi^2 F^2\ms y} \left[ c_m^2\ms m_{a_0}^{}\, K_1(m_{a_0}\ms y)
     - 2\ms d_\eta^2\ms m_\eta^{}\, K_1(m_\eta\ms y) \right] \,,
\nonumber \\[0.2em]
F_{1\ms (\mathrm{R})}^{{SS}} &= 0 \,,
\qquad \quad
F_{1\ms (\mathrm{R})}^{{PP}} = 0 \,,
\qquad \quad
F_{2\ms (\mathrm{R})}^{{SS}}
   = - \frac{1}{2}\ms F_{0\ms (\mathrm{R})}^{{PP}} \,,
\qquad \quad
F_{2\ms (\mathrm{R})}^{{PP}}
   = - \frac{1}{2}\ms F_{0\ms (\mathrm{R})}^{{SS}} \,,
\nonumber \\[0.2em]
F_{1\ms (\mathrm{R})}^{{VV}} &= - F_{1\ms (\mathrm{R})}^{{AA}}
   = \sum_{X = \rho,\ms a_1} \! s_X \, \frac{2 f_X^2\ms m_X^5}{\pi^2 F^2\ms y}
     \left[  K_1(m_X\ms y) + \frac{K_2(m_X\ms y)}{m_X\ms y} \right] \,,
\nonumber \\[0.1em]
F_{2\ms (\mathrm{R})}^{{VV}} &= - F_{2\ms (\mathrm{R})}^{{AA}}
  = -\frac{1}{2}\ms F_{1\ms (\mathrm{R})}^{{VV}}
\end{align}
with signs $s_\rho = 1, s_{a_1} = -1$.
In the numerical evaluation of section~\ref{sec:chiral-comp}, we will take the resonance parameters estimated in \cite{Bruns:2015yto}:
\begin{align}
m_\rho &= 0.8 \gev \,,  & f_\rho &= 0.2 \,,
& & &
m_{a_1} &= 1.25 \gev \,, & f_{a_1} &= 0.1 \,,
\nonumber \\
m_{a_0} &= 1 \gev \,,    & c_m &= 50 \mev \,,
& & &
m_{\eta} &= 600 \mev \,, & d_\eta &= 15 \mev \,,
\nonumber \\
m_\sigma^{} &= 0.5 \gev \,, & c_m^\sigma &= 35 \mev \,.
\end{align}
For the parameters in the chiral sector we will take $m_\pi = 300 \mev$, $F = 100 \mev$ and $B = 2.4 \gev$, which are rounded values of what we extract from our lattice simulations with $L=40$, see \eqref{chiral-params}.

%% file: lattice.tex
\section{Lattice techniques}
\label{sec:lattice}

The lattice computation of the matrix elements \eqref{basic-matel} involves a considerable number of different Wick contractions between the quark fields in the two currents and in the pion source and sink.  This is to be contrasted with, e.g., the computation of single-current matrix elements, where there is only one connected and one disconnected graph.  In the present section, we give details about the lattice contractions for two-current correlators, their relation with the physical pion matrix elements, and their implementation in the lattice simulation.


\subsection{Lattice contractions and their symmetry properties}
\label{sec:symmetries}

\begin{figure*}
\begin{center}
\includegraphics[width=0.9\textwidth]{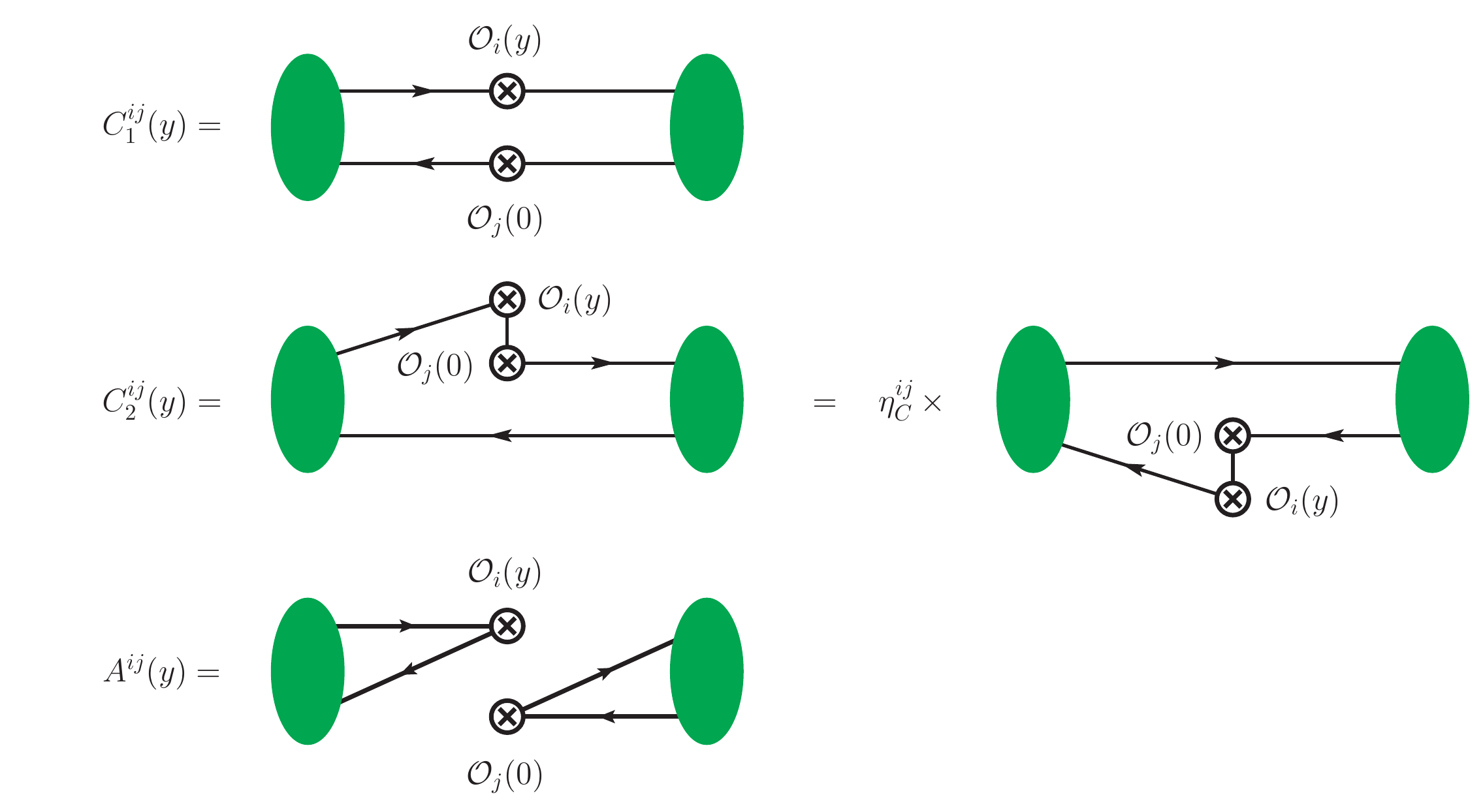} \\[1em]
\includegraphics[width=0.9\textwidth]{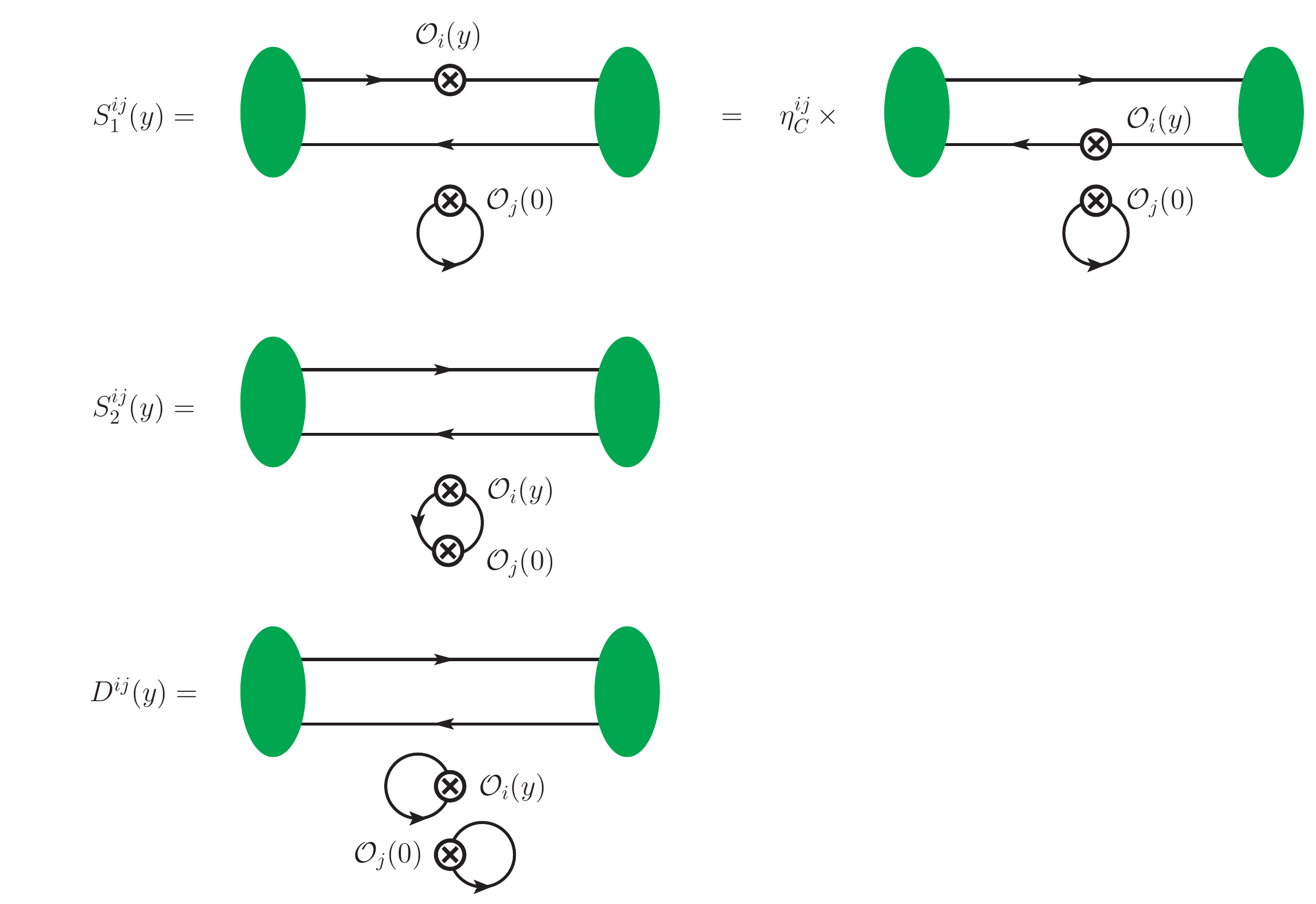}
\end{center}
\caption{\label{fig:contractions} Lattice contractions for the two-current correlation function \protect\eqref{basic-matel} in a pion.  The dependence on the pion momentum $p$ is not indicated for brevity.  The product $\smash{\eta_C^{ij}}$ of charge conjugation parities of the two operators is defined below \protect\eqref{PT-parity}.}
\end{figure*}

The different lattice contractions are pictorially represented in figure~
\ref{fig:contractions}.  We have two fully connected graphs $C_1$ and $C_2$, a graph $A$ in which the pion in annihilated by one current and created again by the second current, two graphs $S_1$ and $S_2$ with one disconnected quark loop, and a doubly disconnected graph $D$ with two quark loops.

The symbols $C_1^{ij}(y)$, $C_2^{ij}(y)$, etc.\ denote contributions to the physical matrix elements \eqref{basic-matel}, i.e.\ it is understood that
\begin{itemize}
\item the contractions shown are averaged over the gauge ensemble and divided by the ensemble average of the pion two-point function,
\item the two currents are inserted at the same time slice $\tau$ (i.e.\ $y^0 = 0$), and the ratio of four-point and two-point functions is evaluated at a plateau in $\tau$.  To use time reversal invariance, the plateau must be symmetric between source and sink times,
\item both source and sink are projected on definite momentum $p$.  By translation invariance one can thus shift their spatial positions by a common amount.
\end{itemize}
Details are given in section~\ref{subsec:sim_details} below.  For brevity, we do not indicate the dependence of $C_1^{ij}(y)$ etc.\ on the pion momentum $p$.

From translation invariance one readily finds
\begin{align}
  \label{transl-rel}
M^{ij}(y) &= M^{ji}(-y) && \text{for $M = S_2, D$} .
\intertext{Charge conjugation invariance gives}
  \label{C-relation}
M^{ij}(y) &= \eta_C^{ij}\, M^{ji}(-y) && \text{for $M = C_1$} .
\end{align}
If $\eta_C^{ij} = -1$ one furthermore obtains $M^{ij}(y) = 0$ for $M = A, S_2, D$.  From $PT$ invariance one gets
\begin{align}
  \label{PT-relations-1}
M^{ij}(y) &= \eta_{PT}^{ij}\ms M^{ij}(-y)
   && \text{for $M = C_1, S_1, S_2, D$} ,
\\
  \label{PT-relations-2}
M^{ij}(y) &= \eta_{PT}^{ij}\ms M^{ji}(y)
   && \text{for $M = A, C_2$} ,
\end{align}
where for the formulation of time reversal in the Euclidean path integral we refer to \cite{Hasenfratz:2005ch}.

Finally, a useful relation for the contractions is obtained by taking the
complex conjugate of the contraction, followed by a parity transformation and charge conjugation.  This gives
\begin{align}
  \label{herm-rel}
\bigl[ M^{ij}(y) \bigr]^* &= \eta_{PT}\ms M^{ij}(-y)
   && \text{for all contractions.}
\end{align}
Combining \eqref{PT-relations-1} with \eqref{herm-rel} we find that $C_1, S_1, S_2$ and $D$ are real valued, whereas combining \eqref{PT-relations-2} with
\eqref{herm-rel} gives
\begin{align}
  \label{re-im-parts}
\re M^{ij}(y) &= \half \bigl[ M^{ij}(y) + M^{ji}(-y) \bigr] \,,
&
i \im M^{ij}(y) &= \half \bigl[ M^{ij}(y) - M^{ji}(-y) \bigr]
\end{align}
for $M = A, C_2$.


\subsection{Physical matrix elements}
\label{sec:phys-matels}

The form of the relation between pion matrix elements and lattice contractions depends on the product of $C$ parities.  Using the shorthand notation
\begin{align}
\label{contr-shorthand}
C_1^{} &= C_1^{ij}(y) \,,
&
C_2^{} &= \half \bigl[ C_2^{ij}(y) + C_2^{ji}(-y) \bigr] \,,
&
A &= \half \bigl[ A^{ij}(y) + A^{ji}(-y) \bigr] \,,
\nonumber \\[0.2em]
S_1^{} &= \half \bigl[ S_1^{ij}(y) + S_1^{ji}(-y) \bigr] \,,
&
S_2^{} &= S_2^{ij}(y) \,,
&
D &= D^{ij}(y) \,,
\end{align}
we have for $\eta_C^{ij} = +1$
\begin{align}
\label{phys-matels}
\mat{+}{+}{uu}{dd} &= C_1 + \bigl[ 2 S_1 + D \bigr] \,,
\nonumber \\
\mat{+}{+}{uu}{uu} &=  \bigl[ 2 C_2 + S_2 \bigr]
   + \bigl[ 2 S_1 + D \bigr] \,,
\nonumber \\
\mat{0}{0}{uu}{dd} &= \bigl[ 2 S_1 + D \bigr] - A \,,
\nonumber \\
\mat{0}{0}{uu}{uu} &=  C_1 + \bigl[ 2 S_1 + D \bigr]
  + \bigl[ 2 C_2 + S_2 \bigr] + A \,,
\nonumber \\
\mat{0}{0}{du}{ud} &= {}- C_1 + \bigl[ 2 C_2 + S_2 \bigr] \,,
\nonumber \\
\mat{-}{+}{du}{du} &= 2 C_1 + 2 A \,,
\nonumber \\
\mat{+}{+}{du}{ud} &= 2 C_2^{ij}(y) + S_2^{} + A^{ij}(y) \,,
\nonumber \\
\sqrt{2}\ms \mat{0}{+}{du}{uu} &= C_1^{}
   + \bigl[ C_2^{ij}(y) - C_2^{ji}(-y) \bigr] + A^{ij}(y) \,.
\end{align}
One can verify that this satisfies the isospin relations in section~\ref{sec:isospin-dec}.  We have three independent real valued combinations, which can for instance be chosen as
\begin{align}
\label{real-combs}
\half F_0(y) &= \mat{+}{+}{uu}{uu} + \mat{+}{+}{uu}{dd}
\nonumber \\
& = C_1 + 2 \ms \bigl[ 2 S_1 + D \bigr] + [2 C_2 + S_2] \,,
\nonumber \\[0.2em]
\half F_1(y) &= \mat{+}{+}{uu}{uu} - \mat{+}{+}{uu}{dd}
\nonumber \\
& = {}- C_1 + \bigl[ 2 C_2 + S_2 \bigr] \,,
\nonumber \\[0.2em]
\half F_2(y) &= \half \mat{-}{+}{du}{du}
\nonumber \\
& = C_1 + A \,,
\end{align}
where $F_{0}$, $F_{1}$ and $F_{2}$ are the isospin amplitudes defined in \eqref{iso-decomp}. These combinations are real valued owing to \eqref{re-im-parts}, as the isospin amplitudes must be.  The imaginary combination  $i F_3$ of matrix elements can e.g.\ be isolated by
\begin{align}
i F_3(y) &= \mat{+}{+}{du}{ud} - \mat{+}{+}{ud}{du}
\nonumber \\
 &= 2 \bigl[ C_2^{ij}(y) - C_2^{ji}(-y) \bigr]
    + \bigl[ A^{ij}(y) - A^{ji}(-y) \bigr] \,.
\end{align}
In the present work, we only study the real valued combinations in \eqref{real-combs}.

For $\eta_C^{ij} =-1$ we have three nonzero lattice contractions, $C_1$, $C_2$ and $S_1$, and two
independent matrix elements, which may be taken as
\begin{align}
\mat{+}{+}{uu}{dd} &= C_1^{} + S_1^{ij}(y) - S_1^{ji}(y) \,,
\nonumber \\
\mat{+}{+}{uu}{uu} &= 2 C_2^{} + S_1^{ij}(y) + S_1^{ji}(y) \,.
\end{align}

Among all contractions, $C_1$ has typically the smallest statistical uncertainties in the lattice simulation and thus plays a special role.  The above equations show that it does not appear in isolation in any physical pion matrix element.  However, $C_1$ can be regarded as ``approximately physical'' in the following sense.  Consider QCD with $n_F=4$ (instead of $n_F=2$) mass degenerate quarks, where SU(4) flavour symmetry is exact.  It is easy to see that in this theory
\begin{align}
\label{eta-matel}
\bra{\pi^{+}} \mathcal{O}_i^{uc}(y)\ms \mathcal{O}_j^{sd}(0) \ket{D_s^+}
  &= C_1^{ij}(y) \,,
\end{align}
where the quark content of the $D_s^+$ meson is $c \bar{s}$.  The correlation function $C_1$ computed in our study does not exactly correspond to this matrix element, because our lattice action has $n_F=2$ and not $n_F=4$ sea quark flavours.  We therefore must interpret \eqref{eta-matel} within the partially quenched approximation.


\subsection{Simulation Details}
\label{subsec:sim_details}

\paragraph{Lattice action and quark mass values.}

We have performed $N_f=2$ lattice simulations with the Wilson gauge action and non-perturbatively improved Sheikholeslami-Wohlert (NPI Wilson-clover) fermions. The gauge configurations were generated by the RQCD and QCDSF collaborations.  As is discussed for instance in \cite{Bali:2014nma,Bali:2014gha}, there are eleven standard ensembles with pion masses down to $150 \mev$.  Two of these are used in the present study, namely ensembles IV and V, with a reduced number of gauge configurations as indicated in table \ref{tab_1}.

\begin{table*}
\begin{center}
\renewcommand{\arraystretch}{1.2}
\begin{tabular}{c|cccccccccccc} \hline \hline
ensemble & $\beta$ &  $a \, [\fm]$ & $\kappa$ & $L^3 \times T$  &
$m_{\pi} \, [\mev]$ & $L m_{\pi}$ & $N_{\text{full}}$ & $N_{\text{used}}$ &
$N_{\text{sm}}$ \\
\hline
IV & 5.29 & 0.071 & 0.13632 &   $32^3\times 64$  &  294.6(14) &  3.42   &
 2023 &  960 & 400  \\
 V & 5.29 & 0.071 & 0.13632 &   $40^3\times 64$  &  288.8(11) &  4.19   &
 2025 & 984 & 400  \\ \hline \hline
\end{tabular}
\end{center}
\caption{\label{tab_1} Details of the ensembles used in this analysis.  $N_{\text{full}}$ is the total number of available gauge configurations and $N_{\text{used}}$ the number of configurations used in the present study.  $N_{\text{sm}}$ indicates the number of Wuppertal smearing iterations in the pion source.  The error on the pion mass combines statistical and systematic errors, see \protect\cite{Bali:2014nma}.  The time difference between pion source and sink is $t = 15 a$ or $t = 32 a$ for both ensembles.}
\end{table*}

We will also study the dependence of the pion matrix elements \eqref{basic-matel} on the quark mass.  To this end, we have performed simulations with ensemble V and  different values of $\kappa$ in the quark propagator, namely
\begin{align}
\label{quark-masses}
& \text{light quarks:}   & \kappa &= 0.13632  & a m_q = 0.00291(3) \,,
\nonumber \\
& \text{strange:} & \kappa &= 0.135616 & a m_q = 0.02195(3) \,,
\nonumber \\
& \text{charm:}   & \kappa &= 0.125638 & a m_q = 0.31475(3) \,,
%
\end{align}
where we also give the values of the bare quark masses $m_q$ in lattice units.  Here ``light quarks'' refers to the $\kappa$ value of ensemble V, whereas the other two values correspond to the physical strange and charm quark masses, as determined in \cite{Bali:2016lvx} and \cite{Bali:2017pdv} by tuning the pseudoscalar ground state mass to $685.8 \mev$ in the first case and the spin-averaged $S$-wave charmonium mass to $3068.5 \mev$ in the second case.  Since our simulations are performed with an $n_F = 2$ fermion action, the strange and charm quarks are partially quenched.


\paragraph{Correlation functions.}

To compute the pion matrix element in \eqref{basic-matel} on a lattice in Euclidean space-time, we need the four-point correlators
\begin{align}
\label{4pt-function}
C^{ij,\mvec{p}}_{4\mathrm{pt}}(\mvec{y},\tau,t)
 &= a^6 \ms \sum_{\mvec{x}, \mvec{z}} e^{- i \mvec{p} \cdot (\mvec{x} - \mvec{z})}\,
    \bra{0} \Pi(\mvec{x},t)\, \mathcal{O}_i(\mvec{y},\tau)\,
    \mathcal{O}_j(\mvec{0},\tau)\, \Pi^{\dagger}(\mvec{z},0) \ket{0} \,,
\end{align}
where $\Pi(x)$ is a pion interpolator.  We use interpolators with Dirac structure $\gamma_5$.  Pion matrix elements are extracted by calculating the following ratio at time slices where excited states should be suppressed:
\begin{equation}
\bra{\pi(p)} \mathcal{O}_i(y)\ms \mathcal{O}_j(0) \ket{\pi(p)}
 = R_{ij}^{\mvec{p}}(\mvec{y})
 = 2 E_{\mvec{p}}\ms V \,
   \frac{C^{ij,\mvec{p}}_{4\mathrm{pt}}(\mvec{y},\tau,t)}{C^{\mvec{p}}_{2\mathrm{pt}}(t)} \; \biggl|_{0 \ll \tau \ll t} \,.
\label{4pt2ptRatio}
\end{equation}
Here $V = L^3\ms a^3$ is the spatial volume and
\begin{align}
C^{\mvec{p}}_{2\mathrm{pt}}(t)
  &= a^6 \ms \sum_{\mvec{x}, \mvec{z}} e^{- i \mvec{p} \cdot (\mvec{x} - \mvec{z})}
      \bra{0} \Pi(\mvec{x},t)\, \Pi^{\dagger}(\mvec{z},0) \ket{0}
\end{align}
the usual pion two-point function.  In analogy to \eqref{4pt-function} and \eqref{4pt2ptRatio}, we use an appropriate ratio of the three-point function $\smash{C^{i,\mvec{p}}_{3\mathrm{pt}}(\tau,t)}$ and $C^{\mvec{p}}_{2\mathrm{pt}}(t)$ to extract the matrix elements $\bra{\pi(p')} \mathcal{O}_i(0) \ket{\pi(p)}$ of the vector and scalar currents.  We thus obtain the vector and scalar form factors for the pion, which are used in sections~\ref{subsec:renorm} and \ref{sec:factorise}.

The pion energy in \eqref{4pt2ptRatio} is computed using the continuum dispersion relation $E_{\mvec{p}} = \sqrt{ \mvec{p}^2 + m_\pi^2 }$, which is well satisfied for the momenta investigated here.  The value of $\mpi$ is obtained from an exponential fit of the two-point function, which gives
\begin{align}
\label{fitted-mpi}
293 \mev & \quad (\text{light}, L = 40) \,, &
299 \mev & \quad (\text{light}, L=32) \,,
\nonumber \\
691 \mev & \quad (\text{strange}, L = 40) \,, &
3018 \mev & \quad (\text{charm}, L = 40) \,.
\end{align}
The statistical errors of the fits are less than 1\%, and since we do not aim at a high-precision analysis, we do not attempt to quantify the uncertainty due to excited states.  We observe that the values in \eqref{fitted-mpi} agree reasonably well with the pion masses given in table \ref{tab_1} for light quarks and quoted after \eqref{quark-masses} for strange quarks.

The source-sink distance is fixed to $t = 15 a \approx 1.07 \fm$ as a default.  To investigate the influence of excited states, we have also calculated graphs $C_1$, $C_2$ and $S_1$ with $t/a = 32$.  To extract the desired matrix elements, we calculate the ratio \eqref{4pt2ptRatio} by fitting or averaging over the plateau around $\tau = t/2$.  For the contractions $C_1$ and $A$, we measure the $\tau$ dependence of $C_{4\mathrm{pt}}$ and fit to a plateau in the $\tau$ ranges specified in \eqref{tau-fit-range}.  For $t/a = 15$ we have $S_2$ data for $\tau/a = 7$ and $8$, which we average, whereas for the remaining contractions $C_2, S_2$ and $D$ we take $\tau/a = 7$ or $8$ for each gauge configuration on a statistical basis.  We recall that the plateau extraction must be symmetric around $\tau = t/2$ in order to respect the time reversal invariance relations of section~\ref{sec:symmetries}.   The data for $C_2$ and $S_1$ with $t/a = 32$ is restricted to $\tau/a = 16$.


\paragraph{Details on the contractions.}

We now give details on the calculation of the different lattice contractions for $C_{4\mathrm{pt}}$.  Let us start with a broad overview, which is pictorially represented in figure~\ref{fig:tech}.  Most graphs are calculated using the one-end trick \cite{Foster:1998vw} at the source, which for $A$ and $C_1$ we are also able to use at the sink.  $C_2$ and $S_1$ have a similar structure, for which the sequential source technique \cite{Martinelli:1988rr} is suitable.  In general, loops are obtained using stochastic insertions, except for the double-insertion loop of $S_2$, where usual point-to-all propagators are used.  The $D$ graph is calculated using point sources only.  For the evaluation of the two-point graph $G_2$ and the connected part $G_3$ of the three-point graph (see figure~\ref{fig:tech}) we alternatively use point sources or the one-end trick.

\begin{figure*}
\begin{center}
\include{graphs/contractions}
\end{center}
\caption{\label{fig:tech} Lattice methods employed for computing the different graphs of the four-point function.  Colours are used to distinguish the different parts of a graph.  Two stochastic sources within a pion source or sink lead always to the application of the one-end trick or two-hand-trick, respectively.}
\end{figure*}

We use stochastic $\mathbb{Z}_2 \otimes i \mathbb{Z}_2$ wall sources, i.e.\ a set of complex random vectors $\eta_{t}^{(\ell)}$ carrying space-time, spin and colour indices (not explicitly written here).  $\eta_{t}^{(\ell)}$ is nonzero only on the time slice $t$, where it has components $(\pm 1 \pm i)/\sqrt{2}$ that are random for each gauge configuration.  Averaged over all realisations, one then has
\begin{align}
\label{stoch-src}
\frac{1}{N^{\mathrm{st}}_{\mathrm{src}}} \sum_{\ell}^{N^{\mathrm{st}}_{\mathrm{src}}}
   \eta_{t}^{(\ell)} \otimes \eta_{t}^{\dagger(\ell)}
 & \to \; \unit_{t} \,,
\end{align}
where the matrix $\unit_t$ is unity if both time indices are equal to $t$ and zero otherwise.  Using this identity as an expression for the pion source represents the one-end trick.

The structure of graphs $C_1$ and $A$ allows us to perform a second one-end trick at the sink, which we refer to as the ``two-hand trick''.  For $C_1$ this trick was applied  already in \cite{Alexandrou:2008ru}.  In a different context it was used in \cite{Yang:2015zja}, where it was dubbed the ``stochastic sandwich method''.  Including the sign of the Wick contraction (easily obtained from the number of fermion loops) we explicitly have\footnote{%
  Note that $C_1^{ij,\mvec{p}}(\mvec{y},\tau,t)$ is a contribution to the four-point correlator $C_{4\mathrm{pt}}$, whereas $C_1^{ij}(y)$ introduced in section~\ref{sec:symmetries} is a contribution to the pion matrix element \protect\eqref{4pt2ptRatio}.  The same holds for the other contractions.}
\begin{align}
\label{C1-lat}
C_1^{ij,\mvec{p}}(\mvec{y},\tau,t)
&= \frac{a^3}{V N^{\mathrm{st}}_{\mathrm{src}}}  \sum_{\ell}^{N^{\mathrm{st}}_{\mathrm{src}}} \sum_{\mvec{x}}
\biggl\langle \left[ \Psi_{t}^{\dagger(\ell),\mvec{0}}(\mvec{x}+\mvec{y},\tau) \ms\gamma_5\ms \Gamma_i \Psi_{0}^{(\ell),-\mvec{p}}(\mvec{x}+\mvec{y},\tau) \right]
\nonumber \\*
 & \quad \times
\left[ \Psi_{0}^{\dagger(\ell),\mvec{0}}(\mvec{x},\tau) \ms\gamma_5\ms \Gamma_j \Psi_{t}^{(\ell),\mvec{p}}(\mvec{x},\tau) \right] \biggr\rangle \,,
\nonumber \\[0.2em]
A^{ij,\mvec{p}}(\mvec{y},\tau,t)  &= -\frac{a^3}{V} \sum_{\mvec{x}} \left\langle B_{0}^{i,-\mvec{p}}(\mvec{x}+\mvec{y},\tau) B_{t}^{j,\mvec{p}}(\mvec{x},\tau) \right\rangle
\nonumber \\
 & \quad
 + \frac{a^3}{V} \sum_{\mvec{x}} \left\langle B_{0}^{i,-\mvec{p}}(\mvec{x}+\mvec{y},\tau) \right\rangle \left\langle B_{t}^{j,\mvec{p}}(\mvec{x},\tau) \right\rangle \,,
\nonumber \\
B_{t}^{j,\mvec{p}}(\mvec{x},\tau)  &= \frac{1}{N^{\mathrm{st}}_{\mathrm{src}}} \sum_{\ell}^{N^{\mathrm{st}}_{\mathrm{src}}} \left[ \Psi_{t}^{\dagger(\ell),\mvec{0}}(\mvec{x},\tau) \ms\gamma_5\ms \Gamma_j\ms \Psi_{t}^{(\ell),\mvec{p}}(\mvec{x},\tau)
\right] \,,
\end{align}
where again $V = L^3\ms a^3$ is the spatial volume.
Here and in the following, the notation $\langle \ldots \rangle$ denotes the average over the gauge ensemble and $[ \ldots ]$ indicates a closed spin-colour structure.  The vector $\Psi^{(\ell),\mvec{p}}_t$ is obtained from an inversion of the Wilson-clover Dirac operator $\mathcal{D}$ on the random source,
\begin{align}
\label{Psi-def}
\mathcal{D}\ms \Psi^{(\ell),\mvec{p}}_{t}
  &= \Phi\ms \mathcal{E}^{\mvec{p}}\ms \eta^{(\ell)}_{t}\,,
\end{align}
where
\begin{align}
\bigl(\, \mathcal{E}^{\mvec{p}} \,\bigr)_{\mvec{x} \bs,\ms \mvec{y}}
  &= e^{-i \mvec{p} \cdot \mvec{x}}\, \delta_{\mvec{x} \mvec{y}}
\end{align}
is a diagonal matrix in the spatial coordinates.  For better legibility, the $(\mvec{x}, t)$-component of $\Psi$ is written as $\Psi(\mvec{x}, t)$ rather than $\Psi_{\mvec{x} t}$ in \eqref{C1-lat}.  We will use the same notation for other quantities that are vectors in space-time.

Source smearing is implemented by $\Phi$, which is a hermitian matrix
acting on spatial coordinates and colour indices and consists of 400 Wuppertal smearing iterations \cite{Gusken:1989qx}.  It turned out that taking $\Phi\ms \mathcal{E}^{\mvec{p}}$ instead of $\mathcal{E}^{\mvec{p}}\ms \Phi$ in \eqref{Psi-def} greatly improves the signal for nonzero momenta. This observation contributed to introducing momentum smearing in \cite{Bali:2016lva}.  The contractions for which we have data with nonzero $\mvec{p}$ are $C_1$ and $A$ on the lattice with $L=40$.  Specifically, $C_1$ was computed for all 24 nonzero momenta satisfying $p^2 \le 3$ and $A$ for all 6 nonzero momenta with $p^2 = 1$.  Here $p$ is given as a multiple of the smallest non-trivial lattice momentum $2\pi /(La)$, which is equal to $437 \mev$ for $L=40$.

As we see in \eqref{C1-lat}, the calculation of $A$ involves the subtraction of vacuum contributions in order to give the fully connected pion matrix element.  For symmetry reasons, these subtractions are only nonzero for the currents $A$ and $P$.

The connected part of graph $S_1$ is obtained by applying the sequential source method either to stochastic sources with the one-end trick (denoted by ``oet'') or to point sources (denoted by ``pt'').  The loop appearing in $S_1$ is calculated by using a stochastic source $\eta_\tau^{(\ell)}$ at time slice $\tau$ with the corresponding solution $\chi_\tau^{(\ell)}$ of the Dirac equation,
\begin{equation}
\mathcal{D} \chi_\tau^{(\ell)} = \eta_\tau^{(\ell)} \,.
\label{eq:stoch_src_sol}
\end{equation}
Specifically, we have
\begin{align}
\label{S1-lat}
S_{1}^{ij,\mvec{p}}(\mvec{y},\tau,t) &= -\frac{a^3}{V} \sum_{\mvec{x}} \left\langle  G_{3}^{i,\mvec{p}}(\mvec{x}+\mvec{y},\tau,t)\, L_1^{j}(\mvec{x},\tau) \right\rangle + \frac{a^3}{V} \sum_{\mvec{x}} \left\langle G_{3}^{i,\mvec{p}}(\mvec{x},\tau,t) \right\rangle \left\langle \! \left\langle L_1^{j}(\tau) \right\rangle \! \right\rangle  \,,
\nonumber \\
L_1^{j}(\mvec{y},\tau) &= \frac{1}{N^{\mathrm{st}}_{\mathrm{ins}}} \sum_{\ell}^{N^{\mathrm{st}}_{\mathrm{ins}}} \left[ \eta_\tau^{\dagger(\ell)}(\mvec{y},\tau)\ms \Gamma_j\ms \chi_\tau^{(\ell)}(\mvec{y},\tau) \right]
\end{align}
and
\begin{align}
\label{average-def}
\left\langle \! \left\langle L_1^j (\tau) \right\rangle \! \right\rangle &= \frac{a^3}{V} \sum_{\mvec{x}} \left\langle L_1^{j}(\mvec{x},\tau) \right\rangle \,.
\end{align}
Note that for symmetry reasons, the vacuum subtraction for $S_1$ in \eqref{S1-lat} is only nonzero for the scalar current $S$.
If the one-end trick is used, then $G_3^{i,\mvec{p}}$ is given by
\begin{align}
\label{G3-def}
G_{3, \mathrm{oet}}^{i,\mvec{p}}(\mvec{x},\tau,t) &= \frac{1}{N^{\mathrm{st}}_{\mathrm{src}}} \sum_{\ell}^{N^{\mathrm{st}}_{\mathrm{src}}} \left[ {X}_{0\ms t, \mathrm{oet}}^{\dagger(\ell),-\mvec{p}}(\mvec{x},\tau) \ms\gamma_5\ms \Gamma_i\ms \Psi_0^{(\ell),\mvec{0}}(\mvec{x},\tau) \right]
\end{align}
with the sequential propagator ${X}_{0\ms t}^{(\ell),\mvec{p}}(\mvec{x},\tau)$ obtained by inversion of
\begin{align}
\Bigl( \mathcal{D} {X}_{0\ms t, \mathrm{oet}}^{(\ell),\mvec{p}} \Bigr)(\mvec{x}',t')
  &= \delta_{t t'} \Bigl( \Phi\, \mathcal{E}^{\mvec{p}}\, \Phi\ms \gamma_5 \ms \Psi_0^{(\ell),-\mvec{p}} \Bigr)(\mvec{x}',t') \,.
\end{align}
We refrain from writing out the corresponding expressions for $G_{3, \mathrm{pt}}$ and $\mathrm{X}_{0 t, \mathrm{pt}}$ with point sources, given that they are identical to those used in standard computations of the pion form factor, see e.g.\ \cite{Brommel:2006ww}.
We find very good agreement between $S_1$ computed with the one-end trick and with point sources, with slightly larger statistical errors for the latter.  In later sections, only the one-end trick results are used.  By contrast, we take the point-source version of $G_3$ to evaluate the connected contributions to the vector and scalar form factors.

The contraction $C_2$ has a similar structure as the connected part of $S_1$, but it requires the calculation of an additional propagator between the two current insertions. For its evaluation we again use a stochastic source at the insertion time slice $\tau$. To reduce statistical noise, we make the following improvements. We consider the hopping parameter expansion of the propagator \cite{Thron:1997iy,Bali:2005fu,Bali:2009hu}, writing $\mathcal{D} = (\unit - H) /(2 \kappa)$ with the hopping term $H$, and make use of the geometric series:
\begin{equation}
\mathcal{M} = \mathcal{D}^{-1} = 2\kappa \left( \unit - H \right)^{-1}
= 2\kappa  \sum_{n=0}^{\infty} H^n
= 2\kappa \!\! \sum_{n = 0}^{n(\mvec{y})-1} H^n
  + 2\kappa \!\! \sum_{n=n(\mvec{y})}^{\infty} H^n \,.
\label{eq:hpe_def}
\end{equation}
\rev{Since for the Wilson-clover action, $H$ involves at most nearest neighbours on the lattice,} one needs at least
\begin{equation}
n(\mvec{y}) = \sum_{i=1}^3 \min\biggl(
  \frac{|y_i|}{a}, L - \frac{|y_i|}{a} \biggr)
\end{equation}
hopping terms to obtain a non-vanishing contribution to the propagator from a point $\mvec{x}$ to $\mvec{x}+\mvec{y}$ on a periodic lattice of spatial size $L a$.  Hence, the first sum on the r.h.s.\ of \eqref{eq:hpe_def} can be omitted, and we get
\begin{align}
\mathcal{M} &= 2\kappa \!\! \sum_{n=n(\mvec{y})}^{\infty} H^n
 = H^{n(\mvec{y})} \; 2\kappa \sum_{n=0}^\infty H^n = H^{n(\mvec{y})} \mathcal{M}
  & \text{for propagation from $\mvec{x}$ to $\mvec{x}+\mvec{y}$,}
\end{align}
\rev{where in the last step we used \eqref{eq:hpe_def} again.}
Taking $H^{n(\mvec{y})} \mathcal{M}$ instead of $\mathcal{M}$ itself, we implicitly omit those terms that do not contribute to the propagation but may add to the stochastic noise.  The expression to be evaluated for the $C_2$ graph is then given by
\begin{align}
\label{C2-lat}
C_2^{ij,\mvec{p}}(\mvec{y},\tau,t) &= \frac{a^3}{V N^{\mathrm{st}}_\mathrm{src}}  \sum_{\ell}^{N^{\mathrm{st}}_\mathrm{src}} \sum_{\mvec{x}} \biggl\langle \left[ {X}_{0\ms t, \mathrm{oet}}^{\dagger(\ell),-\mvec{p}}(\mvec{x},\tau) \ms \gamma_5\ms \Gamma_j \, \xi_\tau^{(\ell) ,n(\mvec{y})}(\mvec{x},\tau) \right]
\nonumber \\
 & \quad \times
 \left[ \eta_\tau^{\dagger(\ell)}(\mvec{x}+\mvec{y},\tau) \ms \Gamma_i\ms \Psi_0^{(\ell),\mvec{0}}(\mvec{x}+\mvec{y},\tau) \right] \biggr\rangle
\end{align}
with
\begin{equation}
\xi_\tau^{(\ell),n} = H^{n} \chi_\tau^{(\ell)} = H^{n} \ms \mathcal{D}^{-1} \eta^{(\ell)}_\tau \,.
\end{equation}

For $D$ and $S_2$, one needs the pion two-point graph $G_2$ and two single insertion loops $L_1$ or one double insertion loop $L_2$, respectively.  For $G_2$ we again use either one-end trick or point sources.  The double insertion loop is evaluated using a point source $S_{x}$ at $x = (\mvec{0}, \tau)$, from which the point-to-all propagator $\mathcal{M}_{x}$ is obtained by inverting
\begin{align}
\mathcal{D} \mathcal{M}_{x} &= S_{x} \,.
\label{eq:pt_all_prop}
\end{align}
Note that $S_{x}$ and hence also $\mathcal{M}_{x}$ is a vector in space-time but a matrix in spin and colour.  The space-time components of the point source are $S_{x}(x') = \unit\ms \delta_{x x'}$.  We then have
\begin{align}
\label{L2-def}
L_2^{ij}(\tau,\mvec{y}) &= \operatorname{tr} \left\{ \gamma_5\ms \mathcal{M}^\dagger_{\mvec{0}\tau}(\mvec{y},\tau) \ms \gamma_5\ms \Gamma_i\, \mathcal{M}_{\mvec{0}\tau}(\mvec{y},\tau)\, \Gamma_j \right\}
\end{align}
with the trace referring to spin and colour indices, and
\begin{align}
\label{S2-lat}
S_2^{ij,\mvec{p}}(\mvec{y},\tau,t) &= -\left\langle G_{2, \mathrm{oet}}^{\mvec{p}}(t)\, L_2^{ij}(\mvec{y},\tau) \right\rangle + \left\langle G_{2, \mathrm{oet}}^{\mvec{p}}(t) \vphantom{ L_2^{ij}} \right\rangle \left\langle L_2^{ij}(\mvec{y},\tau) \right\rangle \,,
\nonumber \\[0.2em]
D^{ij,\mvec{p}}(\mvec{y},\tau,t) &= \frac{a^3}{V} \sum_{\mvec{x}}\, \biggl\{ \left\langle  G_{2, \mathrm{pt}}^{\mvec{p}}(t) \, L_1^{i}(\mvec{x}+\mvec{y},\tau) \, L_1^{j}(\mvec{x},\tau) \right\rangle
\nonumber \\[0.1em]
 &\qquad - \left\langle G_{2, \mathrm{pt}}^{\mvec{p}}(t) \vphantom{ L_2^{ij}} \right\rangle\left\langle L_1^{i}(\mvec{x}+\mvec{y},\tau)\, L_1^{j}(\mvec{x},\tau) \right\rangle
\nonumber \\[0.1em]
 &\qquad - \left\langle G_{2, \mathrm{pt}}^{\mvec{p}}(t)\,  L_1^{j}(\mvec{x},\tau) \right\rangle \left\langle \! \left\langle L_1^{i \phantom{j \!\!}}(\tau) \right\rangle \! \right\rangle
- \left\langle G_{2, \mathrm{pt}}^{\mvec{p}}(t)\, L_1^{i \phantom{j \!\!}}(\mvec{x},\tau) \right\rangle \left\langle \! \left\langle L_1^j (\tau) \right\rangle \! \right\rangle    \biggr\}
\nonumber \\[0.1em]
 &\quad +  2 \left\langle G_{2, \mathrm{pt}}^{\mvec{p}}(t) \vphantom{ L_2^{ij}} \right\rangle \left\langle \! \left\langle L_1^{i \phantom{j \!\!}}(\tau) \right\rangle \! \right\rangle \left\langle \! \left\langle L_1^j (\tau) \right\rangle \! \right\rangle
\end{align}
with $\langle\langle L_1 \rangle\rangle$ defined in \eqref{average-def}.  When evaluated with the one-end trick, the two point graph reads
\begin{align}
G_{2, \mathrm{oet}}^{\mvec{p}}(t) &= \frac{1}{N^{\mathrm{st}}_{\mathrm{src}}} \sum_{\ell}^{N^{\mathrm{st}}_{\mathrm{src}}}  \biggl[ \Psi_0^{\dagger(\ell),\mvec{0}}\ms \Phi \ms \mathcal{E}^{\mvec{p}}\ms \Phi \ms \Psi_0^{(\ell),-\mvec{p}} \biggr] \,.
\end{align}
Note that $\langle L_2^{ij} \rangle$ and $\langle L_1^{i \phantom{j\!\!}}\, L_1^{j} \ms\rangle$ depend on the distance $\mvec{y}$ between the currents and can be nonzero for most combinations of Dirac matrices $\Gamma_i$ and $\Gamma_j$.  This makes vacuum subtractions necessary for $S_2$ and $D$ in most channels.

As indicated in \eqref{S2-lat}, we use only stochastic sources for $S_2$ and only point sources for $D$.  For the pion two-point function $C_{2 \mathrm{pt}}(t) = \langle G_2(t) \rangle$, we compare both methods and find excellent agreement.

\begin{table}
\begin{center}
\renewcommand{\arraystretch}{1.15}
\begin{tabular}{c|ccccc|ccccc}
\hline \hline
      & \multicolumn{5}{c|}{$L = 40$} & \multicolumn{5}{c}{$L = 32$} \\
graph & $N_\mathrm{src}^{\mathrm{st}}$ & $N_{\mathrm{ins},i}^{\mathrm{st}}$ & $N_{\mathrm{ins},j}^{\mathrm{st}}$ & $N_{\mathrm{src}}$ & $N_{\mathrm{ins}}$
      & $N_\mathrm{src}^{\mathrm{st}}$ & $N_{\mathrm{ins},i}^{\mathrm{st}}$ & $N_{\mathrm{ins},j}^{\mathrm{st}}$ & $N_{\mathrm{src}}$ & $N_{\mathrm{ins}}$ \\
\hline
$C_1$ & 1 & -  &  -  &  1 & $L^3$ & 1 & -  &  -  & 1 & $L^3$ \\
$C_2$ & 1 & 10 &  -  &  5 & $L^3$ & 1 & 20 &  -  & 1 & $L^3$ \\
$A$   & 1 & -  &  -  &  4 & $L^3$ & 1 & -  &  -  & 1 & $L^3$ \\
$S_1$ & - & -  & 120 &  4 & $L^3$ & - & -  & 120 & 3 & $L^3$ \\
$S_2$ & 1 & -  &  -  & 16 &   16  & 1 & -  &  -  & 1 &   2   \\
$D$   & - & 60 &  60 &  1 & $L^3$ & - & 60 &  60 & 1 & $L^3$ \\
\hline \hline
\end{tabular}
\end{center}
\caption{\label{tab:statnumbers} Numbers $N^{\text{st}}$ of stochastic noise vectors and numbers of sources and current insertions used for each graph in our simulations for the \rev{two lattice volumes}.  For most graphs, stochastic propagators are connected to the insertion, which implies an average over the entire spatial lattice volume. This is indicated by $N_{\mathrm{ins}} = L^3$.}
\end{table}

To increase statistics, we repeat the calculations at $N_{\mathrm{src}}$ time positions $t_s$ of the source (keeping $\tau$ and $t$ fixed relative to $t_s$) and for several spatial insertion positions $N_{\mathrm{ins}}$ of the current. For most graphs, the latter is implicitly realised by using stochastic wall sources. The corresponding numbers, as well as the size of the stochastic noise vector sets, are given in table~\ref{tab:statnumbers}.  \rev{Note that in some cases these numbers differ between the lattices with $L=40$ and $L=32$ (ensembles V and IV of table~\ref{tab_1}).}


\subsection{Renormalisation}
\label{subsec:renorm}

We convert all lattice currents, which are defined in the lattice
scheme at a lattice spacing $a$, into the $\overline{\mathrm{MS}}$ scheme at the renormalisation scale $\mu = 2 \gev$. The corresponding renormalisation constants depend on the gauge coupling $g^2=6/\beta$, where in our case $\beta=5.29$.  For currents with nonzero anomalous dimension, such as in the scalar and pseudoscalar cases, there is also a dependence on $a\mu$.  The conversion factors used for the correlator \eqref{4pt2ptRatio} of currents $i$ and $j$ read
\begin{align}
R_{ij}^{\msbar} &= \tilde{Z}_i\ms \tilde{Z}_j \ms
  R_{ij}^{\mathrm{lat}}  & \text{ ~with~ }
\tilde{Z}_i^{} &= Z^{\msbar}_i \left( 1 +  a m_q\ms b_i\right) \,,
\end{align}
see e.g.\ \cite{Bali:2014nma}. Here we have included the coefficients $b_i$ for the mass-dependent order $a$ improvement. They become particularly relevant at the charm quark mass. In the vector and axial vector cases, there are additional mass-independent order $a$ improvement terms (accompanied by $c_V$ and $c_A$, respectively), which we ignore in the present work. Note that some of the matrix elements listed in section \ref{sec:phys-matels} receive contributions from flavour singlet current combinations. In these cases our renormalisation procedure should be regarded as approximate, because we do not take into account operator mixing.

The renormalisation factors $Z_i^{\msbar}$ have been obtained in \cite{Gockeler:2010yr} (updated in \cite{Bali:2014nma}) in a two-step procedure: first the currents were matched non-perturbatively from the lattice scheme to the RI'/MOM scheme and then from there to the $\overline{\mathrm{MS}}$ scheme in continuum perturbation theory. The coefficients $b_i = 1 + \mathcal{O}(g^2)$ have been computed in lattice perturbation theory~\cite{Capitani:2000xi,Sint:1997jx,Taniguchi:1998pf}.  Furthermore, in \cite{Fritzsch:2010aw} the coefficient $b_S$ was determined non-perturbatively as
\begin{equation}
\label{b-interpol}
b_S^{\mathrm{np}} = \bigl( 1 + 0.19246 \ms g^2 \bigr) \,
  \frac{1-0.3737 \ms g^{10}}{1-0.5181 \ms g^4}
\end{equation}
with an uncertainty of about 5\%.  We can determine $b_V$ non-perturbatively by evaluating the vector form factor at zero momentum transfer with charm quarks ($a m_q = 0.3147$).  Multiplied by $Z_V^{\msbar} (1 + a m_q\ms b_V^{})$ this must give $1$, owing to charge conservation.  Using the result $F_V(0) = 0.9068(2)$ from our $L=40$ lattice and the value of $Z_V^{\msbar}$ from \cite{Bali:2014nma}, we extract
\begin{align}
\label{b-from-charm}
b_V^{\mathrm{np}} &= 1.586(32) \,,
\end{align}
where the uncertainty is dominated by the uncertainty on $Z_V^{\msbar}$.

\begin{table*}
\begin{center}
\begin{tabular}{c|ccccc} \hline \hline
 & $S$ & $P$ & $V$ & $A$ & $T$ \\
\hline
$Z^{\msbar}$ \rule{0pt}{1.2em} &
0.6153(25) & 0.476(13) & 0.7356(48) & 0.76487(64) & 0.8530(25) \\
$b^{\text{pert}}$ & 1.3453 & 1.2747 & 1.2750 & 1.2731 & 1.2497 \\
$b^{\text{np}}$  & 1.091(55) &  & 1.586(32) &  &  \\
$b^{\text{resc}}$ & 1.673 & 1.586 & 1.586 & 1.584 & 1.555 \\ \hline \hline
\end{tabular}
\end{center}
\caption{\label{tab:renfact} Renormalisation factors $Z_i^{\protect\msbar}$ from \protect\cite{Bali:2014nma} for the different currents \protect\eqref{curr-def} in the $\overline{\mathrm{MS}}$ scheme at scale $\mu = 2 \gev$.  We also list the perturbative estimates of the coefficients $b_i$ from \protect\cite{Bali:2014nma}, the non-perturbative determinations \protect\eqref{b-interpol} and \protect\eqref{b-from-charm}, and the rescaled values \protect\eqref{b-rescaled}.  For the latter, we estimate an error of 34\% as explained in the text.}
\end{table*}

Table \ref{tab:renfact} gives the values for $Z_i^{\msbar}(2 \gev)$, as well as estimates $b_i^{\text{pert}}$ using one-loop perturbation theory with an ``improved'' coupling constant (for details see (26) and (27) in \cite{Bali:2014nma}). In the next row, we give the non-perturbative estimates $b_i^{\text{np}}$ from \eqref{b-interpol} and \eqref{b-from-charm}.  We see that the non-perturbative value for $b_V$ is 24\% larger than the perturbative estimate, while $b_S$ from \cite{Fritzsch:2010aw} is about 19\% smaller than the perturbative result. We remark that different non-perturbative determinations of $b_i$ only need to agree up to order $a$ effects.

Based on our result for $b_V$, we make the naive assumption that all non-perturbative coefficients are larger than the perturbative estimates and define rescaled coefficients
\begin{equation}
\label{b-rescaled}
b^{\text{resc}}_i
  = b^{\text{pert}}_i \, \frac{ b^{\text{np}}_V}{ b^{\text{pert}}_V} \,,
\end{equation}
which are also listed in the table. As is clear from the scalar channel, there is a huge uncertainty in this procedure. We find $| 1 - b_S^{\text{np}}/b_S^{\text{resc}} | = 34\%$ and take this as our uncertainty of the coefficients~$b_i^{\text{resc}}$ with $i \neq V$.  The results in the following chapters are obtained with the coefficients $b_i^{\mathrm{np}}$ for $i=S, V$ and with $b_i^{\mathrm{resc}}$ for all other currents.

%% file: data.tex
\section{Data quality and lattice artefacts}
\label{sec:data}

In this section we investigate the quality of our data and discuss a number of lattice artefacts.  We concentrate on the data for light quarks here, for which we have the largest set of simulations.  We verified that for heavy quarks, the lattice artefacts are not worse than for light ones.

\subsection{Plateaux and excited state contributions}
\label{sec:plateaux}

\begin{figure*}
\begin{center}
  \subfigure[$C_1, \oa, \mvec{y}/a = (0,3,1), p=0, L=40$]{\includegraphics[width=0.48\textwidth,trim=0 0 0 20pt,clip]{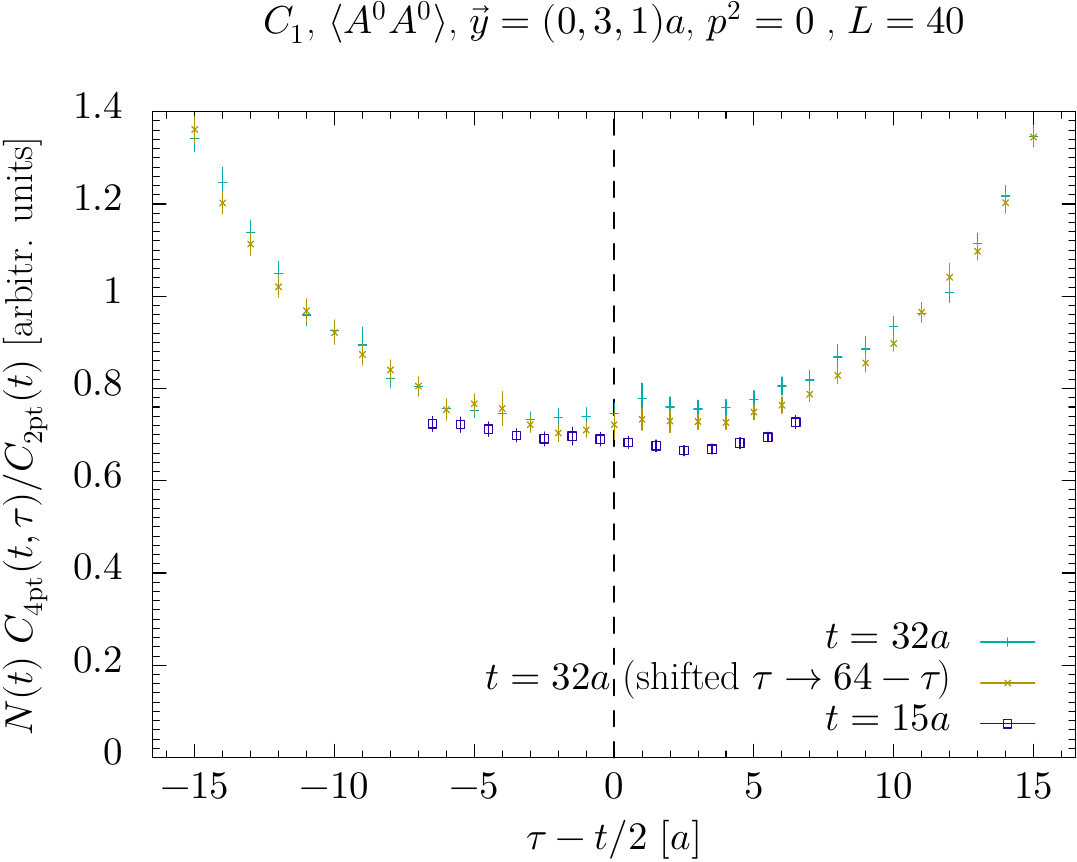}}
  \hfill
  \subfigure[$A, \os, \mvec{y}/a = (0,3,1), p=0, L=40, t=15 \ms a$]{\includegraphics[width=0.48\textwidth,trim=0 0 0 20pt,clip]{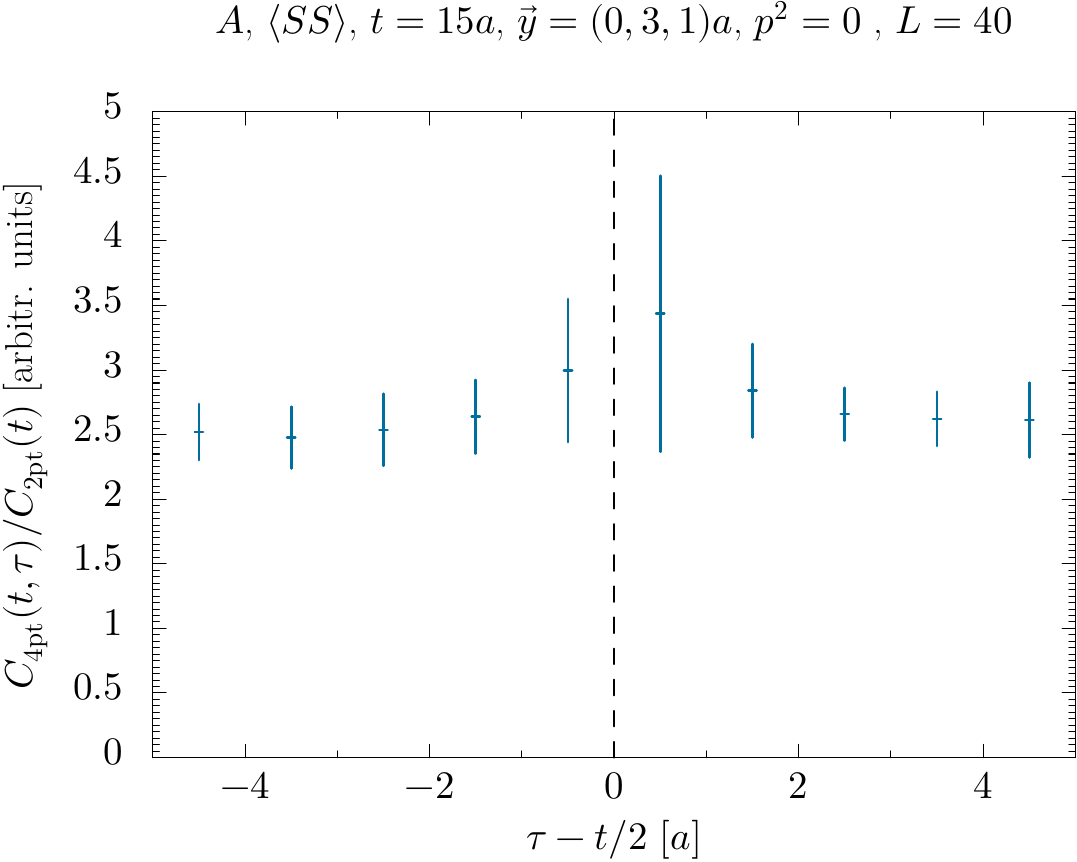}}
  \caption{\label{fig:tau-plateaux} Dependence of the four-point function on the current insertion time $\tau$ at a fixed distance $\mvec{y}$ between the two currents.  Shown are the contributions of the contraction $C_1$ to the correlator $\oa$ and of the contraction $A$ to $\os$.  Here and in all following plots, errors are purely statistical.  The normalisation factor $N(t)$ is specified in the text.  $p$ denotes the pion momentum and $L$ the spatial lattice extent.  Unless mentioned otherwise, all data shown in this paper are for light quarks.}
\end{center}
\end{figure*}

For the contractions $C_1$ and $A$, we have data for the ratio \eqref{4pt2ptRatio} of four-point and two-point functions in a wide range of the current insertion time $\tau$ and can hence verify the existence of a plateau.  Figure~\ref{fig:tau-plateaux} shows this ratio for two cases that are representative of channels with a good signal.  In the case of figure~\ref{fig:tau-plateaux}(a) we have data for sink-source separations of $t/a = 15$ and $32$, which we plot in such a way that the midpoints $\tau = t/2$ coincide in both cases.  Throughout this paper, error bars are purely statistical and determined by the jackknife method.  To eliminate autocorrelations, we take the number of jackknife samples as $1/8$ times the number $N_{\text{used}}$ of gauge configurations (see table~\ref{tab_1}).

The time separation $t/a = 32$ is half the temporal extent of our lattices, so that the pion two-point function $C_{2 \mathrm{pt}}(t)$ at that point has equal contributions from pions that propagate forward or backward from the source at time zero.  To extract the pion matrix element, we must hence multiply $C_{4\mathrm{pt}}(\mvec{y},\tau,t) / C_{2\mathrm{pt}}(t)$ in \eqref{4pt2ptRatio} with a normalisation factor $N(32\ms a) = 2$, which is also done in the figure.  For $t/a = 15$, the contribution from backward propagating pions in the two-point function is below $3\%$. We do not correct for this contamination and take $N(15\ms a) = 1$ in this case.

The data in figure~\ref{fig:tau-plateaux} show a clear plateau around $\tau = t/2$.  In figure~\ref{fig:tau-plateaux}(a) the plateaux for the two source-sink separations are in good agreement with each other, which indicates that contributions from excited states are not very large.  Based on the plots and their analogues for other channels, we extract plateau values by fitting to a constant in the ranges
\begin{align}
\label{tau-fit-range}
& 6 \le \tau/a \le 9 & \text{ ~for~ } t/a = 15 \,,
\nonumber \\
& 14 \le \tau/a \le 18 \text{ ~and~ } 46 \le \tau/a \le 50
  & \text{ ~for~ } t/a = 32 \,.
\end{align}
For $t/a = 32$ we thus combine the two plateaux around $\tau/a = 16$ and $48$.

\begin{figure*}
\begin{center}
  \subfigure[$C_1, \ov, p=0, L=40$]{\includegraphics[width=0.49\textwidth,trim=0 0 0 17pt,clip]{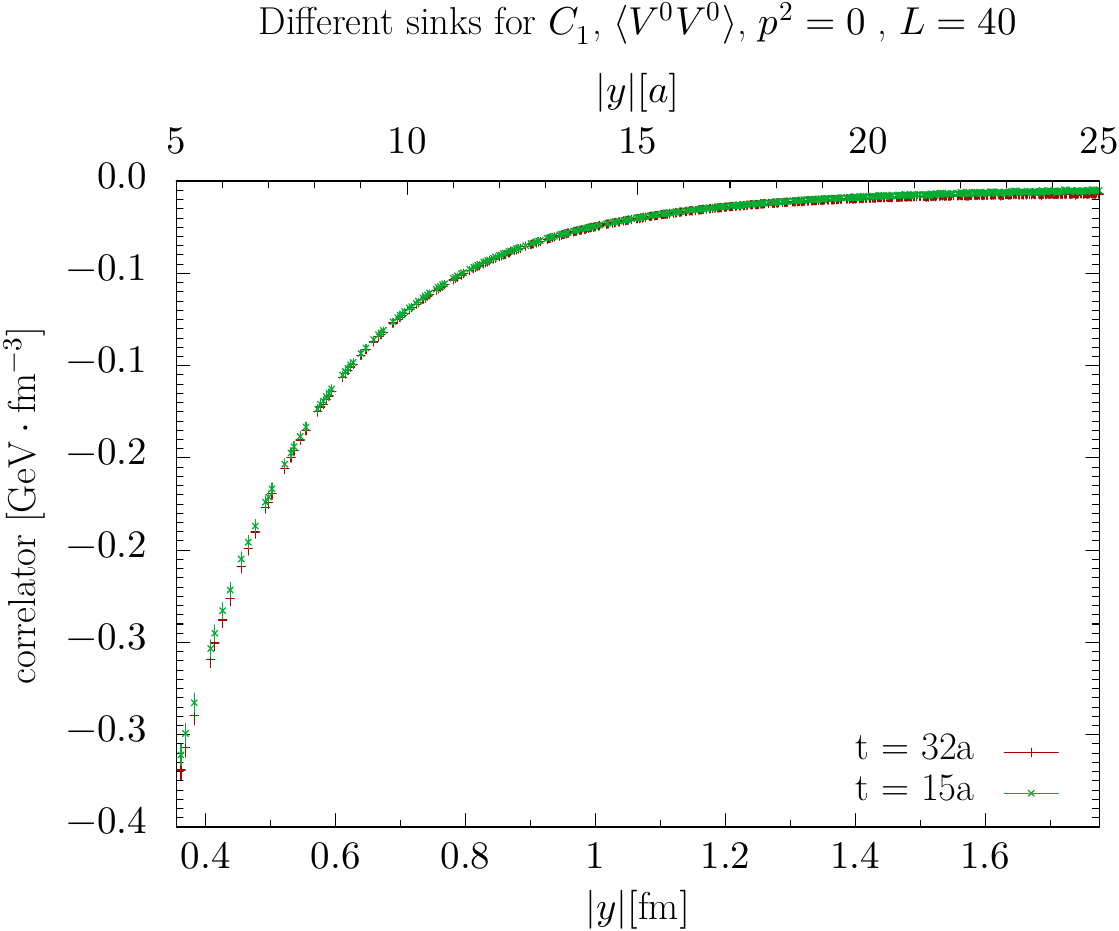}}
  \hfill
  \subfigure[$C_1, \op, p=0, L=40$]{\includegraphics[width=0.49\textwidth,trim=0 0 0 17pt,clip]{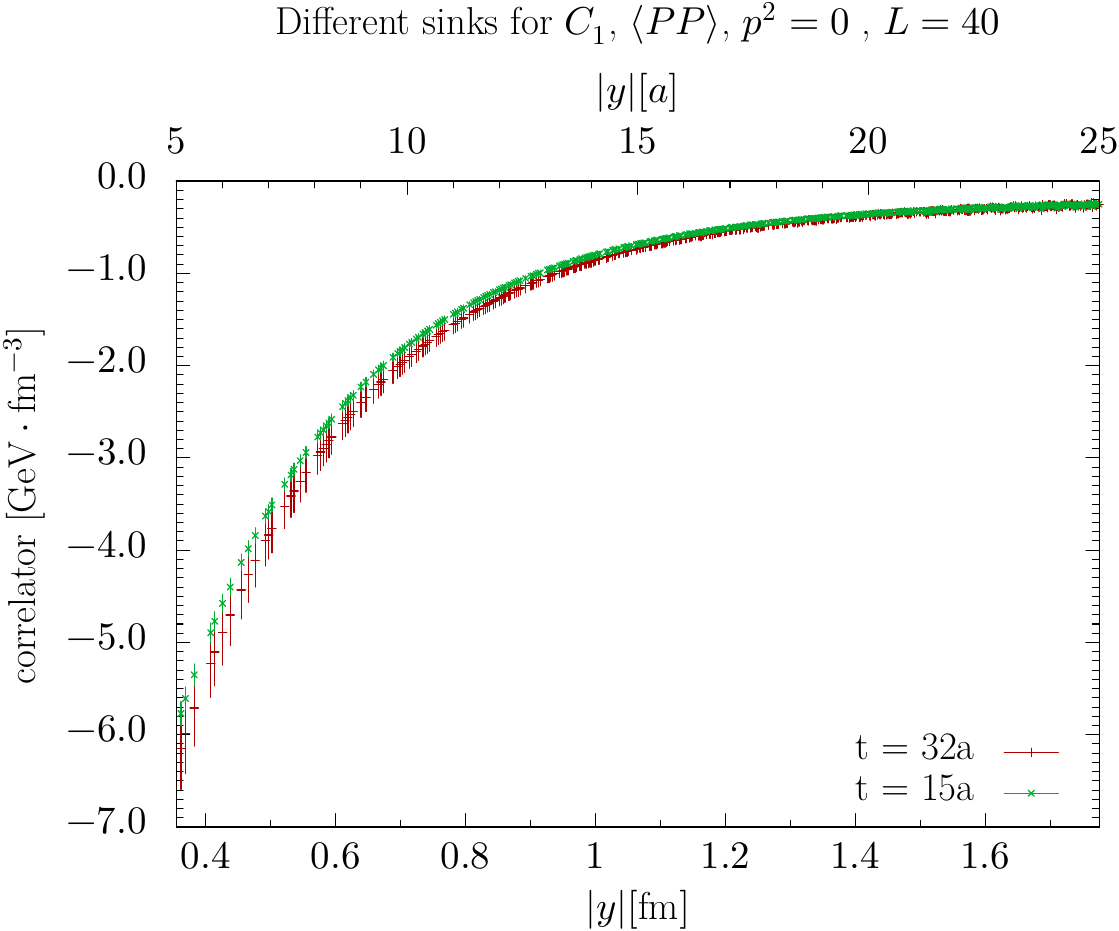}} \\
  \subfigure[$C_2, \os, p=0, L=32$]{\includegraphics[width=0.49\textwidth,trim=0 0 0 17pt,clip]{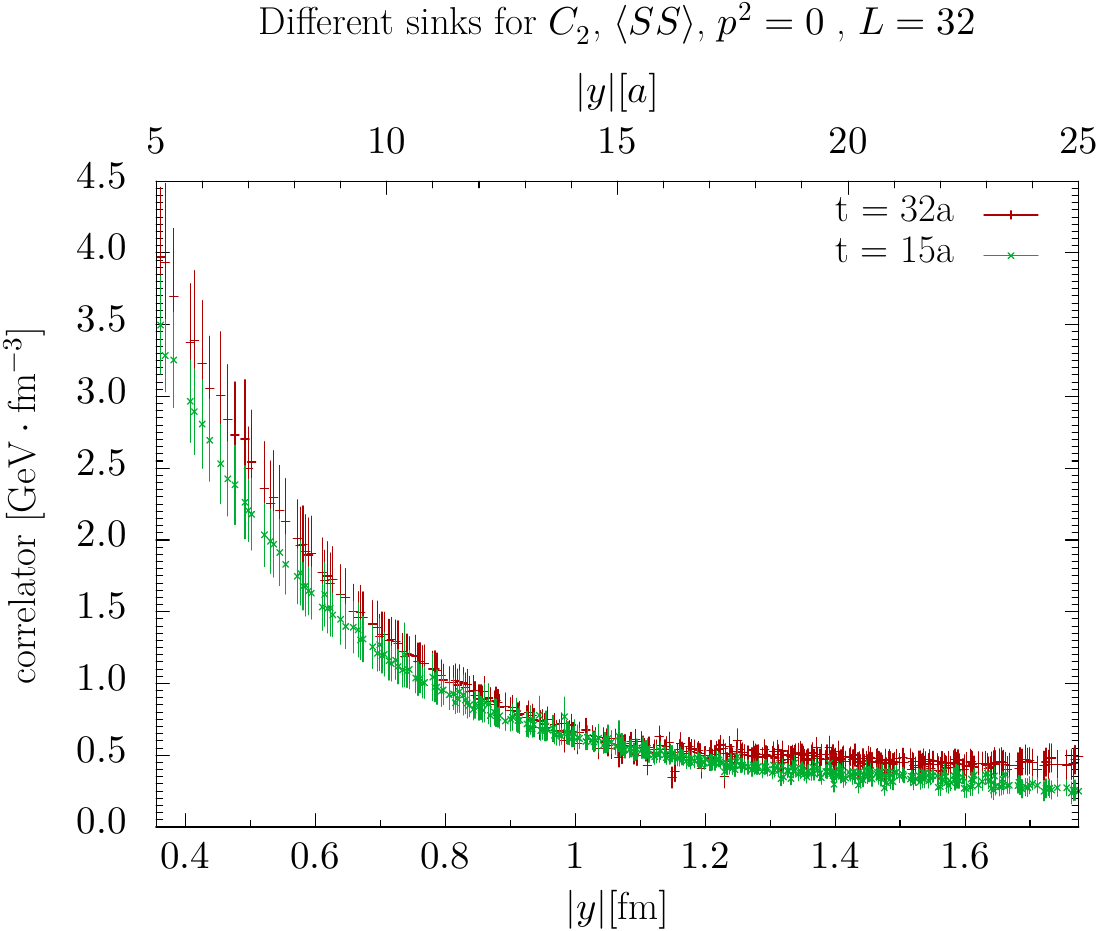}}
  \hfill
  \subfigure[$S_1, \op, p=0, L=32$]{\includegraphics[width=0.49\textwidth,trim=0 0 0 17pt,clip]{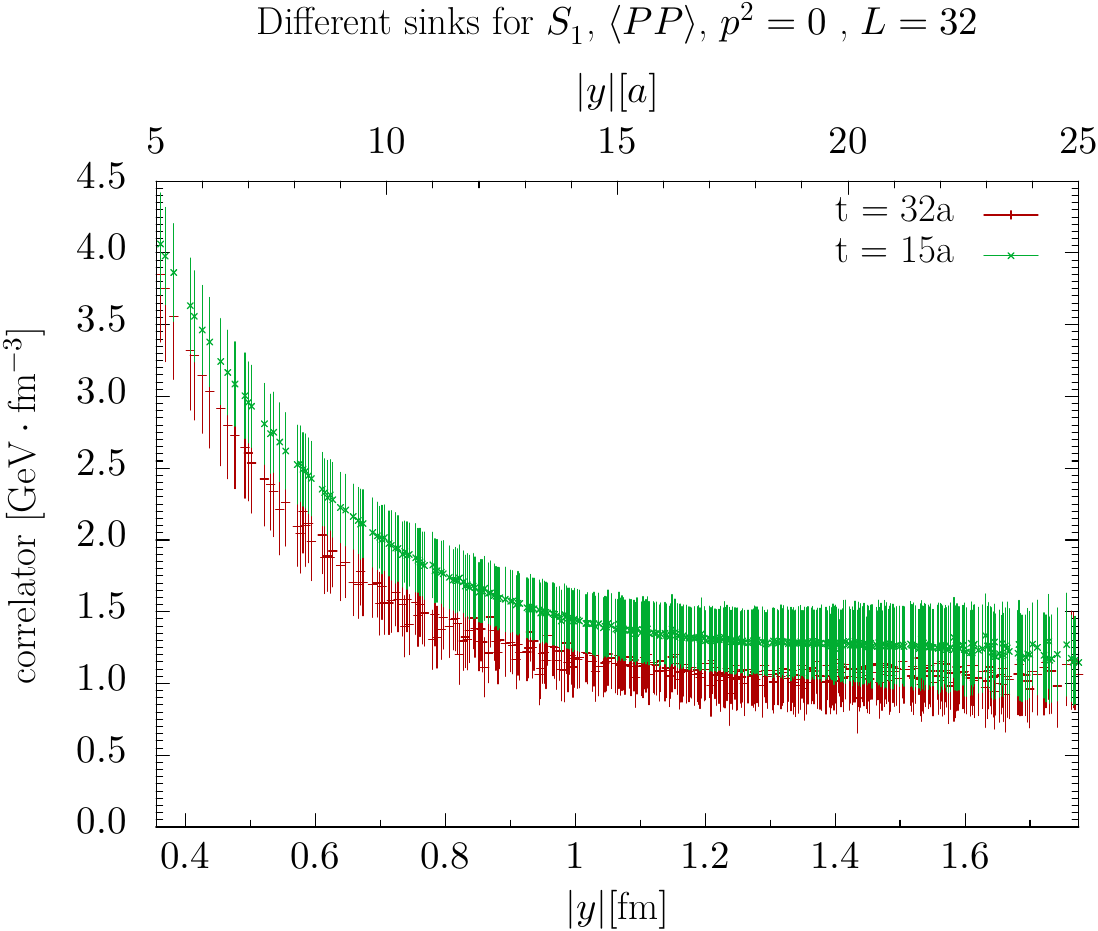}}
  \caption{\label{fig:t32-comp} Comparison between data with $t/a = 15$ and $32$ for different contractions and different lattice sizes.  Shown are only data points satisfying the cut \protect\eqref{cos-cut} discussed in the next section.  The correlation functions in this and all following figures have been converted to the $\overline{\mathrm{MS}}$ scheme at the scale $\mu = 2 \gev$.}
\end{center}
\end{figure*}

Examples for the extracted two-current matrix elements as a function of the distance between the currents are shown in figure~\ref{fig:t32-comp}.  Here and in the following, we write $y = | \mvec{y} |$.  We see that the data for $t/a = 15$ and $32$ are consistent within their statistical uncertainties.  In these plots and in their analogues for the other channels, we do not find any indication for large excited state contamination.

\subsection{Anisotropy effects}
\label{sec:aniso}

In the continuum, the correlation functions $\ov, \oa, \os$ and $\op$ in a pion at rest can only depend on the modulus $y = |\mvec{y}|$.  Lattice regularisation breaks rotational invariance, so that an anisotropy in the $\mvec{y}$ dependence of our correlators is a clear indication of lattice artefacts.

\begin{figure*}
\begin{center}
  \subfigure[$\ov, p=0, L=40$]{\includegraphics[width=0.495\textwidth,trim=0 0 0 17pt,clip]{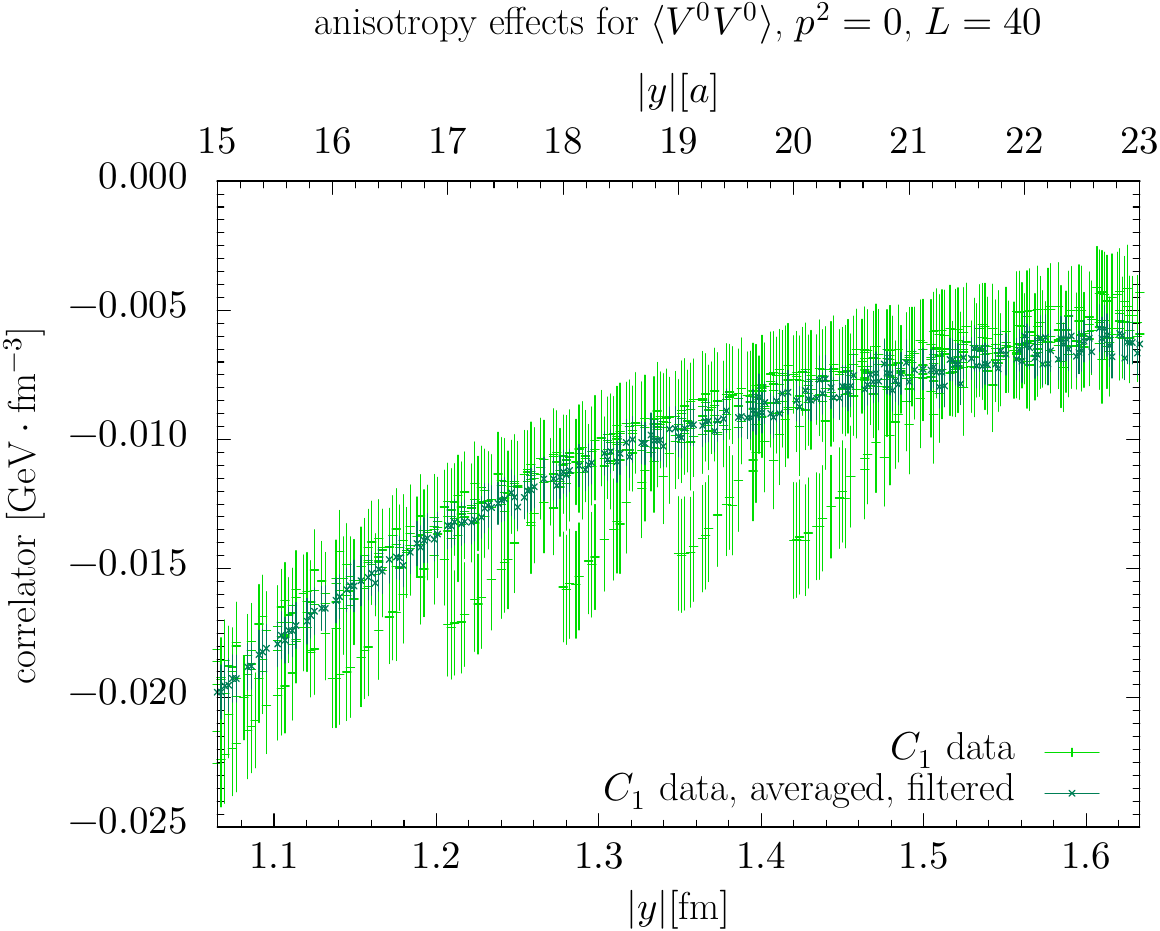}}
  \hfill
  \subfigure[$\os, p=0, L=40$]{\includegraphics[width=0.47\textwidth,trim=0 0 0 17pt,clip]{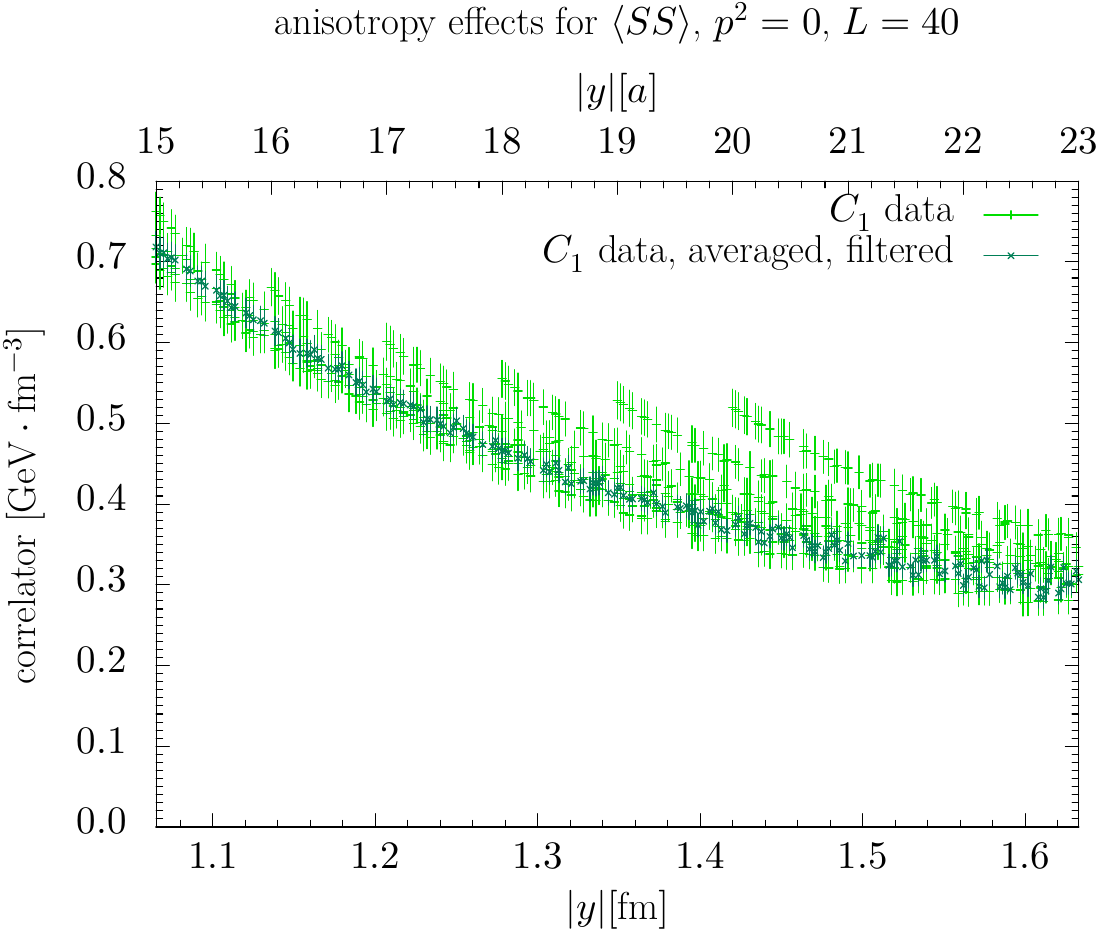}}
\\
  \subfigure[$\op, p=0, L=40$]{\includegraphics[width=0.49\textwidth,trim=0 0 0 17pt,clip]{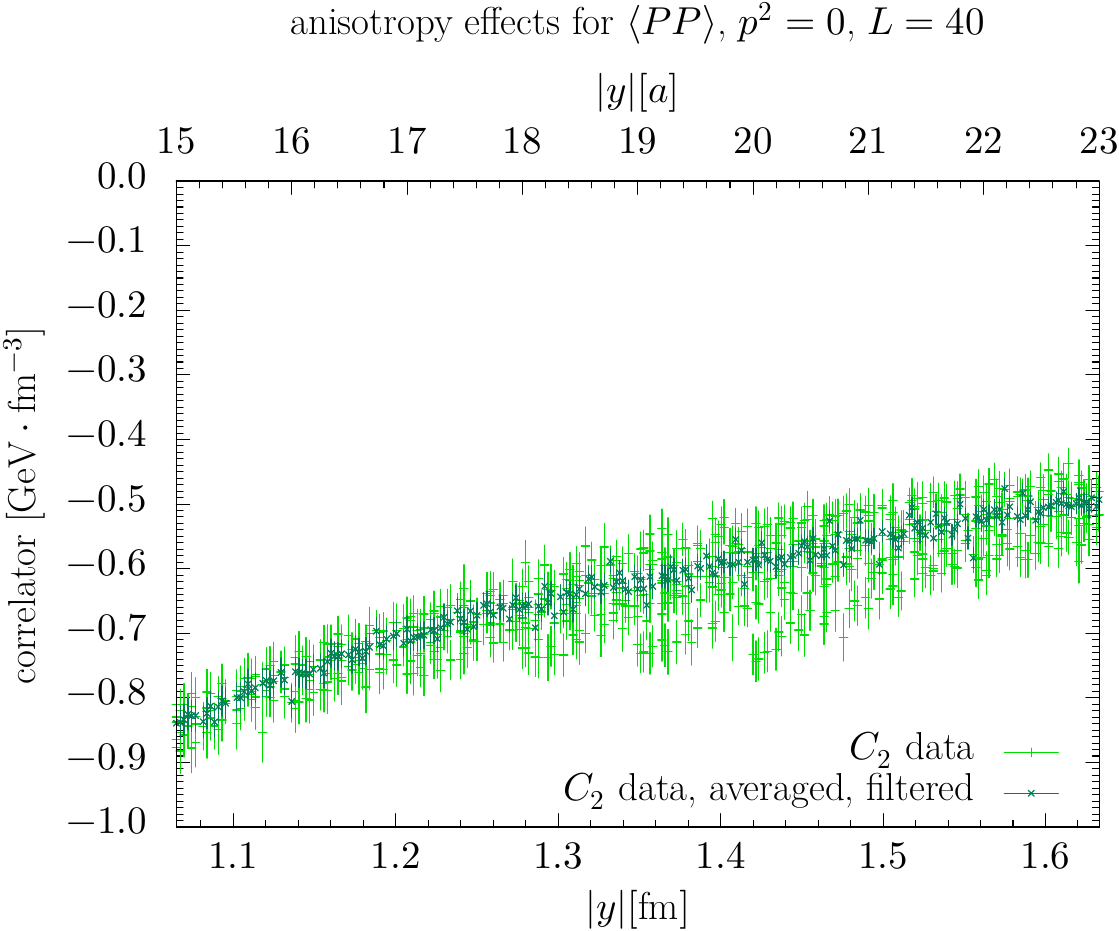}}
  \hfill
  \subfigure[$\op, p=0, L=40$]{\includegraphics[width=0.48\textwidth,trim=0 0 0 17pt,clip]{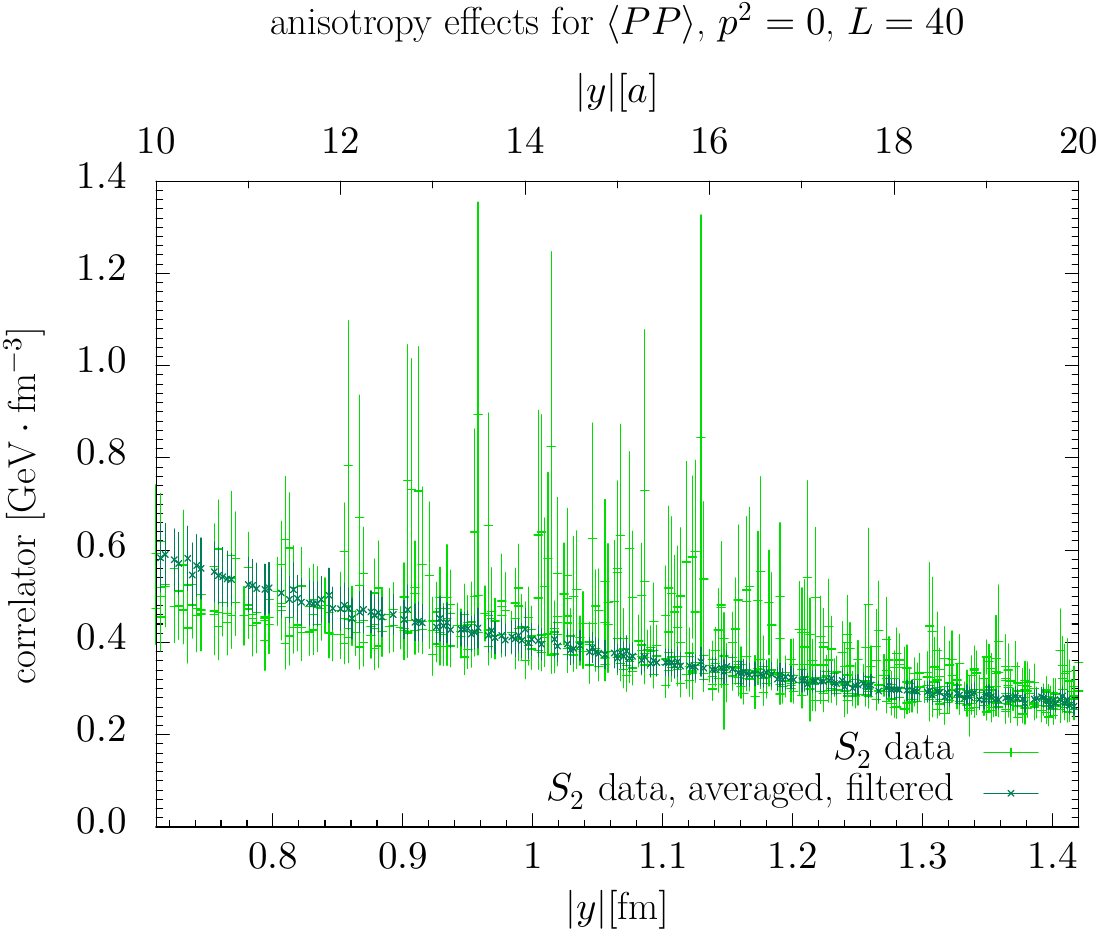}}
  \caption{\label{fig:aniso-large-y} Anisotropy effects in our data at large and intermediate $y$ (light points) and their removal by the cut \protect\eqref{cos-cut} (dark points).}
\end{center}
\end{figure*}

At large distances $y/a \sim L/2$ we see clear signs of anisotropy, as shown in figure~\ref{fig:aniso-large-y}(a) to (c).  As explained in detail in \cite{Burkardt:1994pw}, this can be traced back to the use of a finite lattice with periodic boundary conditions: the two currents are sensitive to the ``images'' of the pion in lattice cells adjacent to the elementary cell of size $L^3$.  At a given distance $y$, this effect is smallest for vectors $\mvec{y}$ with directions close to space diagonals $(\pm 1,\pm 1, \pm 1)$ and largest if $\mvec{y}$ points along one of the coordinate axes.  This gives rise to the distinct ``sawtooth'' pattern seen in our plots, which was also observed in earlier studies \cite{Burkardt:1994pw,Alexandrou:2008ru}.  An analytic investigation of the same phenomenon using a low-energy effective field theory can be found in \cite{Briceno:2018lfj}.  A method to correct for these periodic images was proposed in \cite{Burkardt:1994pw} and employed in \cite{Alexandrou:2008ru}, where simulations were performed on $L=24$ lattices with $a \approx 0.077 \fm$.  The resulting physical length $L a$ was thus 0.81 times the length of the smallest lattice used in our study.

Benefiting from the larger size of our lattices, we take a less sophisticated approach here and consider only data for distances $\mvec{y}$ close to the space diagonals.  To this end, we use exact lattice symmetries to map the vector $\mvec{y}$ into the octant where all its components are non-negative and define $\vartheta(\mvec{y})$ as the angle between that vector and the diagonal $(1,1,1)$.  We then retain only data points satisfying
\begin{align}
\label{cos-cut}
\cos\vartheta(\mvec{y}) & \ge 0.9 \,.
\end{align}
After this cut, we statistically average the correlators over points $\mvec{y}$ of the same length $y$, which greatly decreases statistical errors.  The points thus combined are not necessarily related by an exact lattice symmetry operation, and we checked that restricting the data combination to points that are equivalent on the lattice gives results consistent with those of the more inclusive combination.

As seen in figure~\ref{fig:aniso-large-y}(a) to (c), the above procedure efficiently removes anisotropy effects while retaining enough data points.  It also removes a broad anisotropic structure that we observe for $\op$ in the $S_2$ graph, figure~\ref{fig:aniso-large-y}(d), for which we have no interpretation.

\begin{figure*}
\begin{center}
  \subfigure[$\os, p=0, L=40$]{\includegraphics[width=0.48\textwidth,trim=0 0 0 17pt,clip]{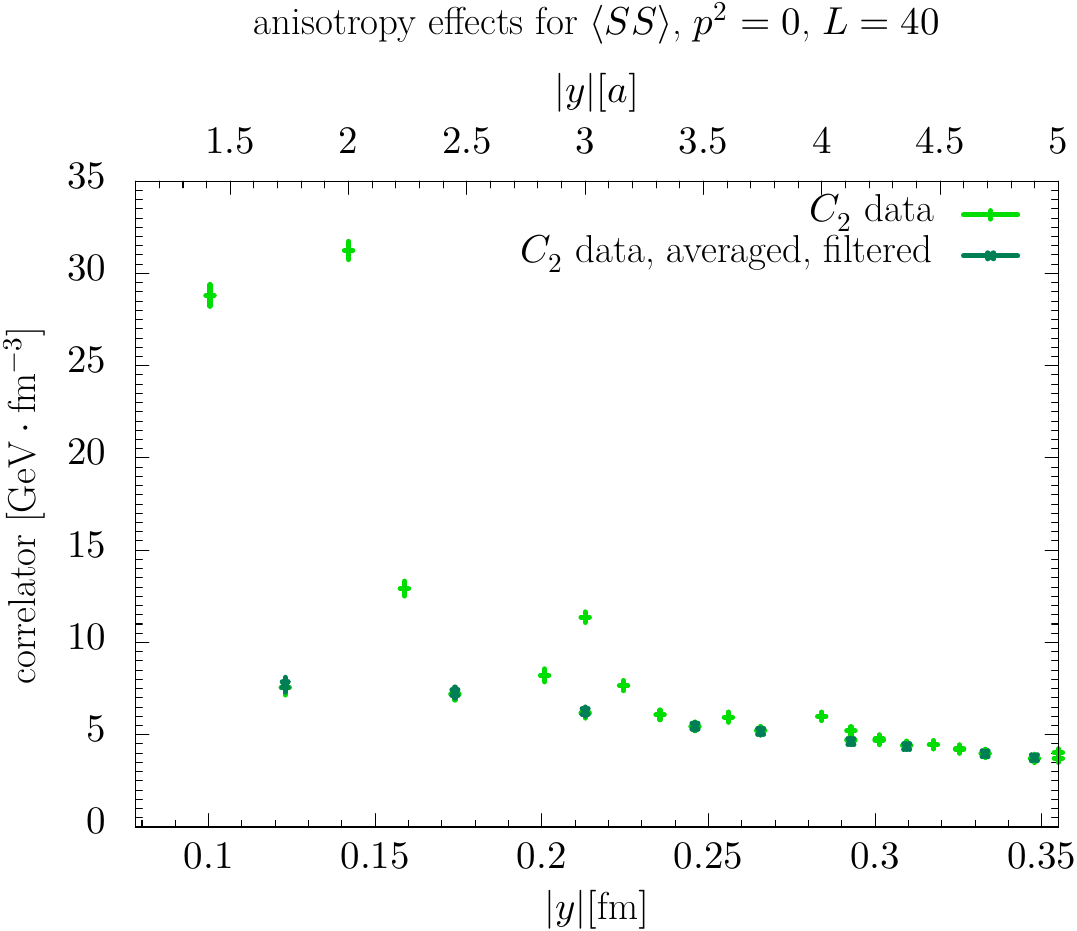}}
  \hfill
  \subfigure[$\op, p=0, L=40$]{\includegraphics[width=0.49\textwidth,trim=0 0 0 17pt,clip]{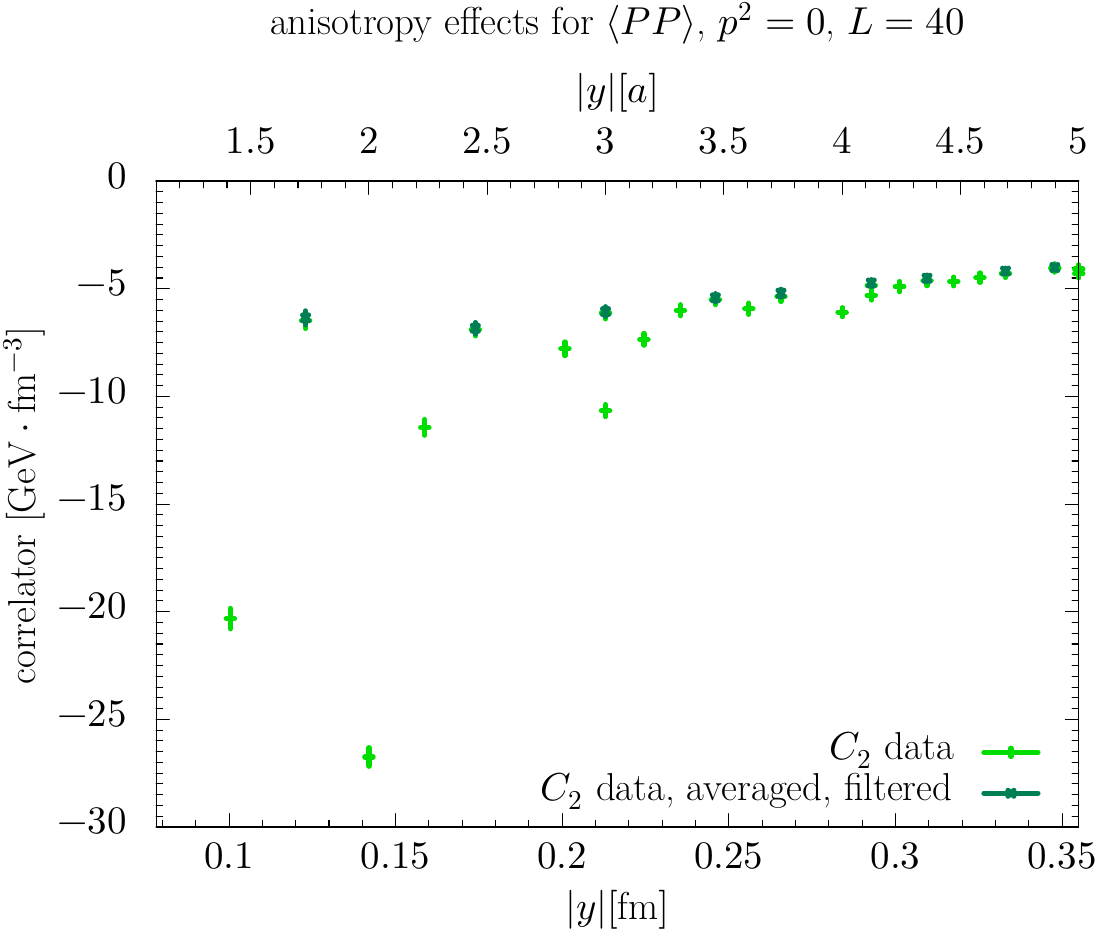}}
\\
  \caption{\label{fig:aniso-small-y} Anisotropy effects for the contraction $C_2$ at small $y$ (light points) and their removal by the cut \protect\eqref{cos-cut} (dark points).}
\end{center}
\end{figure*}

A different type of anisotropy effect appears at small $y$ in our data for the contraction $C_2$: in channels with small statistical errors, we observe a huge dependence of the correlators on the direction of $\mvec{y}$, as shown in figure~\ref{fig:aniso-small-y}.  To understand this, we note that in $C_2$ the two current insertions are directly connected by a fermion propagator, which is hence evaluated for a fixed space-time direction.  The effect we see may thus directly reflect a discretisation effect, namely the anisotropy of the fermion propagator on the lattice.  This anisotropy was studied in \cite{Cichy:2012is} (see figure~1 in that work), where it was shown that lattice artefacts are smallest for distances $y$ close to the diagonal $n = (1,1,1,1)$ in Euclidean space-time.  A cut on $y n$ advocated in \cite{Cichy:2012is} is equivalent to our cut on $\cos\vartheta(\mvec{y})$, given that we always have $y^0 = 0$.  As seen in our figure~\ref{fig:aniso-small-y}, the cut \eqref{cos-cut} indeed removes the anisotropy effects in $C_2$ and yields points that can be connected by a reasonably smooth curve.  One may expect a similar situation for the contraction $S_2$, where the two currents are connected by two lattice propagators, but our statistical errors for $S_2$ are too large to unambiguously identify this effect.

To summarise, we apply the cut \eqref{cos-cut} and the averaging procedure described above to all lattice data shown in the rest of this work.  One must of course bear in mind that the filtered data may still be affected by discretisation effects at small $y$ and by finite size effects at large $y$.  Regarding the former, one should be most careful in the interpretation of data with $y/a$ below, say, $2$ or $3$.  For the latter, we have an additional handle from the comparison of our two lattice volumes, which we discuss next.


\subsection{Finite volume effects}
\label{sec:volume}

\begin{figure*}
\begin{center}
  \subfigure[$|\ms C_1|, \ov, p=0$]{\includegraphics[width=0.44\textwidth,trim=0 0 0 14pt,clip]{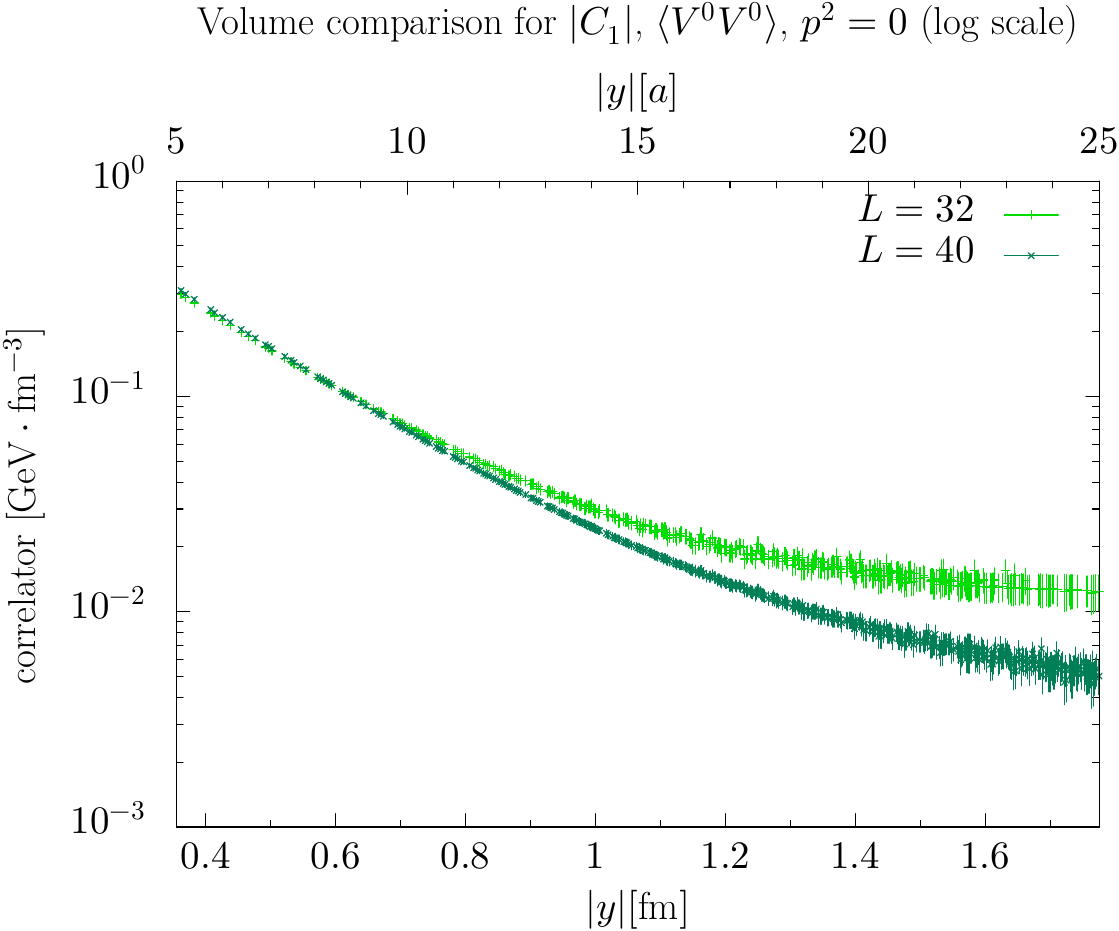}}
  \hfill
  \subfigure[$A, \oa, p=0$]{\includegraphics[width=0.45\textwidth,trim=0 0 0 14pt,clip]{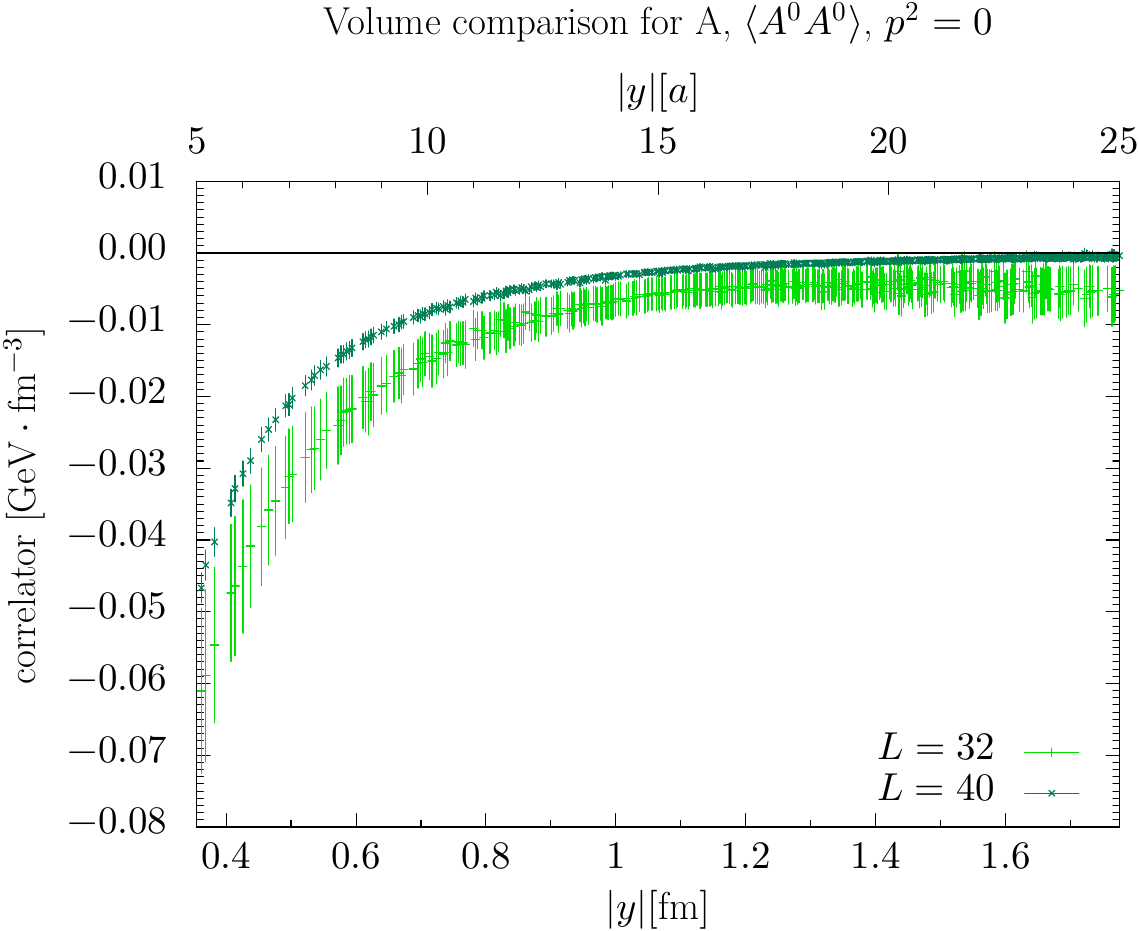}} \\
  \subfigure[$C_2, \os, p=0$]{\includegraphics[width=0.44\textwidth,trim=0 0 0 14pt,clip]{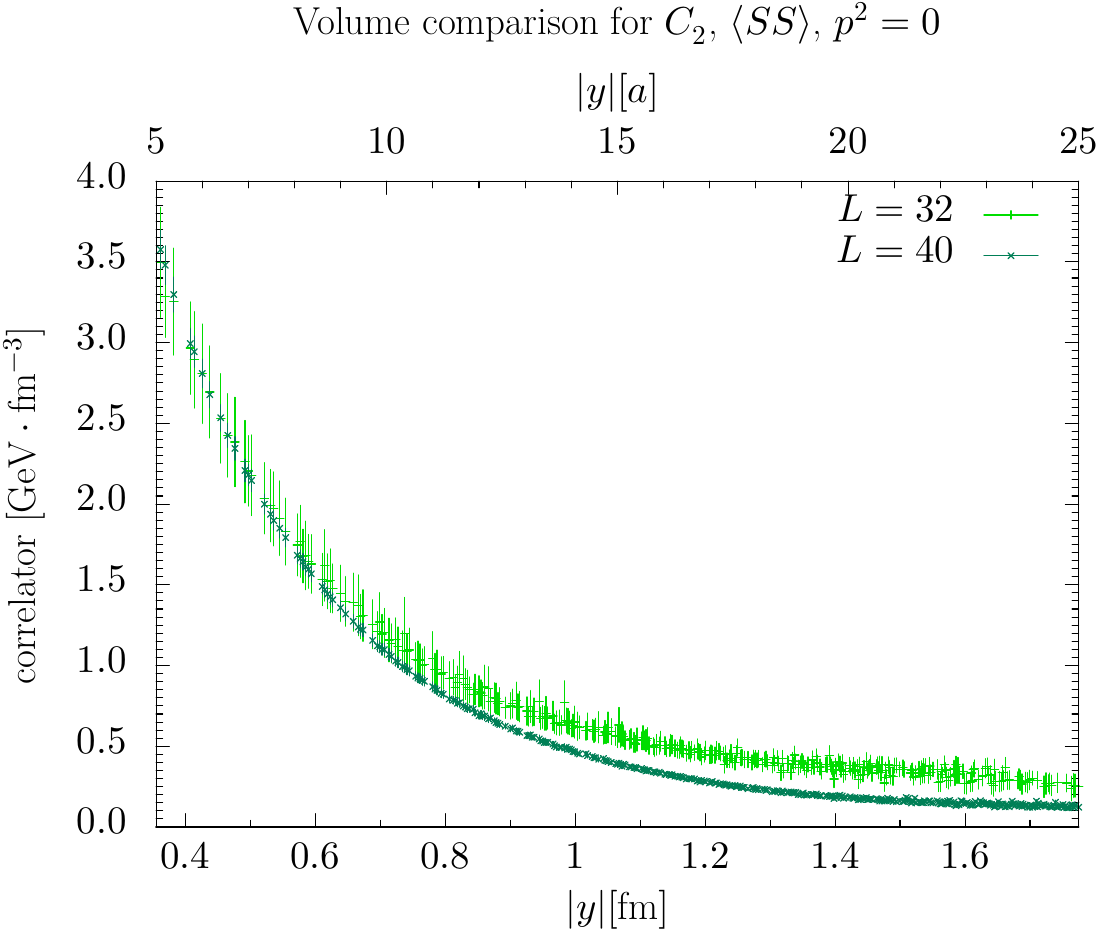}}
  \hfill
  \subfigure[$A, \os, p=0$]{\includegraphics[width=0.45\textwidth,trim=0 0 0 14pt,clip]{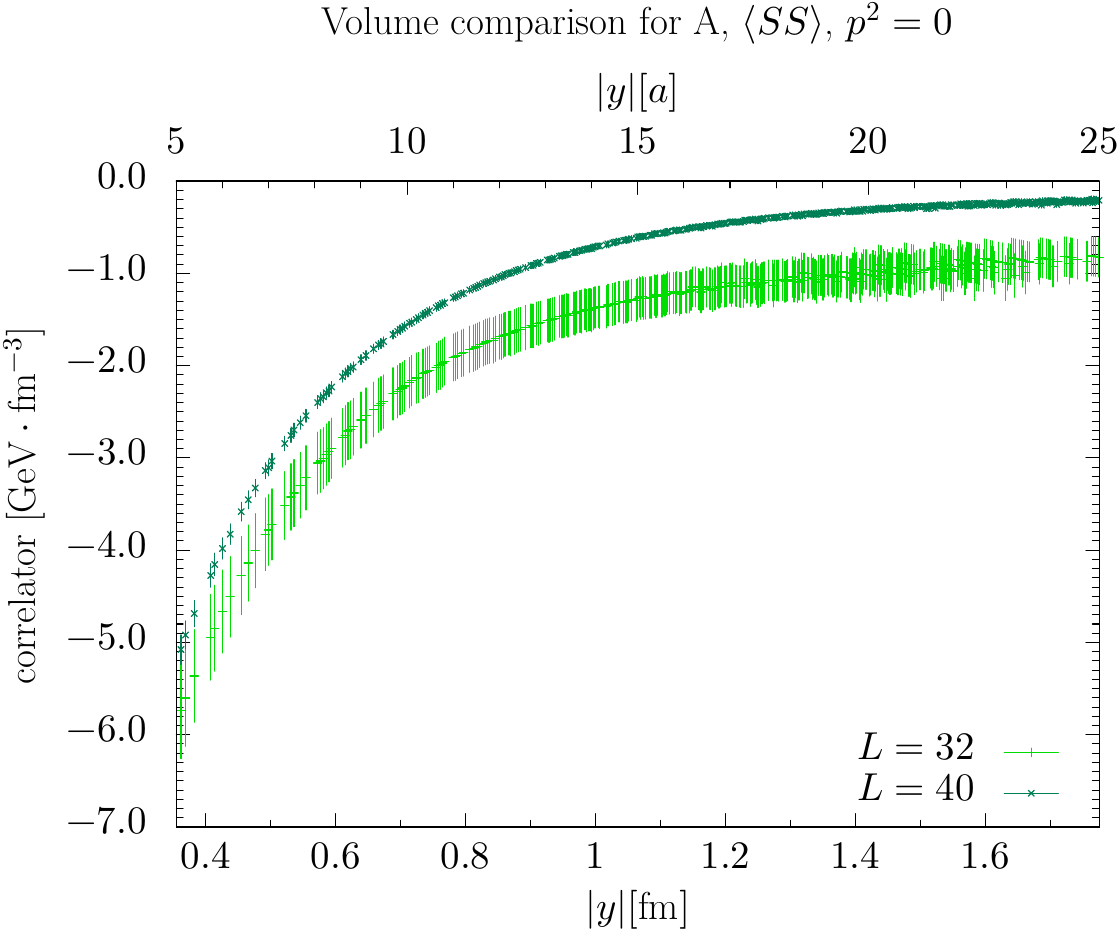}} \\
  \subfigure[$C_1, \op, p=0$]{\includegraphics[width=0.45\textwidth,trim=0 0 0 14pt,clip]{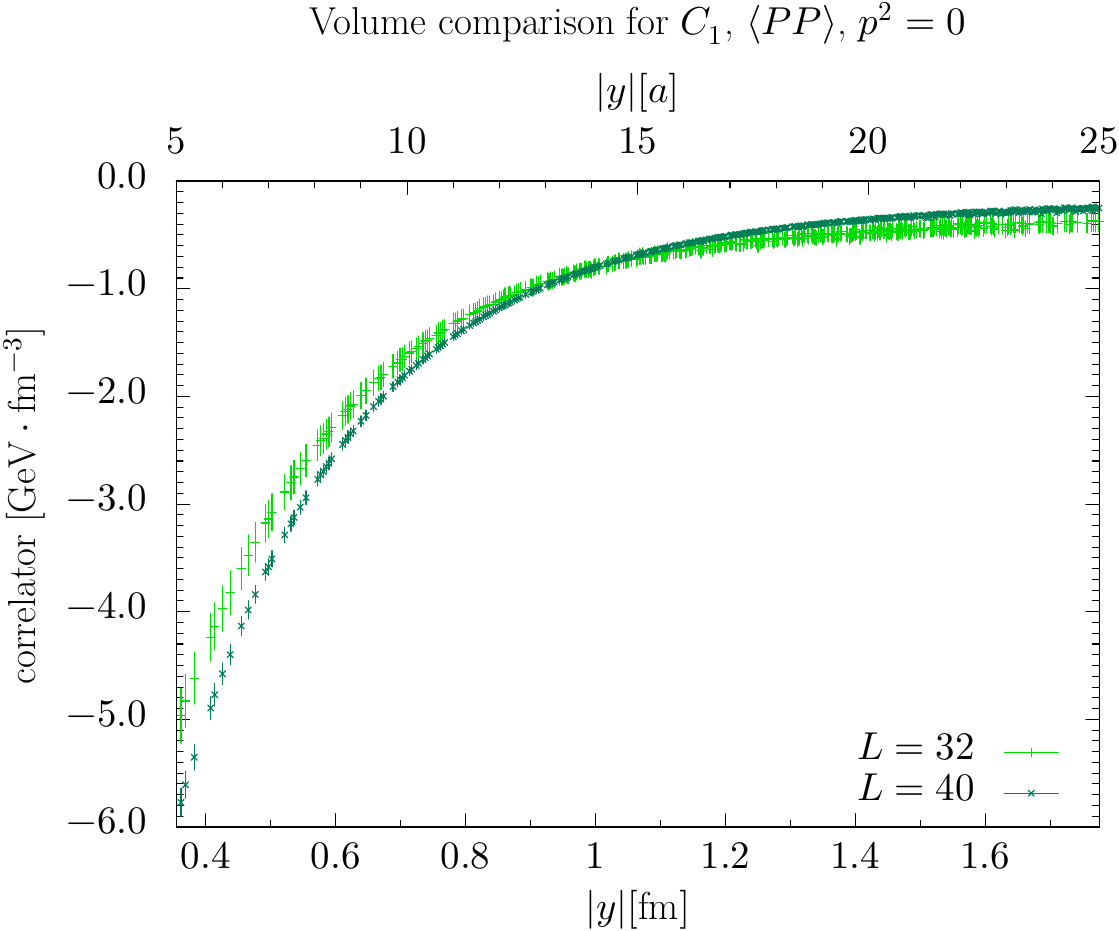}}
  \hfill
  \subfigure[$S_1, \op, p=0$]{\includegraphics[width=0.44\textwidth,trim=0 0 0 14pt,clip]{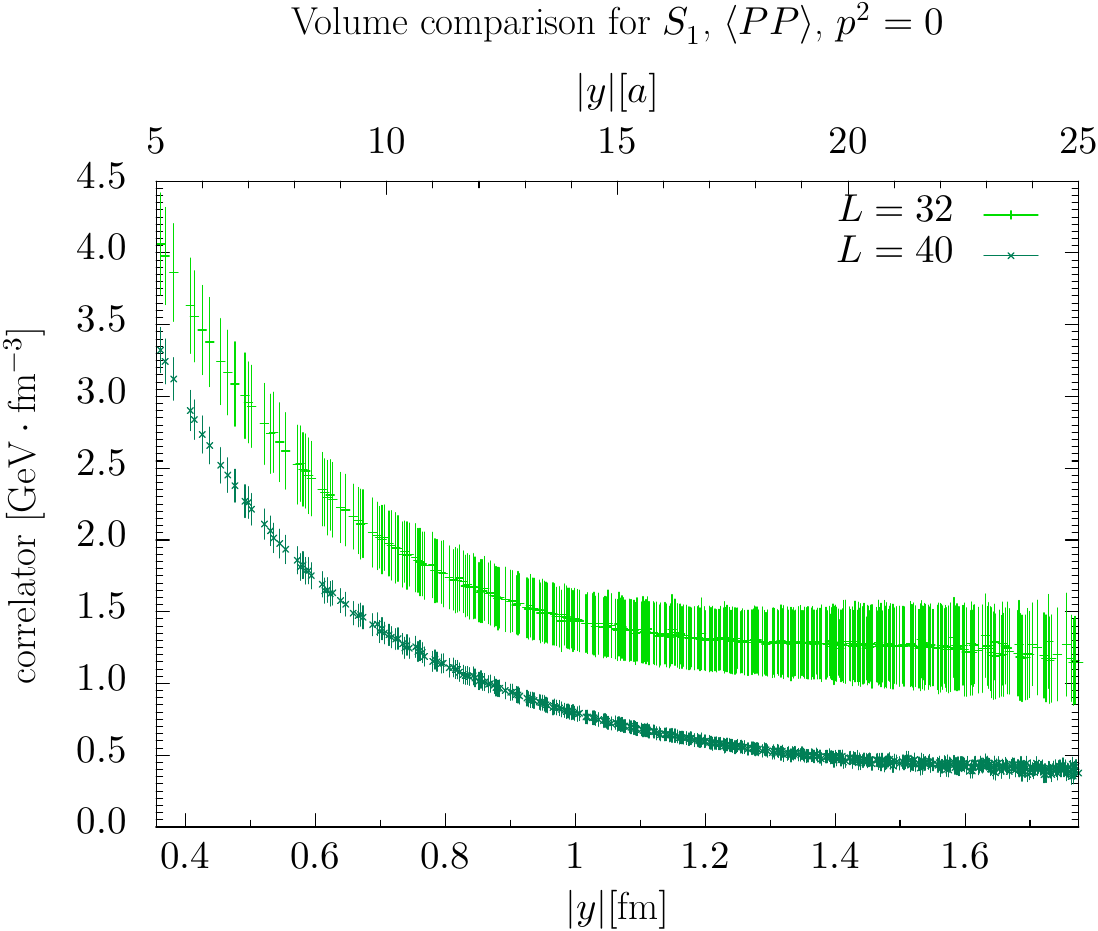}}
\caption{\label{fig:graphs-volcomp} Comparison of correlators for the two spatial volumes used in our simulations.}
\end{center}
\end{figure*}

With lattices of two spatial volumes and otherwise identical parameters, we can investigate finite volume effects on our observables.  Figure~\ref{fig:graphs-volcomp} compares the results of the two lattices with $L=32$ and $L=40$ for a representative selection of contractions and currents.  Here and in the rest of this paper, we show correlators on a linear scale, except in cases where their values cover such a wide range that a logarithmic scale is more advantageous.

As is to be expected, finite size effects are more pronounced at larger distances $y$ between the two currents.  However, we see that, depending on the contraction and on the operator, volume effects may persist down to small $y$.  Clearly, one needs to keep these in mind when comparing our lattice data with predictions e.g.\ from chiral perturbation theory in an infinite volume.  We will come back to this aspect in section~\ref{sec:isospin-amps}.

We notice in figure~\ref{fig:graphs-volcomp} that for the contractions $A$ and $S_1$, the statistical errors are significantly larger for the $L=32$ lattice.  \rev{For $A$ this is plausible from the number of source time insertions, which is $N_{\text{src}} = 4$ for $L=40$ but only $N_{\text{src}} = 1$ for $L=32$ (see table~\ref{tab:statnumbers}).  For $S_1$ the difference in $N_{\text{src}}$ between the two lattices is less pronounced, and we conclude that statistical fluctuations for this disconnected graph are particularly sensitive to the simulation volume.}


\subsection{Finite momentum and boost invariance}
\label{sec:momentum}

\begin{figure*}
\begin{center}
  \subfigure[$C_1, \os, L=40$]{\includegraphics[width=0.48\textwidth,trim=0 0 0 17pt,clip]{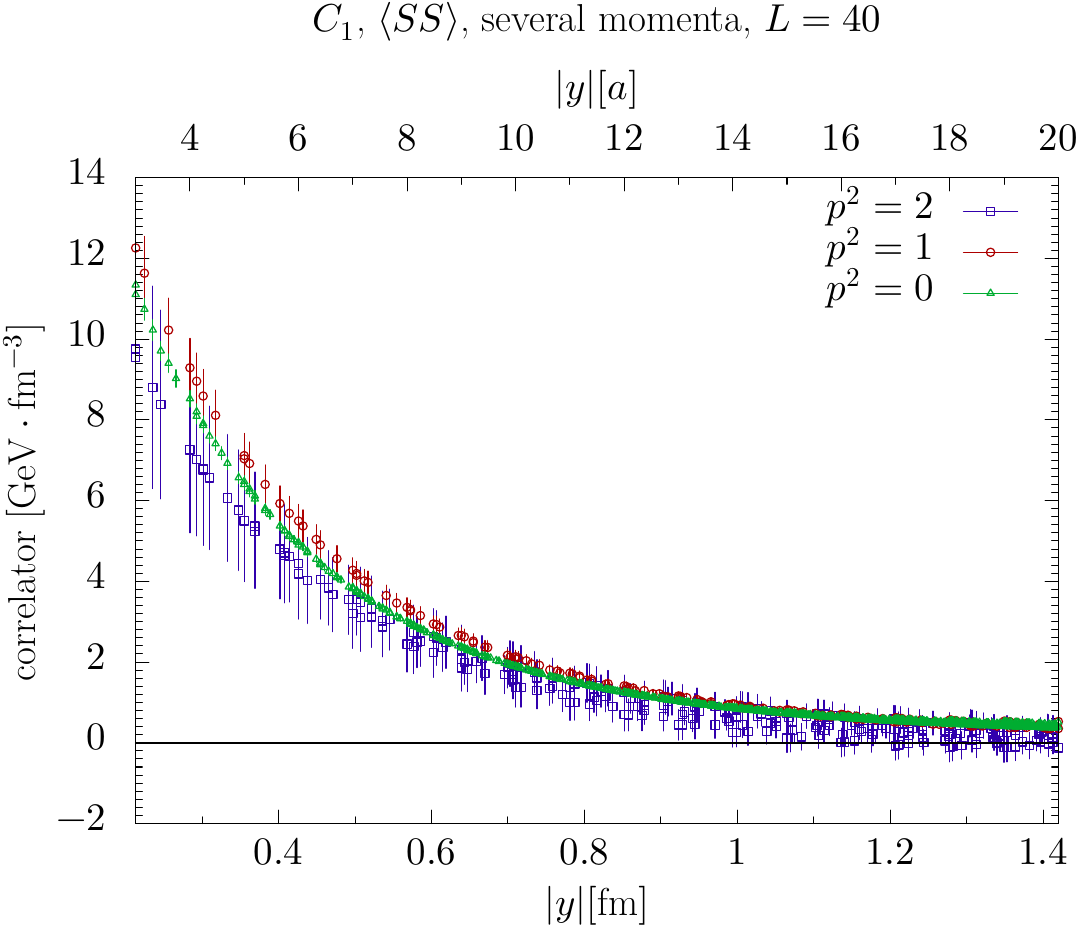}}
  \hfill
  \subfigure[$C_1, \op, L=40$]{\includegraphics[width=0.49\textwidth,trim=0 0 0 17pt,clip]{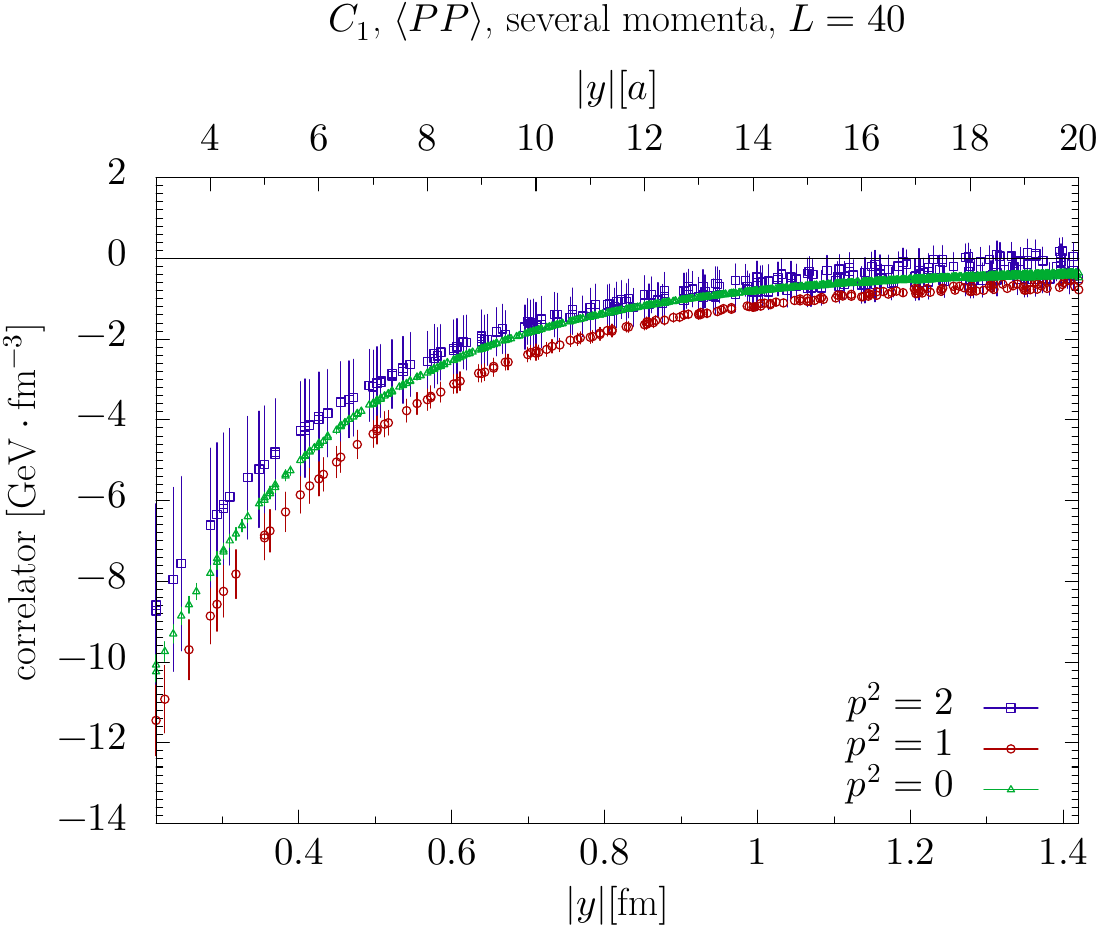}}
  \caption{\label{fig:C1-pcomp} Comparison of the correlator $C_1$ for $\os$ and $\op$ with different pion momenta.  We select distances satisfying $\mvec{p} \cdot \mvec{y} = 0$ and combine lattice data for all $\mvec{p}$ with the same modulus $p = |\mvec{p}|$, which is given in multiples of $2\pi /(La) \approx 437 \mev$.}
\end{center}
\begin{center}
  \subfigure[$F_2, \os, L=40$]{\includegraphics[width=0.49\textwidth,trim=0 0 0 17pt,clip]{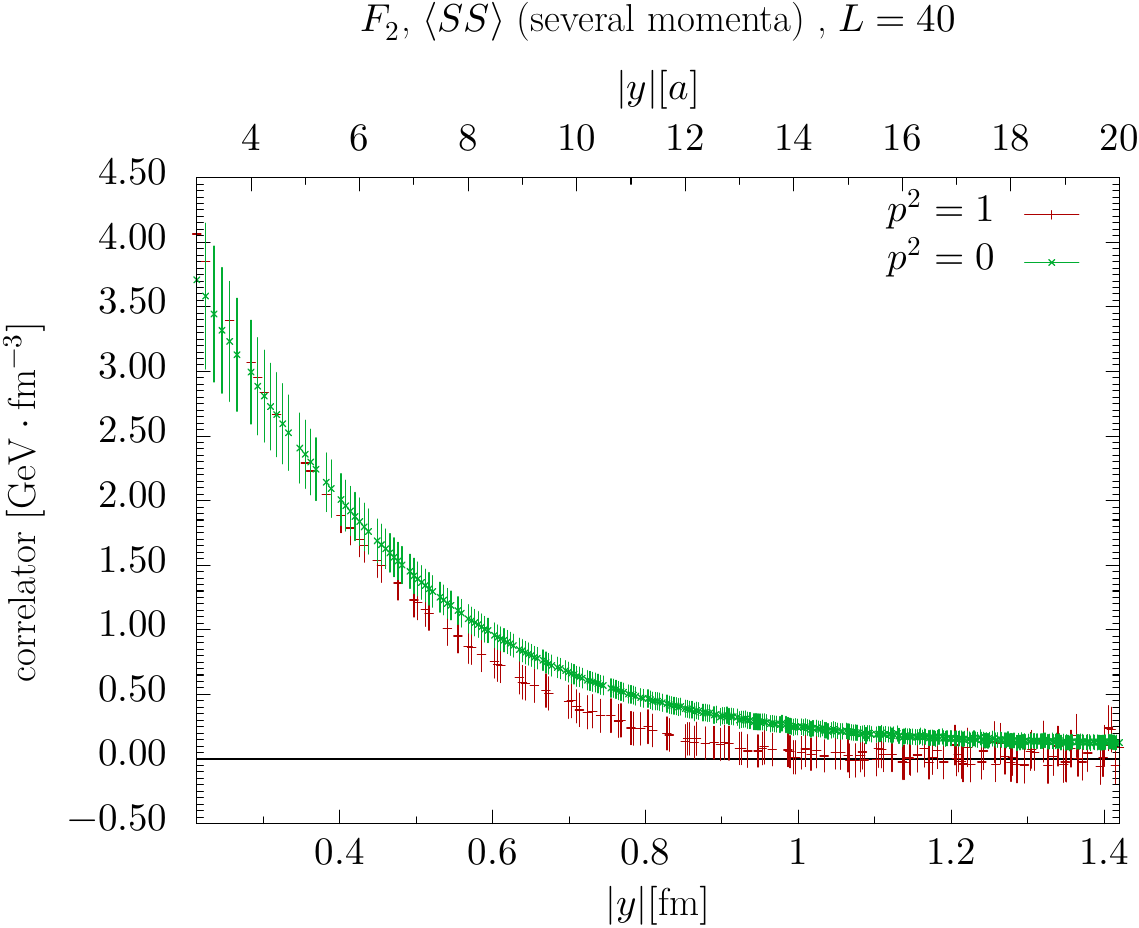}}
  \hfill
  \subfigure[$F_2, \op, L=40$]{\includegraphics[width=0.483\textwidth,trim=0 0 0 17pt,clip]{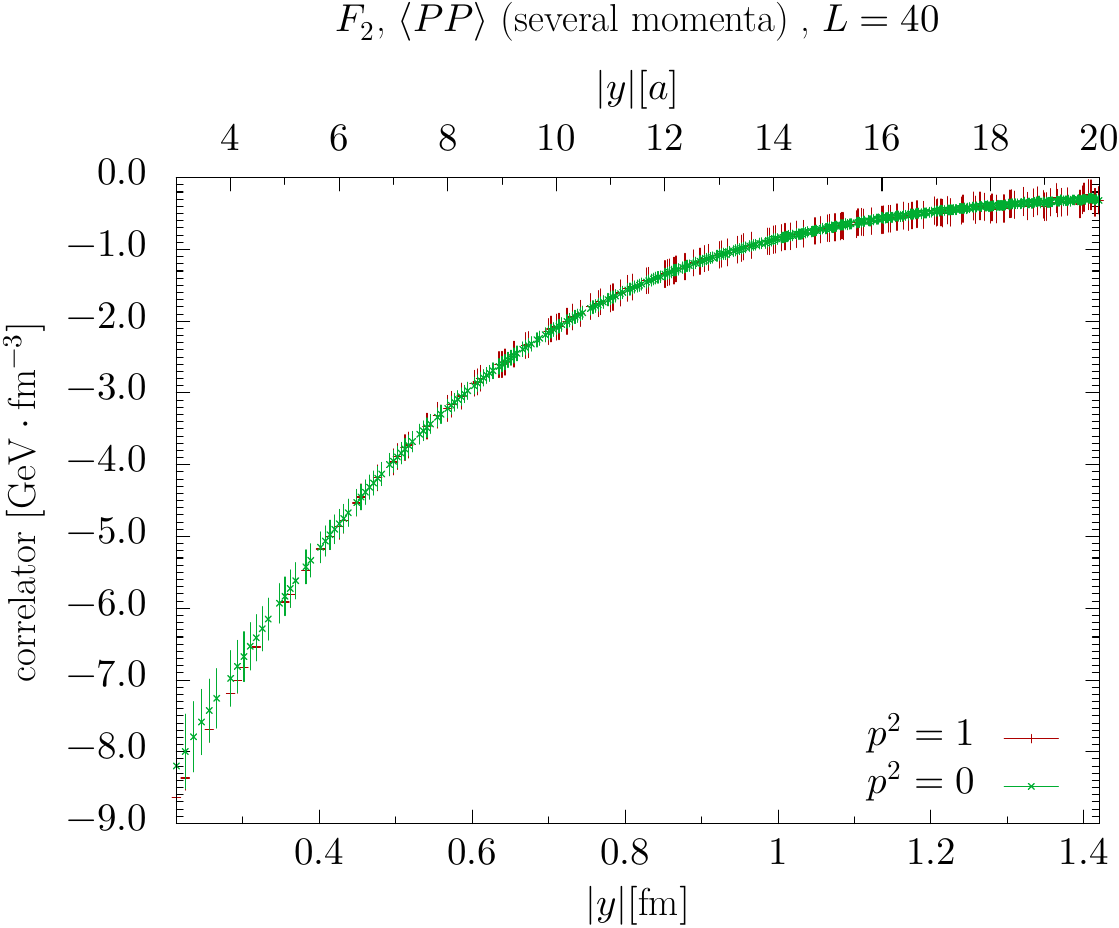}}
  \caption{\label{fig:F2-pcomp} As figure~\protect\ref{fig:C1-pcomp}, but for the  isospin amplitude $F_2 = 2 (C_1 + A)$.}
\end{center}
\end{figure*}

The correlation functions we are studying in this work are matrix elements of two currents at equal time in a pion at rest.  As discussed after equation~\eqref{stat-curr}, we can test that our lattice computation yields boost invariant results using the correlation functions $\os$ and $\op$.  These are Lorentz invariant and hence depend only on the invariant scalar products $y^2$ and $p y$.  Taking always $y^0 = 0$, this means that the correlation functions in these channels must coincide for equal $\mvec{y}^2$ if we select distance vectors with $\mvec{p} \cdot \mvec{y} = 0$.  This comparison is shown for the contraction $C_1$ in figure~\ref{fig:C1-pcomp} with momenta satisfying $\mvec{p}^2 = 0, 1, 2$, where $\mvec{p}$ is given in multiples of $2\pi /(La) \approx 437 \mev$.  We see a good signal for the higher momenta and full agreement between data for different momenta within statistical uncertainties.  This agreement persists for $\mvec{p}^2 = 3$, which is not shown in the figure because the larger errors would obscure the clarity of the plot.

We also have finite momentum data for the contraction $A$, restricted in this case to $\mvec{p}^2 = 1$.  We compare this with zero momentum data in figure~\ref{fig:F2-pcomp} for the combination $F_2 = 2 (C_1 + A)$ of contractions, which according to \eqref{real-combs} represents a physical matrix element in a definite isospin channel.  Again, the data for different momenta are in good agreement with each other.

%% file: graphs.tex
\section{Results for individual lattice contractions}
\label{sec:graphs}

In this section we present results for the individual lattice contractions shown in figure~\ref{fig:contractions}.  Having compared zero and finite pion momenta in section~\ref{sec:momentum}, we restrict our attention to zero momentum data from now on.

\subsection{Light quarks}
\label{sec:light-contr}

\begin{figure*}
\begin{center}
  \subfigure[$\ov, p=0, L=40$]{\includegraphics[width=0.48\textwidth,trim=0 0 0 17pt,clip]{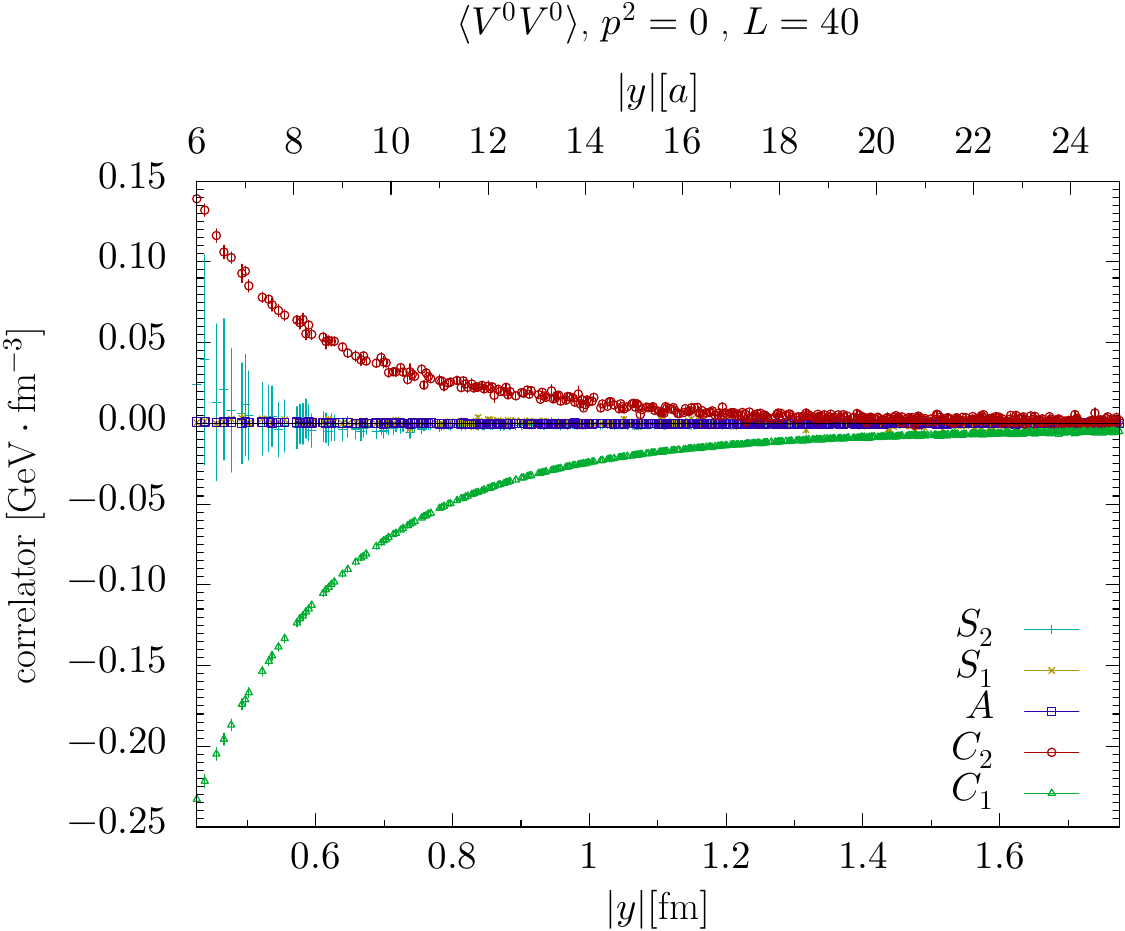}}
  \hfill
  \subfigure[$\oa, p=0, L=40$]{\includegraphics[width=0.48\textwidth,trim=0 0 0 17pt,clip]{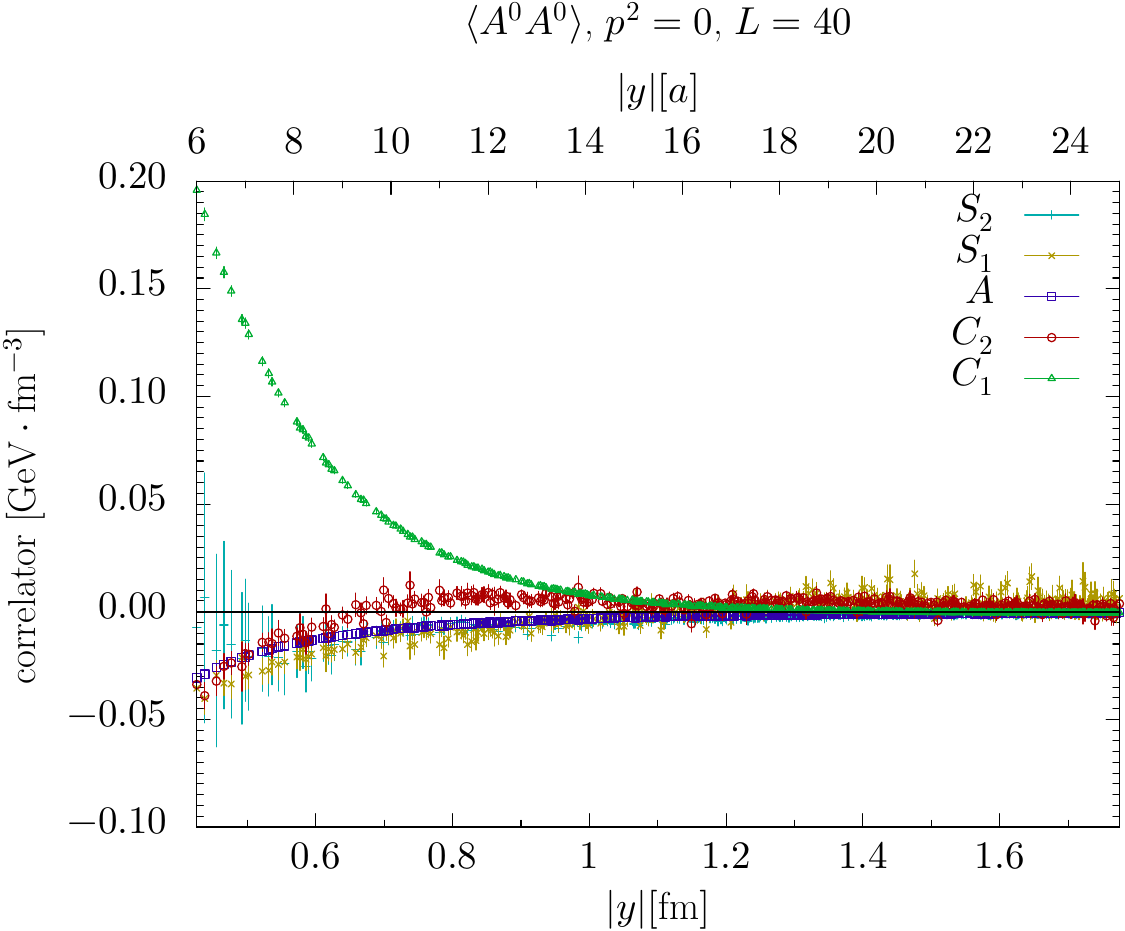}} \\
  \subfigure[$\os, p=0, L=40$]{\includegraphics[width=0.48\textwidth,trim=0 0 0 17pt,clip]{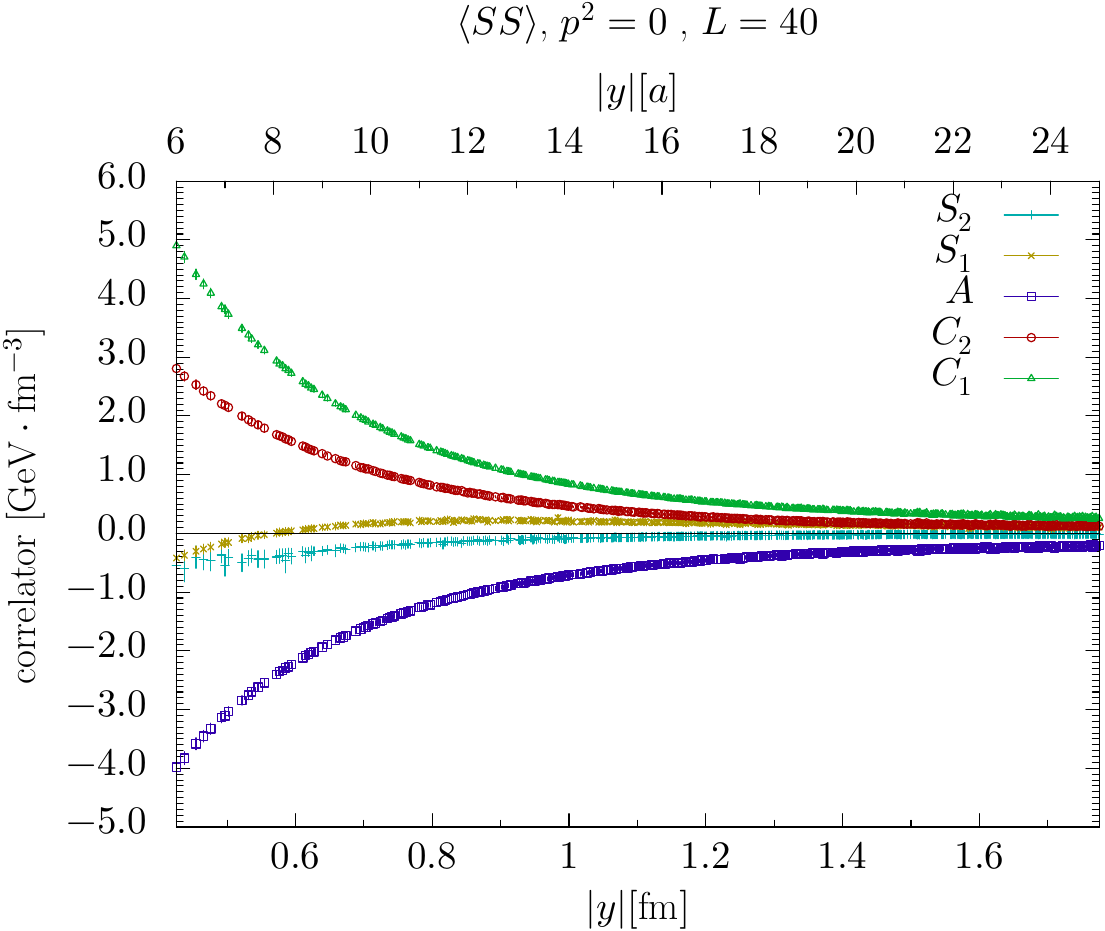}}
  \hfill
  \subfigure[$\op, p=0, L=40$]{\includegraphics[width=0.48\textwidth,trim=0 0 0 17pt,clip]{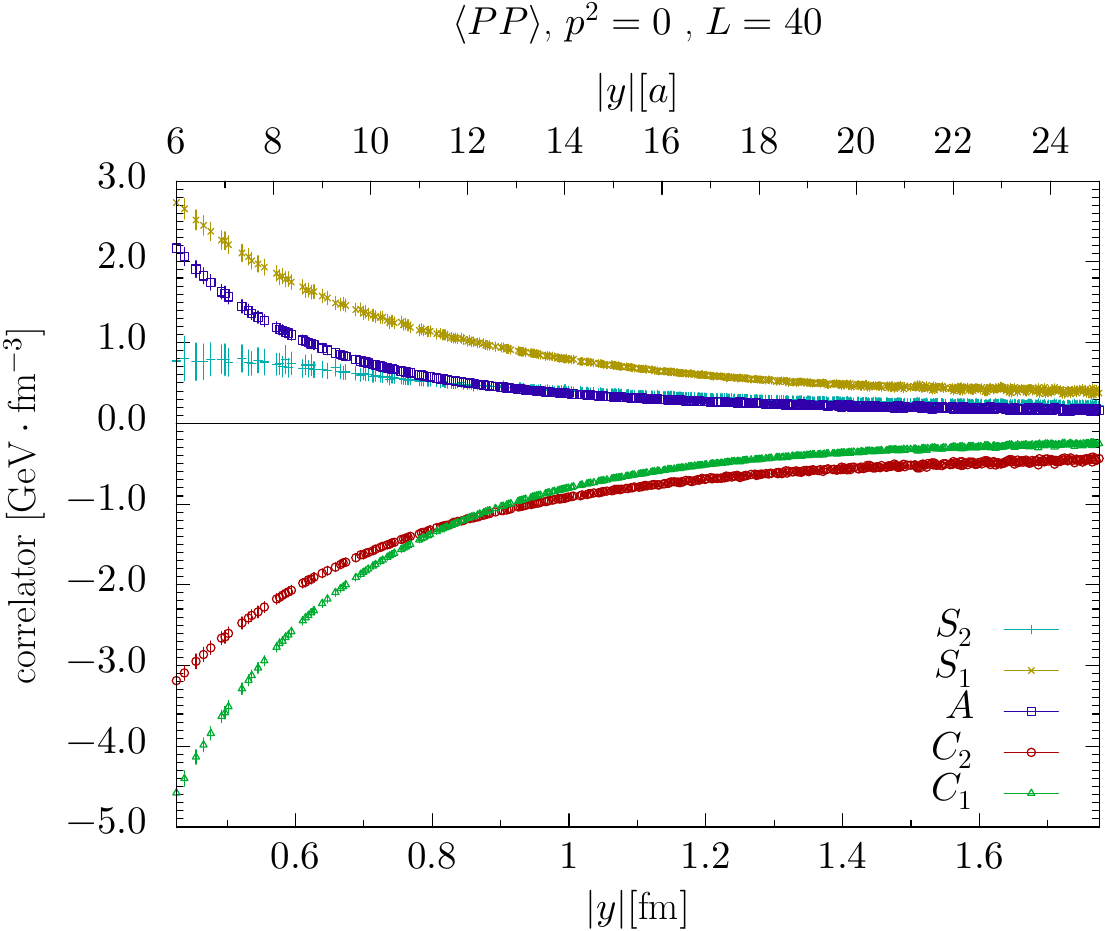}}
  \caption{\label{fig:contr-results}  Correlators of the four current combinations considered in this work.  The contributions of all graphs are shown separately, except for $D$, which has huge statistical errors.}
\end{center}
\end{figure*}

An overview of the different contractions for each current combination is given in figure~\ref{fig:contr-results}.  We show all contractions except for the doubly disconnected graph $D$, where we find huge errors in all channels.  Not only do these errors prevent us from seeing a signal for $D$, but in addition they are large compared with the signal of all other contractions, so that even the upper bounds on $D$ that we could extract from our data would not be very useful.  The large statistical fluctuations of $D$ can be traced back to the vacuum subtraction terms in \eqref{S2-lat}.  For all currents, graph $D$ has a fully connected contribution $\langle G_{2}\, L_1^{i \phantom{j\!\!}}\ms L_1^j \ms\rangle$ and a vacuum subtraction term $\langle G_{2} \rangle \, \langle L_1^{i \phantom{j\!\!}}\ms L_1^j \ms\rangle$ (plus additional subtractions involving $\langle\langle L_1 \rangle\rangle$ in the case of $\os$).  Whilst we see relatively good signals for the separate terms in some cases, at least for smaller distances $y$, the connected and subtraction terms turn out to be almost equal in size and opposite in sign.  Adding them, we lose any signal in the statistical noise.  We will discuss in section~\ref{sec:isospin-amps} how to deal with this situation for physical matrix elements that receive contributions from $D$.

Let us return to figure~\ref{fig:contr-results}.  Given the strong decrease of most correlators with $y$, we start the plots at $y = 6a \approx 0.43 \fm$ to give a clearer picture of the situation at larger distances.  Compared with what is seen in the figure, no qualitative changes occur at smaller $y$.

We observe a marked difference between the different currents as regards the relative importance of different contractions.  For the vector current, the connected graphs $C_1$ and $C_2$ are clearly dominant, whilst the disconnected contributions $S_1$, $S_2$ and the annihilation graph $A$ are much smaller and in fact consistent with zero.  In the axial current case, $C_1$ is clearly dominating over all other contractions at smaller distances; for $y$ above $1 \fm$ the errors on $C_2$ and $S_2$ become too large to draw strong conclusions.  For both $\os$ and $\op$ we obtain a very clear signal for all contractions over a wide $y$ range and see that apart from $C_1 $ and $C_2$ the annihilation graph $A$ is of appreciable size, along with $S_1$ and $S_2$ in the case of $\op$.

We note that the graph $C_1$ is only dominant in the case of $\oa$ and at relatively small distances.  This is among the most striking results of our study, given that one readily associates this graph with the ``valence content'' of a pion and might have expected it to be most prominent in general.


\subsection{Quark mass dependence}
\label{sec:quark-mass}

Let us see how the situation described in the previous subsection changes with increasing quark mass.  The quark and pion mass values of our study are given in \eqref{quark-masses} and \eqref{fitted-mpi}.  With the strange quark mass, we have data only for the contractions $C_1$ and $A$, whilst for charm we have results for all contractions except $S_2$.  The errors on $D$ are again huge in this case, which we will discuss no further here.

\begin{figure*}
\begin{center}
  \subfigure[$\ov, |\ms C_1|, p=0, L=40$]{\includegraphics[width=0.475\textwidth,trim=0 0 0 20pt,clip]{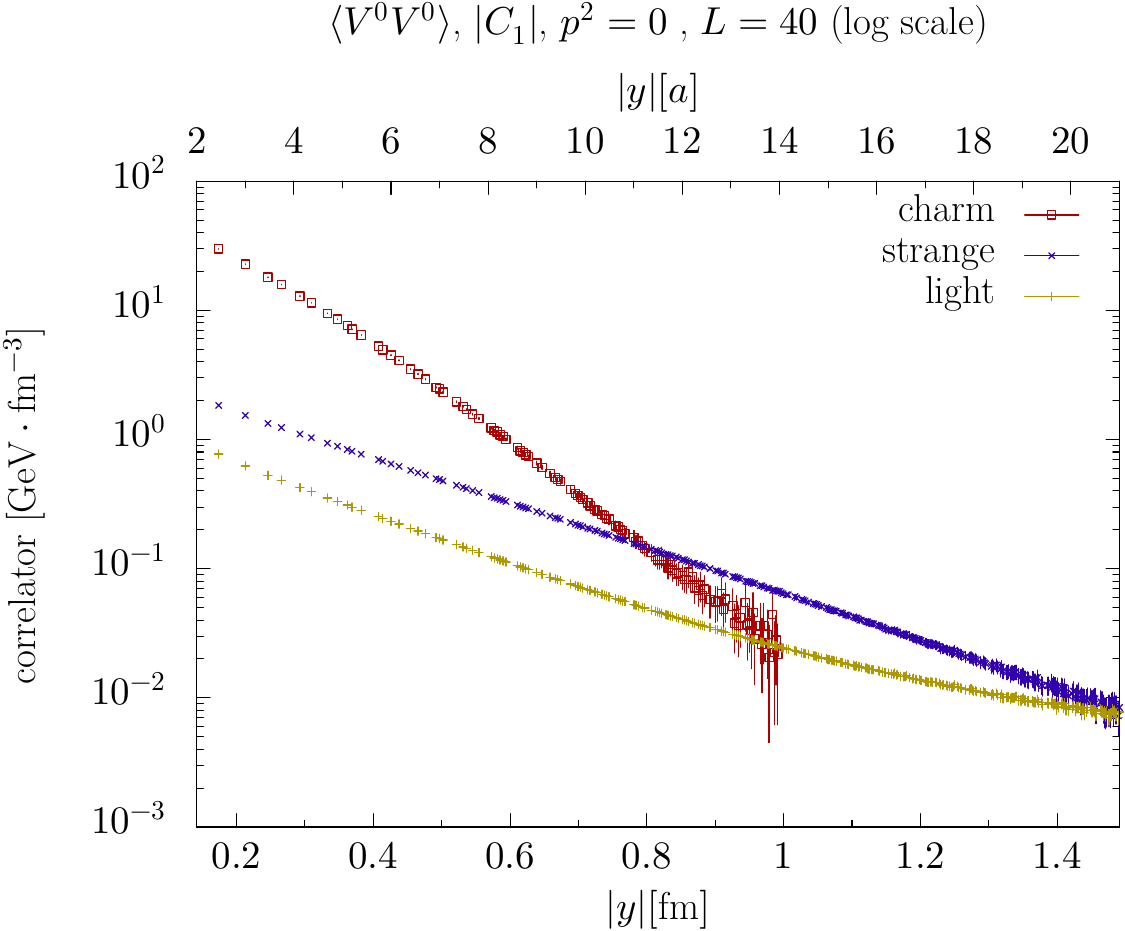}}
  \hfill
  \subfigure[$\oa, |\ms C_1|, p=0, L=40$]{\includegraphics[width=0.475\textwidth,trim=0 0 0 20pt,clip]{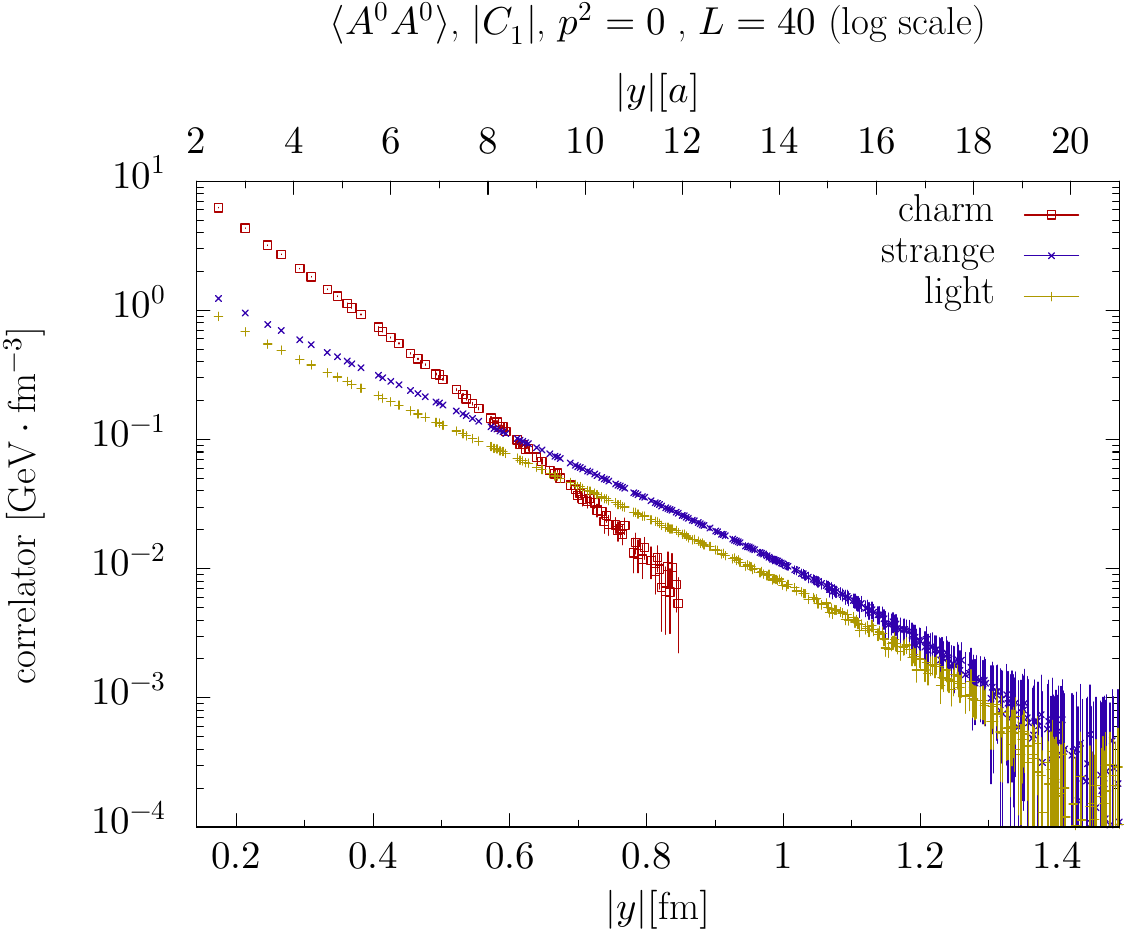}}
\\
  \subfigure[$\os, |\ms C_1|, p=0, L=40$]{\includegraphics[width=0.475\textwidth,trim=0 0 0 17pt,clip]{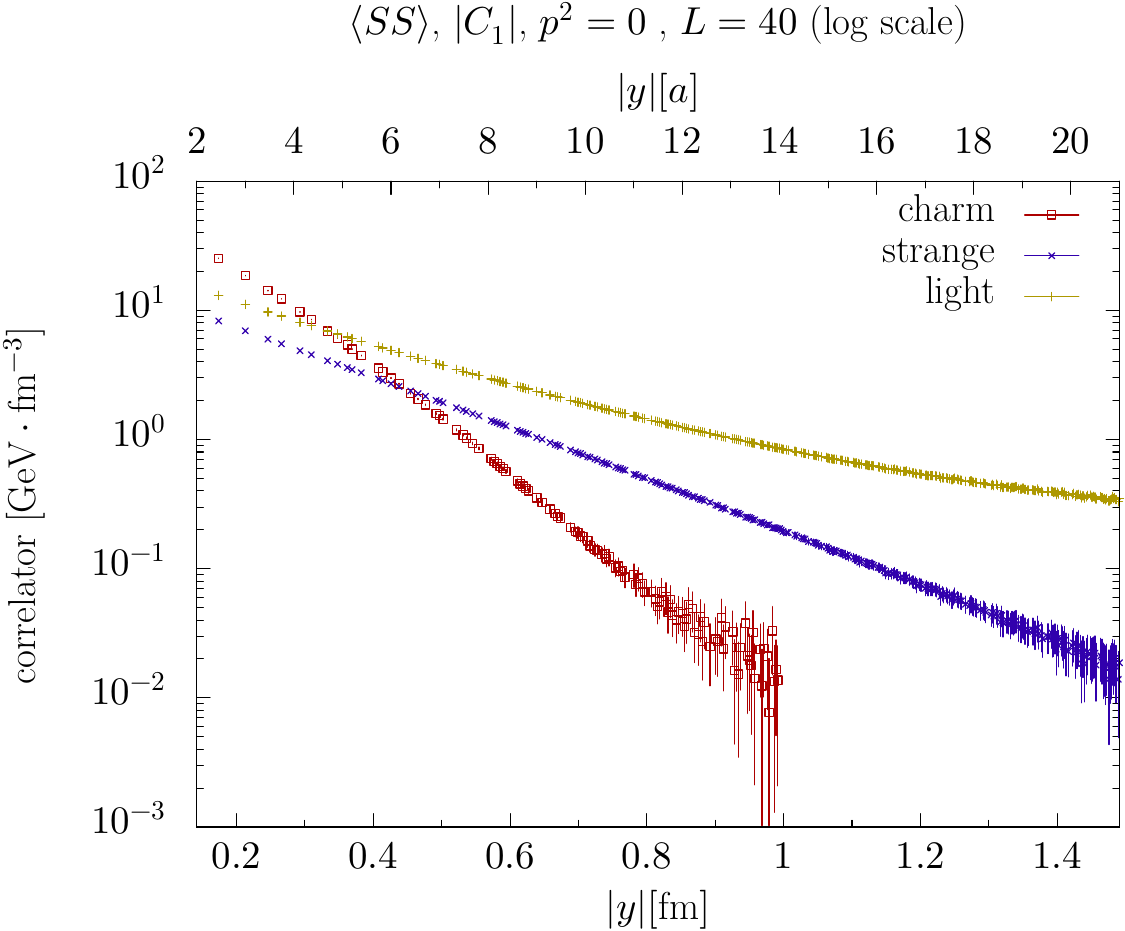}}
  \hfill
  \subfigure[$\op, |\ms C_1|, p=0, L=40$]{\includegraphics[width=0.475\textwidth,trim=0 0 0 17pt,clip]{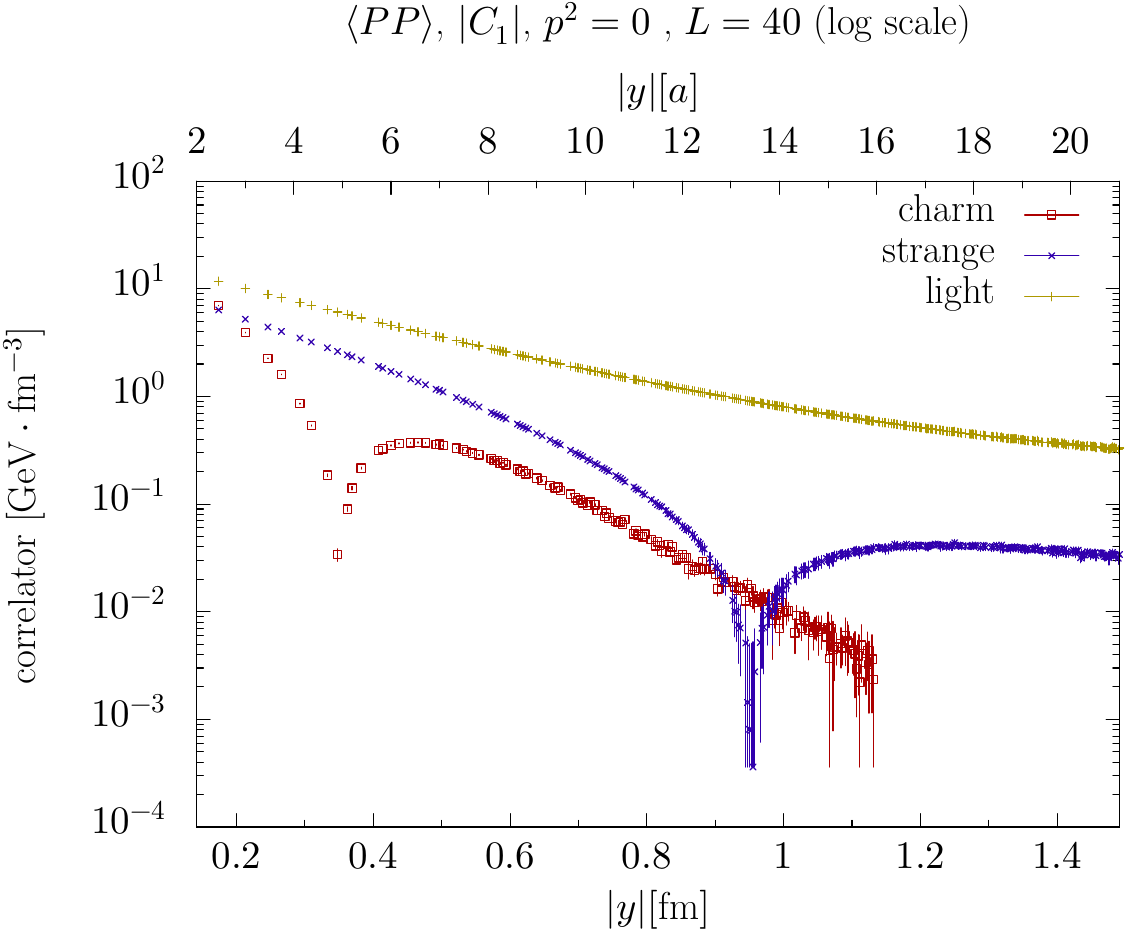}}
  \caption{\label{fig:C1-quark-masses}  Comparison of $|\ms C_1|$ for the three different quark masses of our study.  For clarity, we have omitted data points for which the statistical error is too large compared with the central value.  The $\op$ correlator undergoes a sign change for strange and charm quarks.}
\end{center}
\end{figure*}

We compare the correlator $C_1$ for our three quark masses in figure~\ref{fig:C1-quark-masses}.  As is to be expected, the behaviour for charm quarks differs more strongly from the two other cases than light and strange quarks do from each other.  The steeper decrease with $y$ for charm quarks is also expected and indicates that a ``pion'' made from heavy quarks is more compact.  \rev{We will return to this in section~\ref{sec:rms}.}  In the case of $\op$, we observe that $C_1$ crosses zero (seen as a sharp dip in the logarithmic plot).  This happens around $0.95 \fm$ for strange quarks and around $0.35 \fm$ for charm.

\begin{figure*}
\begin{center}
  \subfigure[$\ov, p=0, L=40$, charm mass]{\includegraphics[width=0.48\textwidth,trim=0 0 0 17pt,clip]{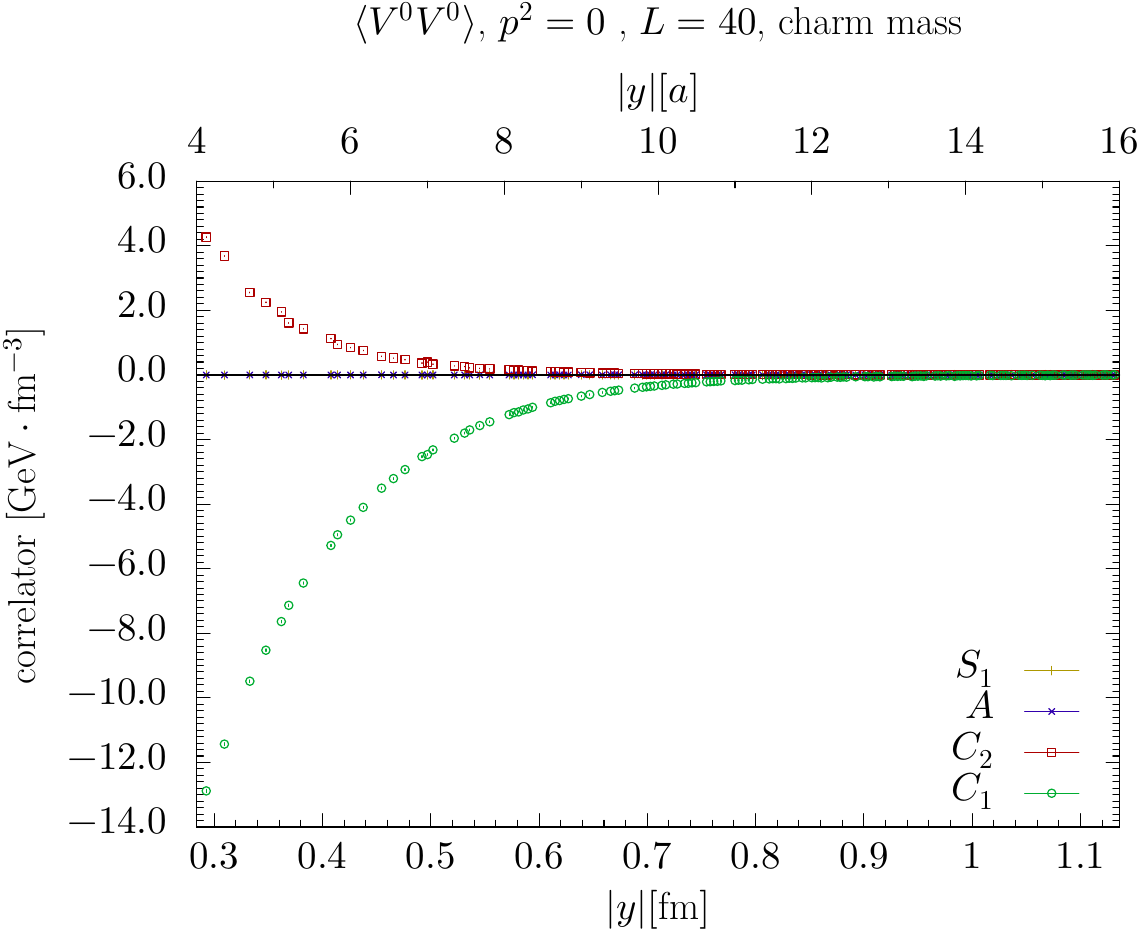}}
  \hfill
  \subfigure[$\oa, p=0, L=40$, charm mass]{\includegraphics[width=0.48\textwidth,trim=0 0 0 17pt,clip]{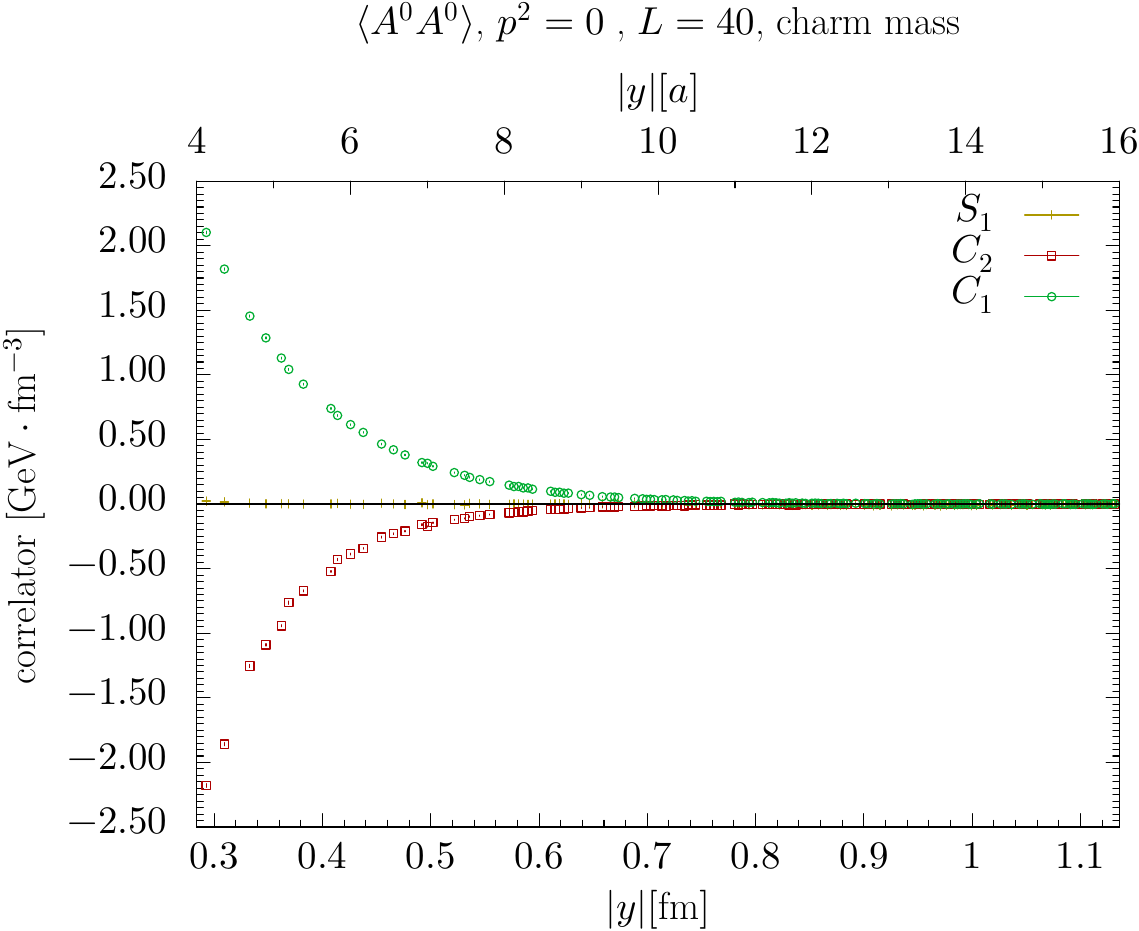}}
\\
  \subfigure[$\os, p=0, L=40$, charm mass]{\includegraphics[width=0.48\textwidth,trim=0 0 0 17pt,clip]{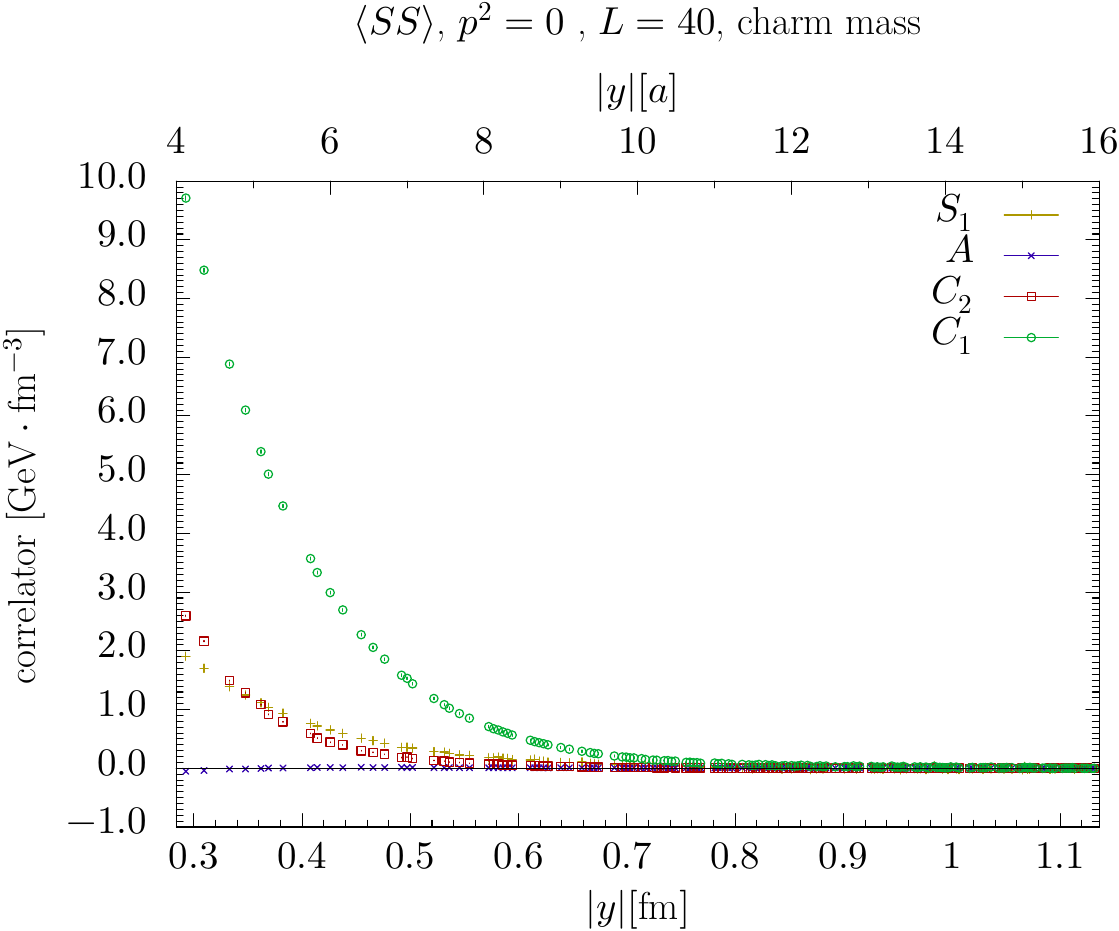}}
  \hfill
  \subfigure[$\op, p=0, L=40$, charm mass]{\includegraphics[width=0.49\textwidth,trim=0 0 0 17pt,clip]{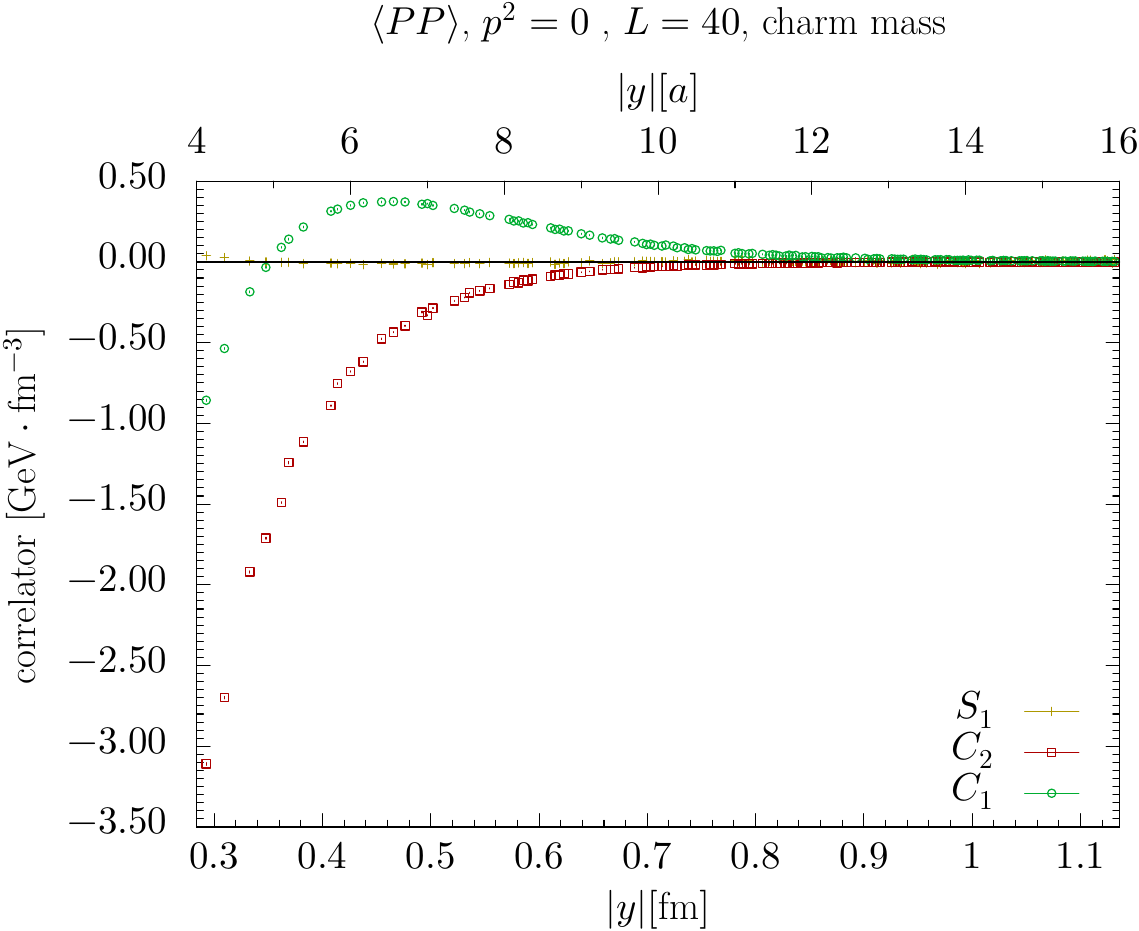}}
  \caption{\label{fig:charm}  As figure~\protect\ref{fig:contr-results} but for charm quarks.  The contribution from the annihilation graph $A$ is not shown for $\oa$ and $\op$, where its relatively large errors would obscure the plots.}
\end{center}
\end{figure*}

Not shown in our plots is the annihilation contribution for strange quarks.  It is very small compared with $C_1$, except in the $\os$ channel, where the ratio $-A/C_1$ is about $40 \%$ at $y \sim 0.5 \fm$ and decreases with $y$.  The annihilation mechanism thus significantly loses importance when going from light to strange quark masses.  With charm quarks, we find $A$ to be negligible in all channels.

In figure~\ref{fig:charm} we show different contractions for charm quarks.  Compared with the case of light quarks in figure~\ref{fig:contr-results}, we see that not only the importance of $A$ but also the one of $S_1$ is considerably reduced for charm.  In turn, $C_2$ remains important compared with $C_1$ for both $\oa$ and $\op$.  The expectation that $C_1$ is the dominant contraction in each channel is thus not even realised for charm quarks.

Looking at the largest contraction in each current combination, we observe that for charm quarks $\oa$ is small compared to $\ov$ and $\op$ relatively small compared to $\os$.  By contrast, for light quarks $\oa$ is comparable in size to $\ov$, and $\op$ is comparable in size to $\os$.  Our results for charm quarks confirm that in the heavy-quark limit, matrix elements of the spin dependent currents $A^\mu$ and $P$ are suppressed compared with their spin independent counterparts $V^\mu$ and $S$.  Seen from this angle, the contraction $C_1$ is indeed dominant for charm because it is the most prominent contribution to $\ov$ and $\os$.


\subsection{Test of factorisation}
\label{sec:factorise}

The two-current matrix elements $\ov$ and $\os$ quantify the distribution of vector or scalar charges at two separate points in the pion.  By contrast, the corresponding single-current matrix elements $\bra{\pi(p')} V^0 \ket{\pi(p)}$ and $\bra{\pi(p')} S \ket{\pi(p)}$ give the vector or scalar charge distribution at one point in the pion if one performs a Fourier transform from the momentum transfer $\mvec{p} - \mvec{p}'$ to three-dimensional space.\footnote{%
As is well known and discussed e.g.\ in \cite{Burkardt:2000za}, this simple interpretation receives relativistic corrections, which become relevant for distances of order $1/\mpi$ or below.  Our treatment here is fully relativistic and not affected by this limitation.}
In the absence of correlations, the two-current distribution can be computed from the single-current one.  A detailed analysis of this has already been given in \cite{Burkardt:1994pw}, with some focus on the non-relativistic limit.  A way to express the expectation for the two-current distribution in the absence of correlations is to insert a complete set of intermediate states between the two currents and to keep only the pion ground state \cite{Burkardt:1994pw,Diehl:2011yj}.  In momentum space, we have
\begin{align}
\label{matels-delta}
\mathcal{M}_{ii}(\mvec{q}^2) &=
\int d^3{y}\; e^{i \mvec{q} \mvec{y}}\,
   \bra{\pi^+(p)} \mathcal{O}_i^{uu}(y) \ms \mathcal{O}_i^{dd}(0) \ket{\pi^+(p)}
\nonumber \\
 & \overset{?}{=}
   \int d^3{y}\; e^{i \mvec{q}\mvec{y}}
   \int \frac{d^3{p}'}{(2\pi)^3\ms 2p'^{\ms 0}}
     \bra{\pi^+(p)} \mathcal{O}_i^{uu}(y) \ket{\pi^+(p')}\,
     \bra{\pi^+(p')} \mathcal{O}_i^{dd}(0) \ket{\pi^+(p)}
\nonumber \\
 &= \frac{\eta^i_C}{2 E_{q}}\;
   \bigl| \bra{\pi^+(E_{q}, - \mvec{q})}
      \mathcal{O}_i^{uu}(0) \ket{\pi^+(p)} \bigr|^2 \,,
\end{align}
where $E^2_{q} = \mpi^2 + \mvec{q}^2$ and it is understood that $\mvec{p} = \mvec{0}$.  The question mark above the equality sign indicates an assumption, which will be tested in the following.
In the last step of \eqref{matels-delta} we used the relation  $\bra{\pi^+} \mathcal{O}_i^{dd} \ket{\pi^+} = \eta^i_C\, \bra{\pi^+} \mathcal{O}_i^{uu} \ket{\pi^+} $ with the $C$ parity  $\eta^i_C$ defined in \eqref{C-parity}.  Defining vector and scalar form factors as
\begin{align}
\label{ff-defs}
\bra{\pi^+(p')} V_{uu}^\mu(0) \ket{\pi^+(p)} & = (p' + p)^\mu\ms F_V(Q^2) \,,
&
\bra{\pi^+(p')} S_{uu}(0) \ket{\pi^+(p)} & = F_S(Q^2)
\end{align}
with $Q^2 = - (p - p')^2$, we get
\begin{align}
\label{stat-mat-els}
{}- \mathcal{M}_{V^0 V^0}(\mvec{q}^2) & \overset{?}{=} {}
  \frac{(\mpi + E_{q})^2}{2 E_{q}} \,
    \Bigl[\ms F_V\bigl( 2 \mpi E_{q} - 2 \mpi^2 \bigr) \ms\Bigr]^2 \,,
\nonumber \\
  \mathcal{M}_{SS}(\mvec{q}^2) & \overset{?}{=} {}
  \frac{1}{2 E_{q}} \,
  \Bigl[\ms F_S\bigl( 2 \mpi E_{q} - 2 \mpi^2 \bigr) \ms\Bigr]^2 \,.
\end{align}
In momentum space, the absence of correlations thus implies that the two-current matrix element factorises into the product of two single-current form factors (and a kinematic prefactor).  To compare the matrix elements in position space, we analytically Fourier transform the r.h.s.\ of \eqref{stat-mat-els} using a parameterisation of the form factors.
We note in passing that in the limit $\mvec{q}^2 \ll \mpi^2$ the argument of the form factors in \eqref{stat-mat-els} becomes $\mvec{q}^2$, as appropriate for a non-relativistic treatment.  For our study with $\mpi \approx 300 \mev$, this limit is however not relevant.

It follows from \eqref{phys-matels} that the two-current correlators in \eqref{matels-delta} receive contributions from different contractions in the combination $C_1 + 2 S_1 + D$.  Given the poor quality of our data for $D$, we formulate a version of the factorisation hypothesis that requires only $C_1$ for the two-current correlator and the connected three-point graph $G_3$ for the elastic form factors.  To this end, we use the partially quenched scenario of $n_F = 4$ mass-degenerate quarks $u$, $d$, $s$ and $c$, described at the end of section~\ref{sec:phys-matels}.  In this case, one starts from the matrix element $\bra{\pi^+} \mathcal{O}_i^{uc}(y)\ms \mathcal{O}_i^{sd}(0)  \ket{D_s^+}$ and inserts a full set of intermediate states with  quark content $c \bar{d}$.  Retaining only the ground state term $\ket{D^+} \bra{D^+}$ and using SU(4) flavour symmetry to relate the two single-current matrix elements to each other, one obtains the analogue of \eqref{stat-mat-els}.  Disconnected contributions are in this case excluded by the quantum numbers of the $uc$ and $sd$ currents.  Our lattice computation can be understood as giving the corresponding matrix elements, up to the effect of partial quenching due to the lattice action with $n_F = 2$.


\paragraph{Factorisation at $\boldsymbol{q = 0}$.}

Let us first test the factorisation hypothesis \eqref{stat-mat-els} at $\mvec{q} = \mvec{0}$, where it reads
\begin{align}
\label{fact-del-zero}
{}- (2 \mpi)^{-1} \int d^3{y}\; \ov(\mvec{y}) & \overset{?}{=} 1 \,,
&
2 \mpi \int d^3{y}\; \os(\mvec{y}) & \overset{?}{=} \bigl[ F_S(0) \ms\bigr]^2 \,.
\end{align}
For the vector current we used the normalisation condition $F_V(0) = 1$.
We evaluate the integrals on the l.h.s.\ of \eqref{fact-del-zero} as discrete sums over all lattice sites.\footnote{%
In this case we do not use the cut in \protect\eqref{cos-cut}.  As discussed in section~\protect\ref{sec:aniso}, this leaves us with lattice artefacts from images at large $y$, but this region is not important in the sum over $\mvec{y}$.}
The result for the vector correlator is
\begin{align}
{}- (2 \mpi)^{-1}\, a^3 \, \smash{\sum_{\mvec{y}}} \ov(\mvec{y})
   &= 0.975(33) \text{ ~~(light quarks), }
\nonumber \\
   &= 0.985(31) \text{ ~~(strange), }
\nonumber \\
   &= 1.053(36) \text{ ~~(charm)}
\end{align}
for the three quark masses used in our study.  In all cases, the agreement with factorisation is excellent, despite the possibility of substantial lattice spacing effects for the charm quark.

In the scalar channel we obtain
\begin{align}
\biggl[ 2 m_\pi\ms a^3
  \sum_{\mvec{y}}\; \os(\mvec{y}) \biggr]^{1/2} &=  3.14(4) \gev
\end{align}
for light quarks, which is to be compared with the value $F_S(0) = 2.2 \gev$ we extract from our scalar form factor data (see table~\ref{tab:ff-fits} below).  While this is of the same order of magnitude, factorisation clearly fails in this case.

Let us give a heuristic argument why  the factorisation hypothesis works so well in the vector channel.  $V_{qq}^\mu(y)$ is a conserved current in QCD, so that the associated Noether charge $\rho_q = \int d^3{y}\; V_{qq}^\mu(y)$ corresponds to a good quantum number.  For a positive pion one  has $\bra{\pi^+} \rho_u = \bra{\pi^+}$ and thus obtains
\begin{align}
\int d^3{y}\, \bra{\pi^+} V_{uu}^{0}(y)\ms V_{dd}^{0}(0) \ket{\pi^+}
  &= \bra{\pi^+} \rho_u^{}\ms V_{dd}^{0}(0) \ket{\pi^+}
   = \bra{\pi^+} V_{dd}^{0}(0) \ket{\pi^+} = {}- 2 \mpi
\end{align}
for zero pion momentum.  To turn this argument into a theorem, one would need to discuss possible short-distance singularities of the integrated two-current matrix element at $\mvec{y} = \mvec{0}$.  We shall not attempt this here.
It is clear that the argument just given cannot be extended to the scalar charge, which is not the Noether charge of a conserved current.


\paragraph{Factorisation as a function of ${\boldsymbol y}$.}

We have extracted the vector form factor and the connected part of the scalar form factor from our lattice simulations, using in this case the full number  of 2025 gauge configurations available for our lattice with $L=40$.  To Fourier transform the r.h.s.\ of \eqref{stat-mat-els} to position space, we fit the form factor data to a power law
\begin{align}
\label{ff-pole}
F(Q^2) &= \frac{F(0)}{\bigl( 1 + Q^2 /M^2 \bigr)^{n} \rule{0pt}{1.05em}} \,.
\end{align}
For each form factor, we take two fit variants with different powers $n$, so as to have a handle on the bias of the extrapolation to large $Q^2$, where we have no data.  Such an extrapolation bias is inevitable when we Fourier transform to position space.

\begin{table}
\begin{center}
\renewcommand{\arraystretch}{1.1}
\begin{tabular}{cccccc} \hline \hline
 fit & form factor & $F(0)$ & $M [\gev]$ & $n$ \\
\hline
$1$ & $F_V$ & $1 \text{~(fixed)}$ & $0.777(12)$ & $1 \text{~(fixed)}$ \\
$2$  & & $1 \text{~(fixed)}$ & $0.872(16)$ & $1.173(69)$ \\
\hline
$1$ & $F_S$ & $2.222(19) \gev$ & $1.314(39)$ & $1 \text{~(fixed)}$ \\
$2$ & & $2.212(19)\gev$ & $2.023(50)$ & $2 \text{~(fixed)}$ \\
\hline \hline
\end{tabular}
\end{center}
\caption{\label{tab:ff-fits}  Parameters obtained in fits of our form factor data to \protect\eqref{ff-pole}.}
\end{table}

\begin{figure*}
\begin{center}
  \subfigure[fits of $F_V(Q^2)$, $L=40$]{\includegraphics[width=0.49\textwidth,trim=0 0 0 17pt,clip]{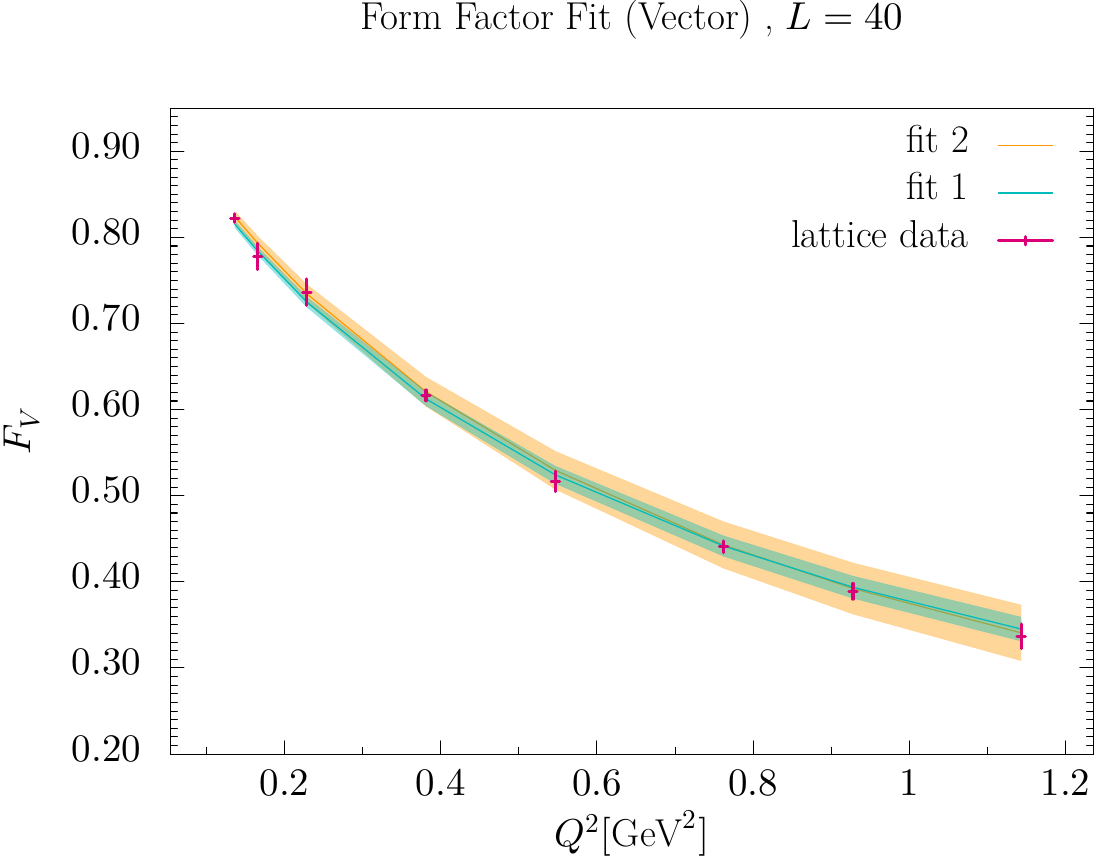}}
  \hfill
  \subfigure[fits of $F_S(Q^2)$, $L=40$]{\includegraphics[width=0.49\textwidth,trim=0 0 0 17pt,clip]{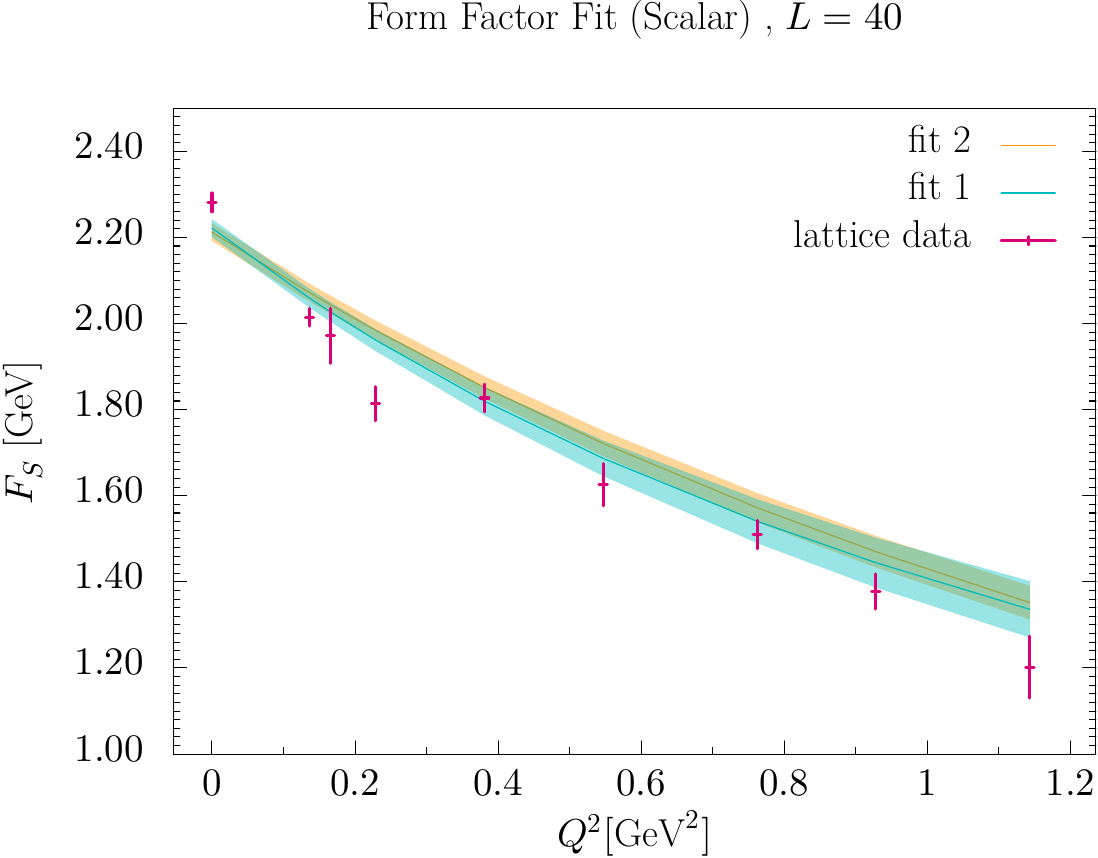}}
  \caption{\label{fig:form-factors} Data for the vector and scalar pion form factors extracted from our simulations, together with the fits specified in table~\protect\ref{tab:ff-fits}.  $F_S$ is given in the $\overline{\mathrm{MS}}$ scheme at scale $\mu = 2 \gev$.}
\end{center}
\begin{center}
  \subfigure[factorisation test for ${}- \ov, L=40$]{\includegraphics[width=0.48\textwidth,trim=0 0 0 17pt,clip]{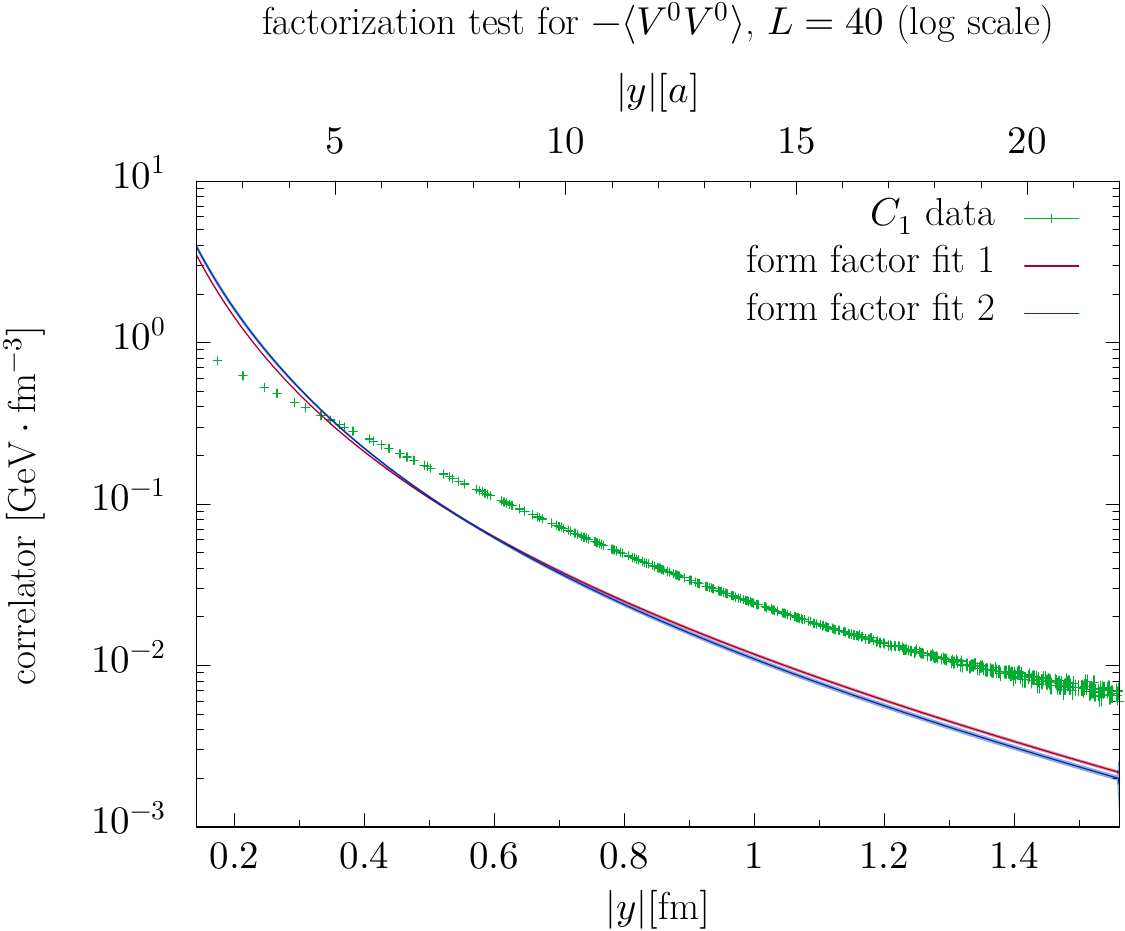}}
  \hfill
  \subfigure[factorisation test for $\os, L=40$]{\includegraphics[width=0.48\textwidth,trim=0 0 0 17pt,clip]{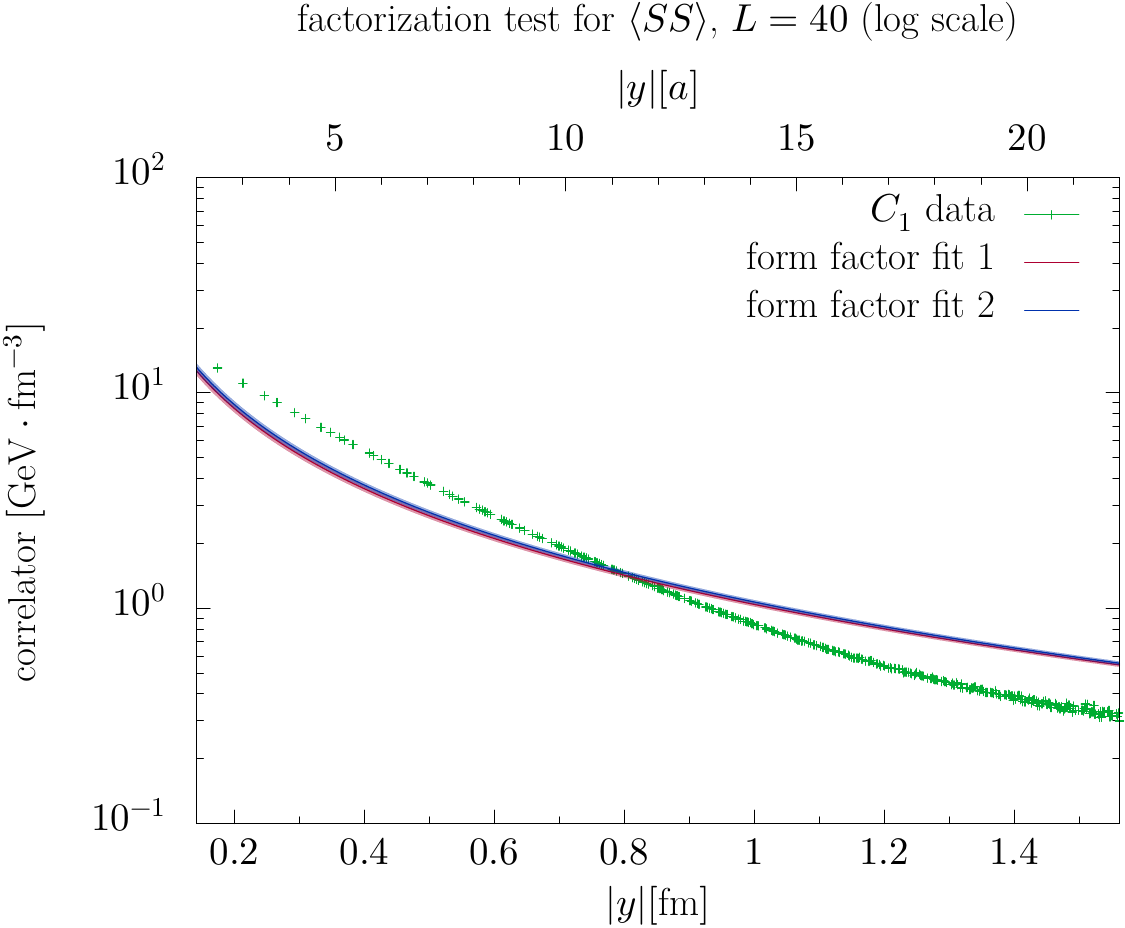}}
\\
  \caption{\label{fig:fact-test} Test of the factorisation hypothesis for the two-current correlators $\ov$ and $\os$.  The lattice data corresponds to l.h.s.\ of \protect\eqref{stat-mat-els} in position space, and the curves with error bands are obtained by Fourier transforming the r.h.s.\ of \protect\eqref{stat-mat-els} accordingly, using the form factor parameterisations specified in \eqref{ff-pole} and table~\protect\ref{tab:ff-fits}.}
\end{center}
\end{figure*}

The extracted form factor data and the results of the fits are shown in figure~\ref{fig:form-factors}, and the fitted parameters are given in table~\ref{tab:ff-fits}.  We see that the fits describe the $F_V$ data very well.  For $F_S$, the quality of the data is less good, and so is the agreement between our fits and the data points, but we consider this sufficient for our present study.

Having determined the form factors, we can evaluate the Fourier transforms of \eqref{stat-mat-els} and compare the result with our lattice data for $C_1$ as a function of $y$.  We see in figure~\ref{fig:fact-test} that the Fourier transforms of the two form factor parameterisations agree very well in the $y$ range shown.  For both currents, the factorisation hypothesis fails very clearly over a wide range of $y$, indicating significant correlation effects for the distribution of both vector and scalar charge in a pion of mass $300 \mev$.


\subsection{Root mean square radii}
\label{sec:rms}

In figure~\ref{fig:C1-quark-masses} we noted a clear difference between the decrease of the two-current correlators for light and heavy quarks, reflecting the different size of the pseudoscalar bound state in the two cases.  In the present section, we make this observation more quantitative.  We focus on the current combinations $\ov$ and $\oa$ and on graph $C_1$, which can be interpreted as a physical matrix element in the sense discussed at the end of section~\ref{sec:phys-matels}.
Following \cite{Wilcox:1986ge} and later work, we quantify the characteristic length scale of the two-current correlators via root mean square (rms) radii $r_{VV}$ and $r_{AA}$, defined by
\begin{align}
\label{rms-def}
r^2_{VV} &= \frac{\int d^3 y\;{y}^2\, C_{1}^{\ms VV}(\mvec{y})}{
   \int d^3 y\; C_{1}^{\ms VV}(\mvec{y})}
\end{align}
for $\ov$ and in analogy for $\oa$.  For the sake of legibility, we omit superscripts $0$ in the labels $VV$ and $AA$ in the remainder of this section.

One might think of evaluating \eqref{rms-def} from the ratio
\begin{align}
\label{naive-rms}
\sum_{\mvec{y} \in E(\mvec{0})} y^2\, C_{1}^{\ms VV}(\mvec{y}) \Bigg/
\sum_{\mvec{y} \in E(\mvec{0})} C_{1}^{\ms VV}(\mvec{y})
\end{align}
of lattice data, with the sum running over all $\mvec{y}$ in the elementary lattice cell $E(\mvec{0})$ defined by $-L/2 < y^i/a \le L/2$ for $i=1,2,3$.  However, the weight $y^2$ in the numerator of \eqref{naive-rms} enhances the importance of large distances to the extent that for our $L=40$ lattice, the contribution of points with $y/a > L/2$ to the sum over all $\mvec{y}$ amounts to 39\% for light and to 16\% for strange quarks.  At such distances, the effect of periodic images discussed in section~\ref{sec:aniso} is considerable.  Moreover, the slow decrease of $y^2\, C_{1}^{\ms VV}(\mvec{y})$ with $y$ implies that the numerator of \eqref{naive-rms} misses important contributions from distances $y/a > \sqrt{3}\ms L /2$ that are not included in the elementary cell.

In the following we present a method to deal with this situation.  Let us assume that the correlation function computed on the lattice has the form
\begin{align}
\label{mirror-img}
C_1^{\text{lat}}(\mvec{y}) &=
  \sum_{\mvec{n} \in \mathbb{Z}_3} C_1^{}(\mvec{y} + \mvec{n} L a) \,,
\end{align}
where $C_1(\mvec{y})$ is the correlation function in the physical limit and the terms with $\mvec{n} \neq \mvec{0}$ are due to the periodic boundary conditions on the lattice \cite{Burkardt:1994pw}.  The rms radii can be computed from the Fourier transform
\begin{align}
\label{cont-FT}
\widetilde{C}_1(\mvec{q}) &= \int d^3 y\; e^{i \mvec{q} \mvec{y}}\, C_1(\mvec{y})
\end{align}
as
\begin{align}
\label{rms-deriv}
r^2 &= {}- 6 \ms \bigl[ \widetilde{C}_{1}(\mvec{0}) \bigr]^{-1} \,
  \frac{d\ms \widetilde{C}_1(\mvec{q})}{d\ms q^2} \biggl|_{q^2 = 0} \,,
\end{align}
where we used that $\widetilde{C}_1$ depends on $\mvec{q}$ only via $q = |\mvec{q}|$.  On the lattice we can evaluate the discrete Fourier transform
\begin{align}
\label{discr-FT}
\widetilde{C}_1^{\text{lat}}(\mvec{q}) &= a^3 \!\! \sum_{\mvec{y} \in E(\mvec{0})}
  e^{i \mvec{q} \mvec{y}}\, C_1^{\text{lat}}(\mvec{y})
& \text{ for } \mvec{q} = \mvec{k} p_0
\end{align}
with $\mvec{k} \in \mathbb{Z}_3$ and $p_0 = 2\pi / (L a)$.  We have
\begin{align}
\widetilde{C}_1^{\text{lat}}(\mvec{q}) &=
  a^3 \sum_{\mvec{n} \in \mathbb{Z}_3^{}} \; \sum_{\mvec{y} \in E(\smash{\mvec{0}})}
  e^{i \mvec{q} \mvec{y}}\, C_1^{}(\mvec{y} + \mvec{n} L a)
= a^3 \sum_{\mvec{n} \in \mathbb{Z}_3^{}} \; \sum_{\mvec{y} \in E(\mvec{n})}
  e^{i \mvec{q} \mvec{y} - i \mvec{q} \mvec{n} L a}\, C_1(\mvec{y})
\nonumber \\[0.4em]
&= a^3 \!\! \sum_{\mvec{y}/a \in \mathbb{Z}_3} e^{i \mvec{q} \mvec{y}}\,
   C_1(\mvec{y}) \,,
\end{align}
where $E(\mvec{n})$ denotes the shifted lattice cell defined by $-L/2 < y^i/a + n^i \le L/2$ for $i=1,2,3$, and in the last step we have used the condition $\mvec{q} = \mvec{k} p_0$.  We thus find that for $a\to 0$ the discrete Fourier transform \eqref{discr-FT} becomes equal to the infinite-volume expression \eqref{cont-FT}.  The periodic images included in $\widetilde{C}_1^{\text{lat}}(\mvec{q})$ provide the contribution of the infinite-volume correlator $C_1(\mvec{y})$ at distances outside the elementary cell $E(\mvec{0})$.

To evaluate the rms radii from \eqref{rms-deriv}, we need to construct a smooth function of $\mvec{q}$ out of $\widetilde{C}_1(\mvec{q})$ at the points $\mvec{q} = \mvec{k} p_0$.  Whilst direct computation of \eqref{rms-def} would require us to \emph{extrapolate} $C_1(\mvec{y})$ to large values of $y$ and to remove the contributions from periodic images, we now need to \emph{interpolate} between discrete values of $\widetilde{C}_1(\mvec{q})$ in the vicinity of $\mvec{q} = \mvec{0}$.  The reliability of this interpolation is of course higher for a higher density of points $\mvec{k} p_0 = 2\pi \mvec{k} /(L a)$ and thus for larger physical lattice size $L a$.

\begin{figure*}
\begin{center}
\includegraphics[width=0.5\textwidth]{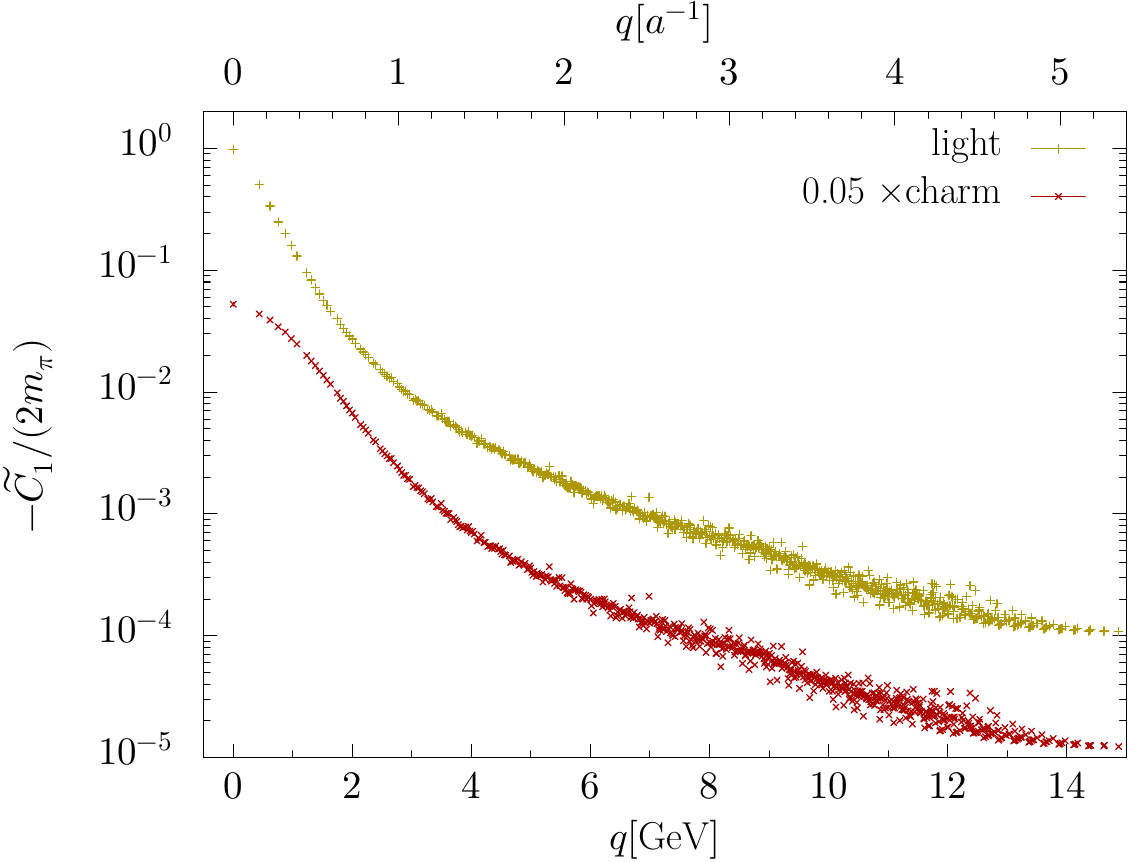}
\caption{\label{fig:FFT-all-q} The Fourier transform of the correlator $C_1(\mvec{y})$, evaluated using \protect\eqref{discr-FT}.  The values for charm quarks are scaled by $0.05$ for the sake of clarity.  Statistical uncertainties on the data points are not visible at this scale.   We divide the Fourier transform by $- 2 m_\pi$, so that according to our findings in section~\ref{sec:factorise} we have a simple normalisation $- \widetilde{C}_{1}^{\ms VV}(\mvec{0}) / (2 m_\pi) = 1$.}
\end{center}
\end{figure*}

In figure~\ref{fig:FFT-all-q} we show $\widetilde{C}_{1}^{\ms VV}(\mvec{q})$ obtained from our lattice data for light and charm quarks.  The results for strange quarks and those for $\widetilde{C}_{1}^{\ms AA}(\mvec{q})$ look qualitatively similar.
At values $q \sim \pi/a$ we see a clear anisotropy in the Fourier transform (which in the continuum limit depends on the length but not on the direction of $\mvec{q}$).  This is to be expected and reflects discretisation effects in $C_1(\mvec{y})$ at distances $y$ of a few lattice units $a$.  With decreasing $q$, the anisotropy gradually disappears and we have a very clear and smooth signal.

\begin{figure*}
\begin{center}
\subfigure[$\ov, p=0$]{\includegraphics[width=0.49\textwidth]{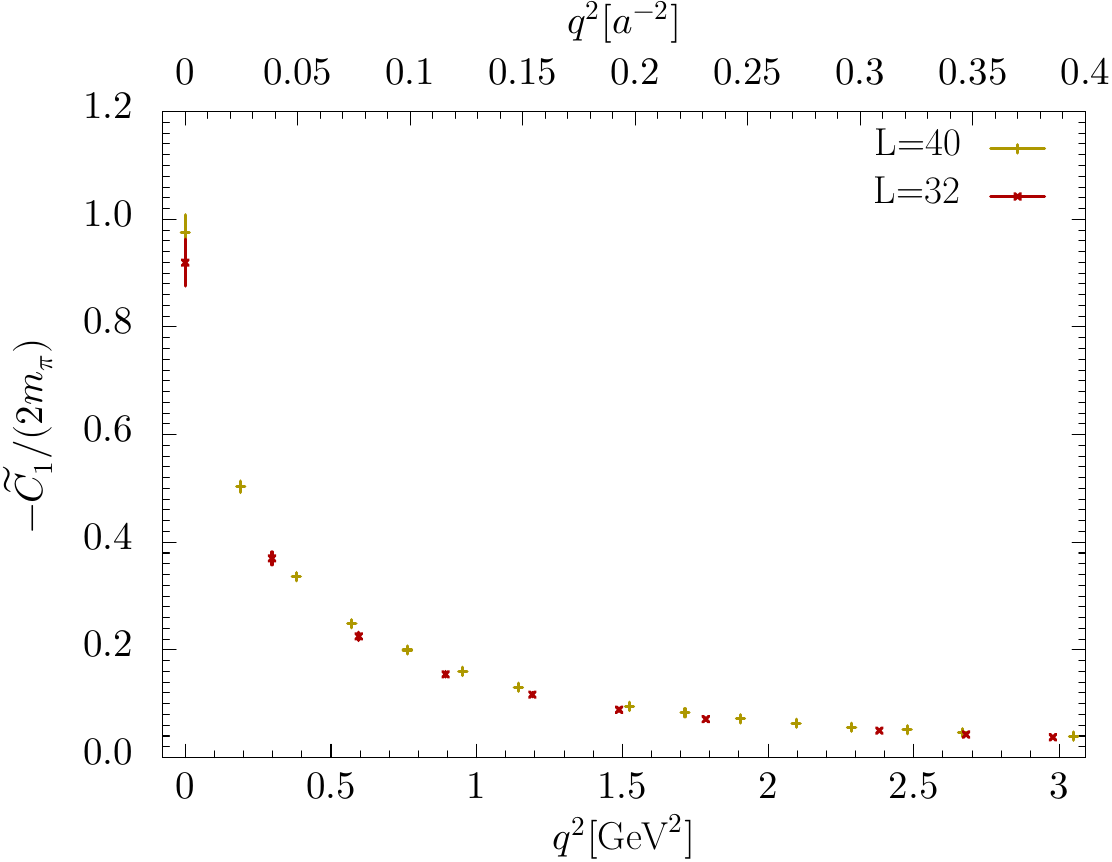}}
\hfill
\subfigure[$\oa, p=0$]{\includegraphics[width=0.49\textwidth]{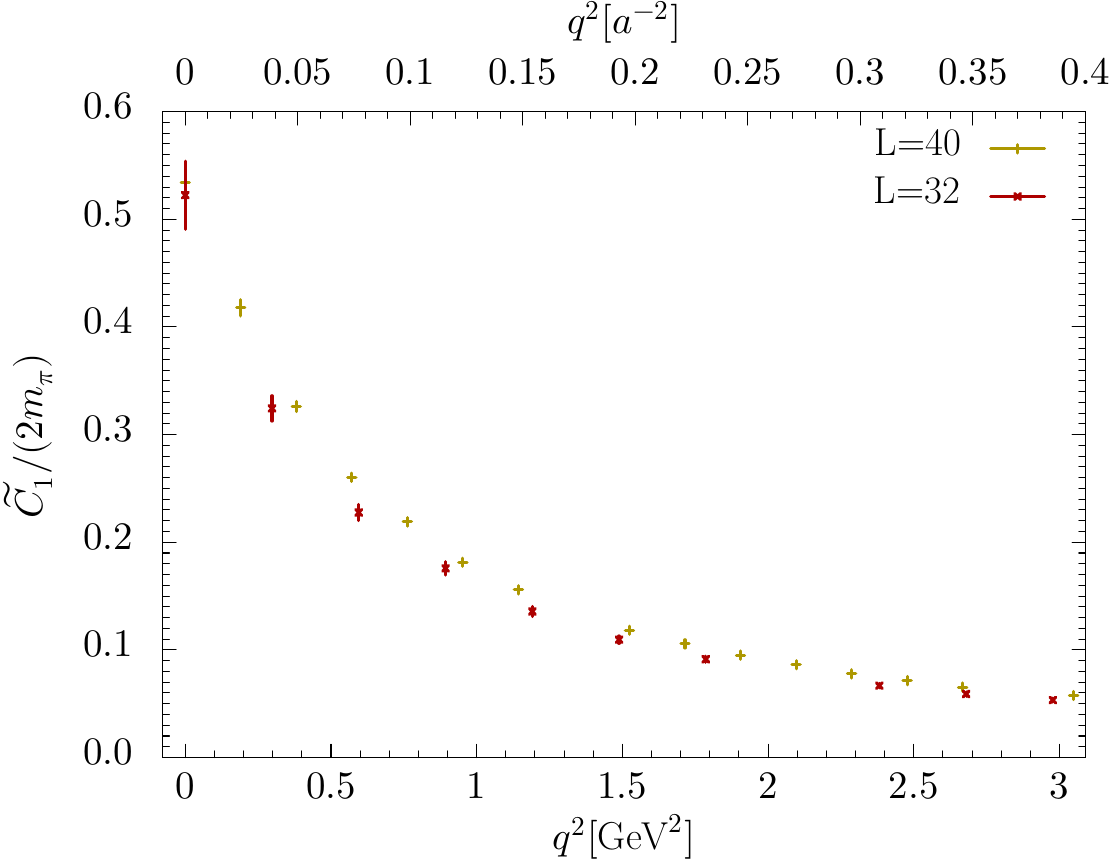}}
\caption{\label{fig:FFT-vols} Volume comparison of the Fourier transform $\widetilde{C}_1$ for light quarks.}
\end{center}
\end{figure*}

In figure~\ref{fig:FFT-vols} we focus on the small $q$ region and compare our results for light quarks on the lattices with $L=32$ and $L=40$.  Although there is some indication for a systematic shift of $\widetilde{C}_1$ between the two lattices, we find the agreement very satisfactory.  This indicates that the main finite size effect in the two-current correlators is indeed due to periodic images, which was the hypothesis underlying our arguments following~\eqref{mirror-img}.

\begin{figure*}
\begin{center}
\subfigure[$\ov, p=0, L=40$, light quarks]{\includegraphics[height=0.374\textwidth]{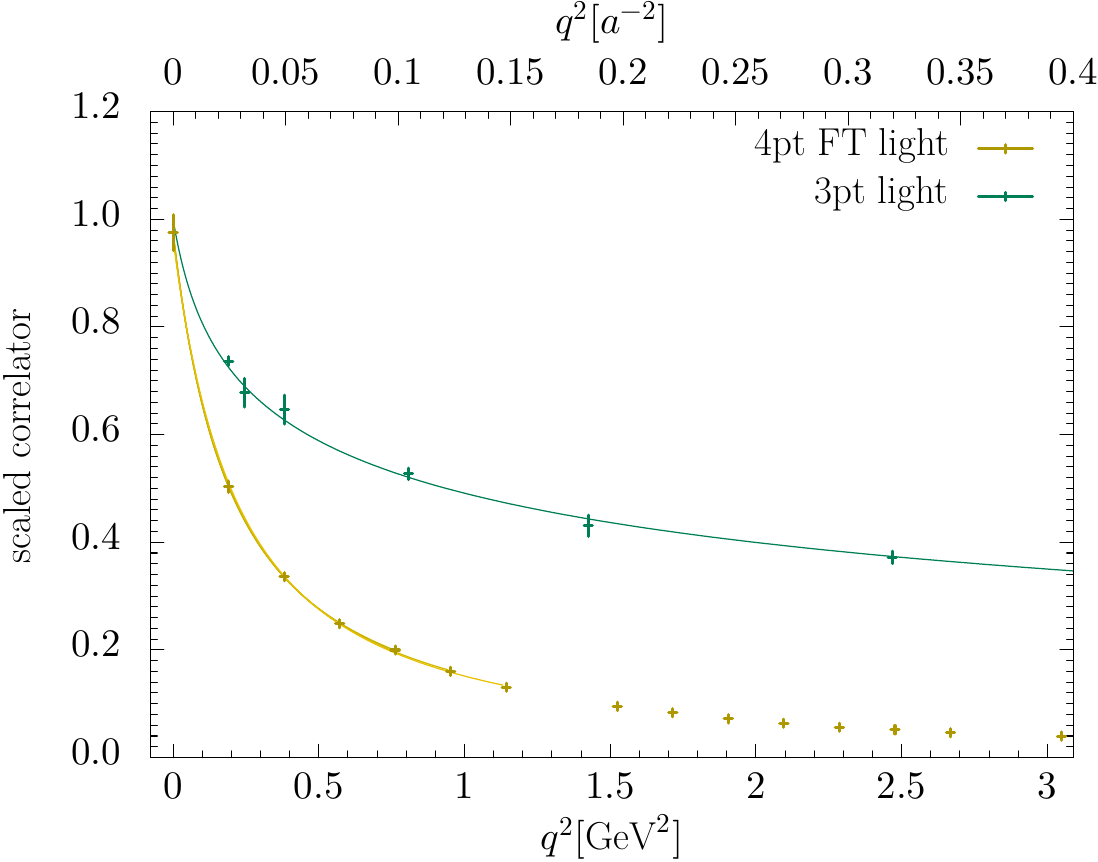}}
\hfill
\subfigure[$\ov, p=0, L=40$, strange and charm]{\includegraphics[height=0.375\textwidth]{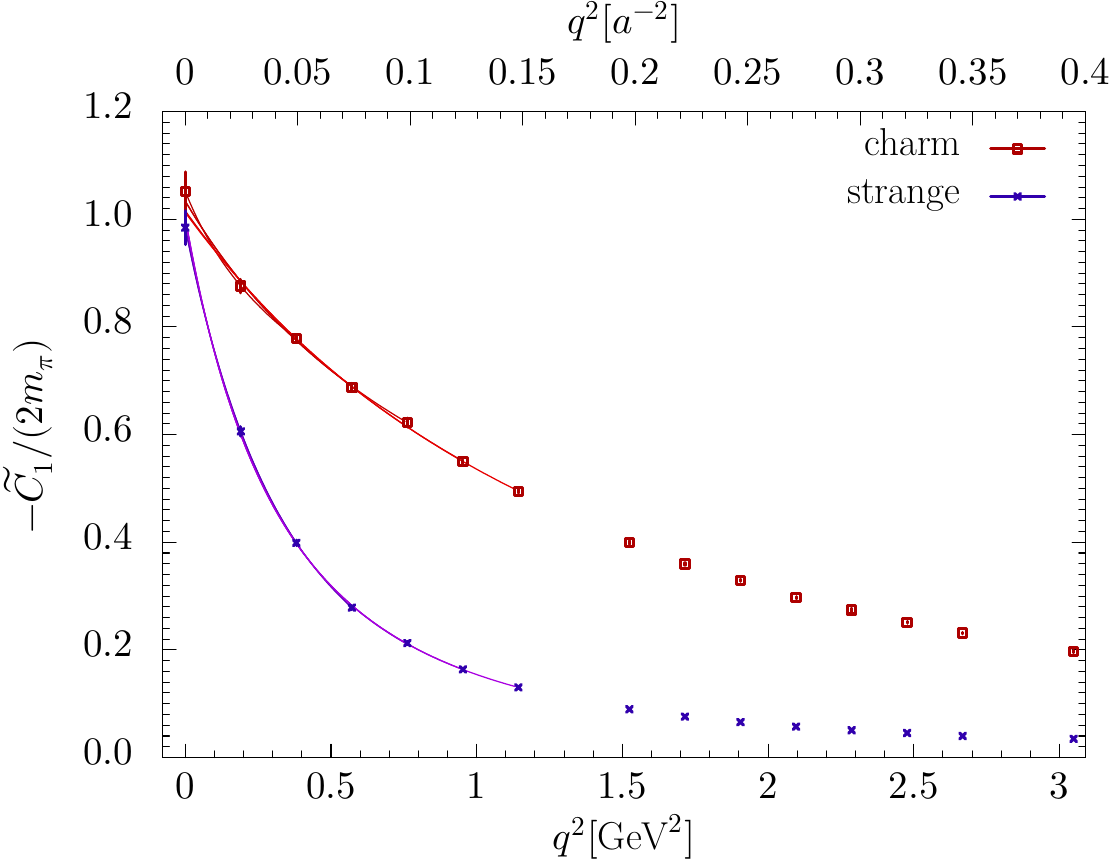}}
\\
\subfigure[$\oa, p=0, L=40$]{\includegraphics[height=0.375\textwidth]{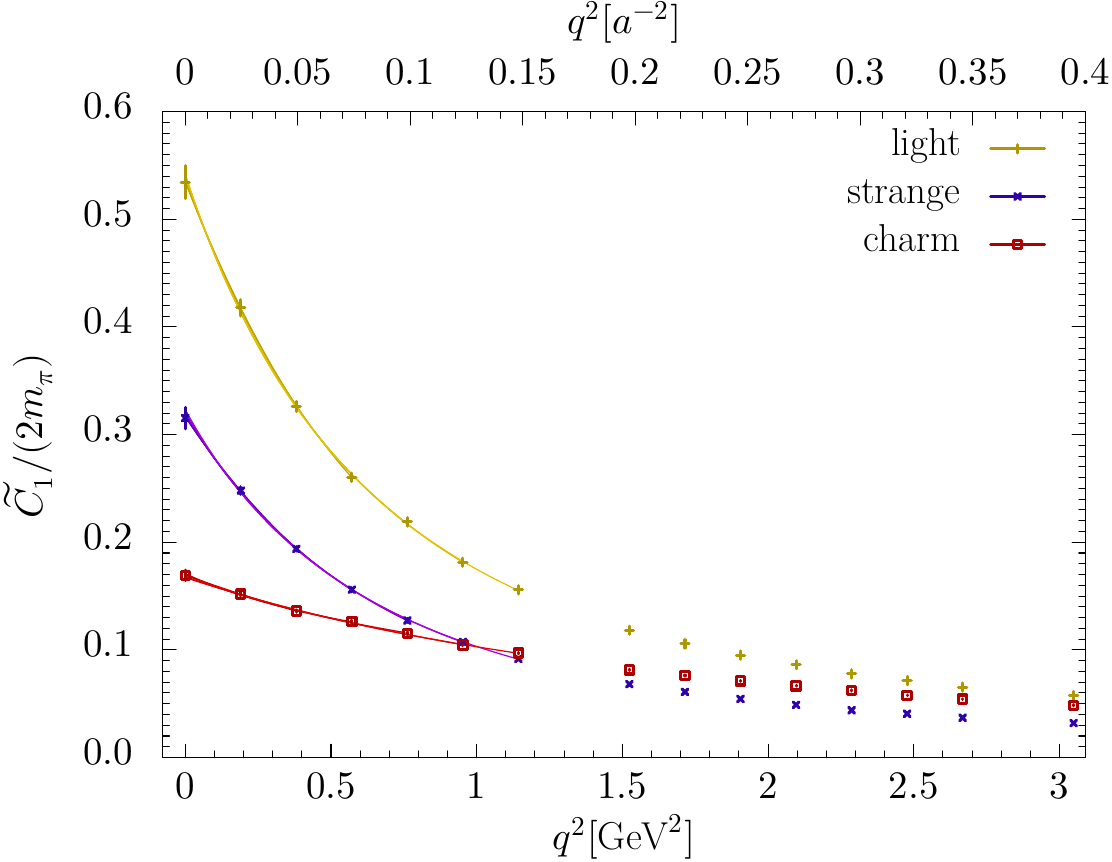}}
\caption{\label{fig:FFT-fits} The Fourier transform $\widetilde{C}_1$ at low values of $q^2$.  The curves (without error bands) show the series of fits discussed in the text.  In panel (a) the results for $- \widetilde{C}_1 /(2 m_\pi)$ are labelled by ``4pt FT light'', whereas the points and curve labelled by ``3pt light'' show the r.h.s.\ of \protect\eqref{fact-vec} evaluated with our data for the vector form factor $F_V$ and fit~1 of table~\protect\ref{tab:ff-fits}.}
\end{center}
\end{figure*}

\begin{figure*}
\begin{center}
\subfigure[$\ov, p=0, L=40$]{\includegraphics[width=0.48\textwidth,trim=0 0 0 25pt,clip]{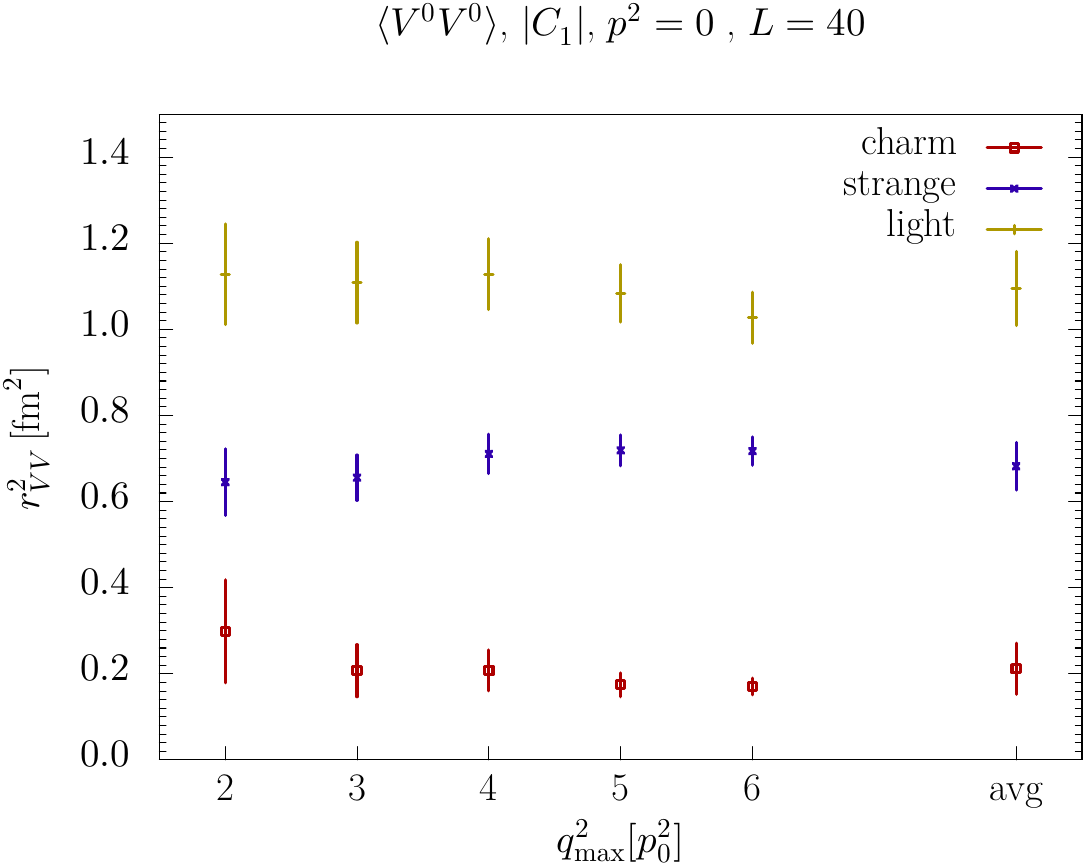}}
\hfill
\subfigure[$\oa, p=0, L=40$]{\includegraphics[width=0.48\textwidth,trim=0 0 0 25pt,clip]{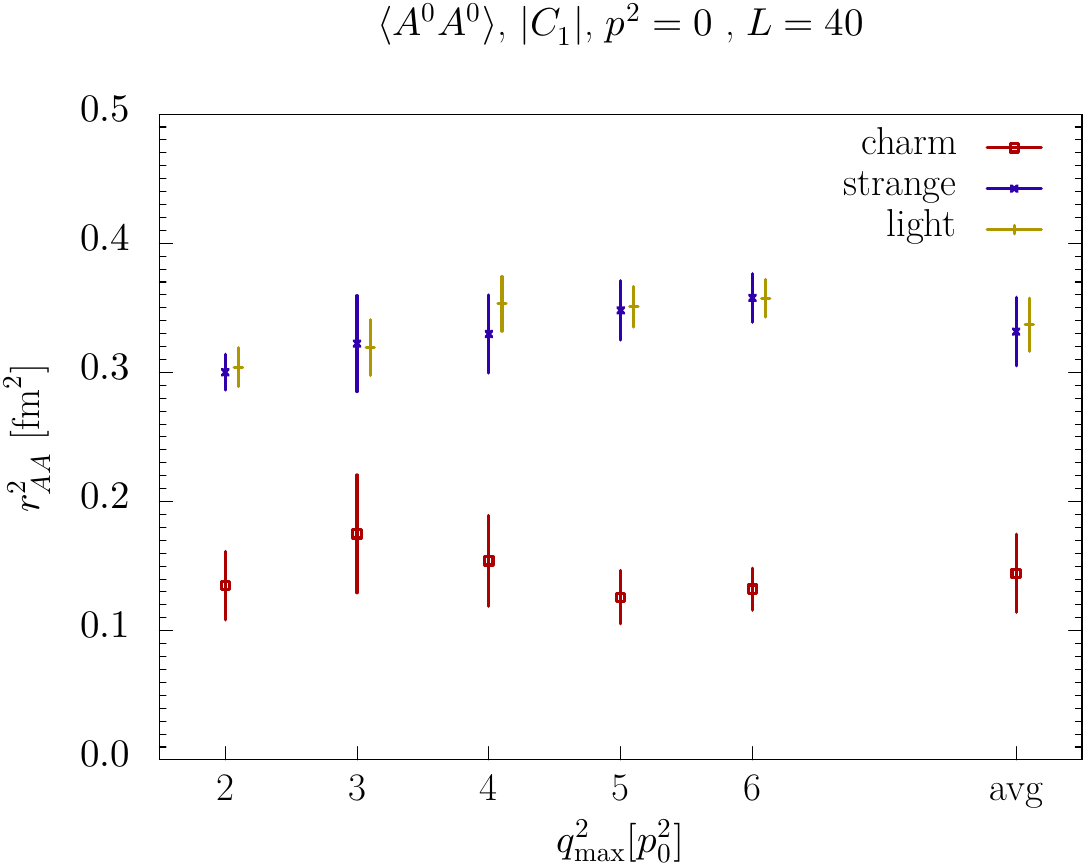}}
\caption{\label{fig:fitted-rms} Squared rms radii $r^2$ extracted from fitting $\widetilde{C}_1$ to the form \protect\eqref{fourier-fit} in the interval $0 \le q^2 \le q^2_{\text{max}}$, together with their jackknife errors.  The entries labelled ``avg'' give the averaged values corresponding to table~\ref{tab:rms}, with statistical and systematic errors added in quadrature.}
\end{center}
\end{figure*}

In order to compute the derivative in $q^2$ around $q=0$, we fit our results to the form
\begin{align}
\label{fourier-fit}
\widetilde{C}_{1}(\mvec{q})
  &= \frac{\widetilde{C}_1(\mvec{0})}{\bigl[ 1 + r^2 q^2 / (6 n) \bigr]^n}
\end{align}
in the range $0 \le q^2 \le q^2_{\text{max}}$, varying the value of $q^2_{\text{max}}$ between $2 \ms p_0^2$ and $6 \ms p_0^2$.  For the lowest value of $q^2_{\text{max}}$ we have as many data points as parameters in \eqref{fourier-fit}, whereas the highest value is motivated by the fact that there is no point with $q^2 = 7 p_0^2$ because $7$ cannot be written as the sum of squares of three integers.  Fits extending to even higher $q^2$ put less and less emphasis on the fit quality around $q^2=0$ and are therefore less well suited for determining the derivative at that point.  Figure~\ref{fig:FFT-fits} shows the low $q^2$ data together with the different fit curves.  The latter are in general very close to each other and can barely be distinguished individually.  In figure~\ref{fig:fitted-rms} we show the fitted values of $r^2_{VV}$ and $r^2_{AA}$.  The dependence of $r^2$ on $q^2_{\text{max}}$ is in general mild, although for light or strange quarks there seems to be a systematic increase of $r^2_{AA}$ with $q^2_{\text{max}}$.  We do not regard the fits with lower or higher $q^2_{\text{max}}$ as intrinsically more reliable around $q^2 = 0$ and therefore determine an overall value and error for each radius in the following way.  The averaged value of $r^2$ is obtained as the arithmetic mean of the fitted values $r^2(q^2_{\text{max}})$, and the root of the variance of these values is used as systematic error on $r^2$ due to our fitting procedure.  The statistical error on $r^2$ is taken as the arithmetic mean of the jackknife errors on the fit results $r^2(q^2_{\text{max}})$.  The outcome of this procedure is given in table~\ref{tab:rms}, converted to values and errors for $r$ instead of $r^2$.

\begin{table}[bh]
\begin{center}
\renewcommand{\arraystretch}{1.2}
\begin{tabular}{c|ccc} \hline \hline
radius [fm] & light quarks & strange & charm \\ \hline
$r_{VV}$ & 1.046(40)(09) & 0.826(32)(09) & 0.460(60)(25) \\
$r_{AA}$ & 0.580(15)(09) & 0.576(21)(08) & 0.380(38)(12) \\
\hline \hline
\end{tabular}
\end{center}
\caption{\label{tab:rms} Rms radii $r$ for the correlator $C_1(y)$.  The first error is statistical and the second one due to the fitting procedure, as specified in the text.}
\end{table}

We see that $r_{VV}$ decreases with increasing quark mass, as expected for a quantity that characterises the size of the pion.  The ratio $r_{AA} / r_{VV}$ increases with the quark mass, being clearly below $1$ for light and strange quarks and consistent with $1$ for charm.  A dynamical interpretation of this interesting finding is beyond the scope of the present work.

The factorisation hypothesis for $C_{1}^{\ms VV}$ discussed in section~\ref{sec:factorise} gives the relation
\begin{align}
\label{fact-vec}
{}- \frac{\widetilde{C}_1(\mvec{q})}{2 \mpi}  & \overset{?}{=}
  \frac{(\mpi + E_{q})^2}{4 m_{\pi} E_{q}} \,
    \Bigl[\ms F_V\bigl( 2 \mpi E_{q} - 2 \mpi^2 \bigr) \ms\Bigr]^2
\end{align}
according to \eqref{stat-mat-els}.  We see in figure~\ref{fig:FFT-fits}a that this is clearly ruled out, which confirms our finding in position space in figure~\ref{fig:fact-test}a.

For rms radii, the hypothesis \eqref{fact-vec} implies $r_{VV}^2 \overset{?}{=} 2\ms r_V^2$, where $r_V$ is the radius associated with the vector form factor $F_{V}$, i.e.\ the conventional charge radius.  Interestingly, the kinematic prefactor on the r.h.s.\ of \eqref{fact-vec} does not contribute to $d/d q^2$ at $q^2 = 0$, so that for the radii one obtains the same result as in the non-relativistic approximation, where the prefactor is $1$.
Quantitatively, we obtain $\sqrt{2}\, r_V = 0.878(14) \fm$ with fit 1 and $\sqrt{2}\, r_V = 0.849(12) \fm$ with fit 2 of table~\ref{tab:ff-fits}, which is clearly below the value $r_{VV} = 1.046(40)(09) \fm$ for light quarks in table~\ref{tab:rms}.  A detailed discussion of the difference between $r_{VV}$ and $\sqrt{2}\, r_{V}$ can be found in \cite{Burkardt:1994pw}.


\subsection{Subtraction term for the annihilation graph}
\label{sec:sub-ann}

The vacuum subtraction term for the annihilation graph $A$ in \eqref{C1-lat} involves the same three-point functions $\langle B_t \rangle$ that are necessary to compute the matrix elements of the axial or pseudoscalar current between a pion state and the vacuum.  As a by-product of our simulations, we can thus extract these matrix elements.  We can then compute the pion decay constant from the relation
\begin{align}
\bigl|\ms \bra{\pi^+} A^0_{ud}(0) \ket{0} \bigr| &= \sqrt{2}\ms \mpi\ms F_\pi
\end{align}
for a pion at rest.  Using $\partial_\mu A^{\smash{\mu}}_{ud}(x) = 2 m_q\ms P_{ud}(x)$, where $m_q$ is the average of $u$ and $d$ quark masses, we have for a pion at rest
\begin{align}
\biggl| \frac{\bra{\pi^+} P_{ud}(0)
   \ket{0}}{\bra{\pi^+} A^0_{ud}(0) \ket{0}} \biggr|
 &= \frac{m_\pi}{2 m_q} = \frac{B}{\mpi} \,,
\end{align}
where the last relation holds at leading order in chiral perturbation theory.  The quark mass $m_q$ and the chiral symmetry breaking parameter $B$ are understood to be renormalised in the $\overline{\mathrm{MS}}$ scheme at the scale $\mu = 2 \gev$, corresponding to the renormalisation constants given in table~\ref{tab:renfact}.  From the data of our two lattices, we obtain
\begin{align}
\label{chiral-params}
F_\pi &= (100.2 \pm 0.4 ) \mev \,,  &  B &= ( 2.41 \pm 0.02 ) \gev
  & \text{ ~for~ } L = 40 & & (\mpi = 293 \mev) \,,
\nonumber \\
F_\pi &= ( 98.2 \pm 1.0 ) \mev \,,  &  B &= ( 2.50 \pm 0.05 ) \gev
  & \text{ ~for~ } L = 32 & & (\mpi = 299 \mev) \,,
\nonumber \\
\end{align}
where we also recall the pion masses fitted from our two-point correlation functions.  The quoted errors are purely statistical.

The above values of $F_\pi$ for $\mpi \approx 300 \mev$ agree within a few percent  with chiral perturbation theory at NLO, see figure~10 in \cite{Durr:2014oba}.  From equations~(46), (65) and table~14 in the FLAG review \cite{Aoki:2013ldr}, we obtain $B = \Sigma /F^2 \approx 2.6 \text{~to~} 2.7 \gev$ for $n_F = 2$ quark flavours in the chiral limit.  This agrees reasonably well with our values in \eqref{chiral-params}, given that we use lowest-order chiral perturbation theory for their extraction and have not attempted to correct our data for lattice artefacts.

%% file: isospin.tex
\section{Results for isospin amplitudes}
\label{sec:matrix-elms}

In this section we present our results for the isospin amplitudes $F_0$, $F_1$ and $F_2$ defined in \eqref{iso-decomp}.  We compare them with the predictions of chiral perturbation theory that were computed in \cite{Bruns:2015yto} and recalled in section~\ref{sec:chpt-pred} of the present paper.  We consider the case of light quarks throughout.

\subsection{Isospin amplitudes}
\label{sec:isospin-amps}

\begin{figure*}
\begin{center}
  \subfigure[$F_1, \ov, p=0$]{\includegraphics[width=0.44\textwidth,trim=0 0 0 17pt,clip]{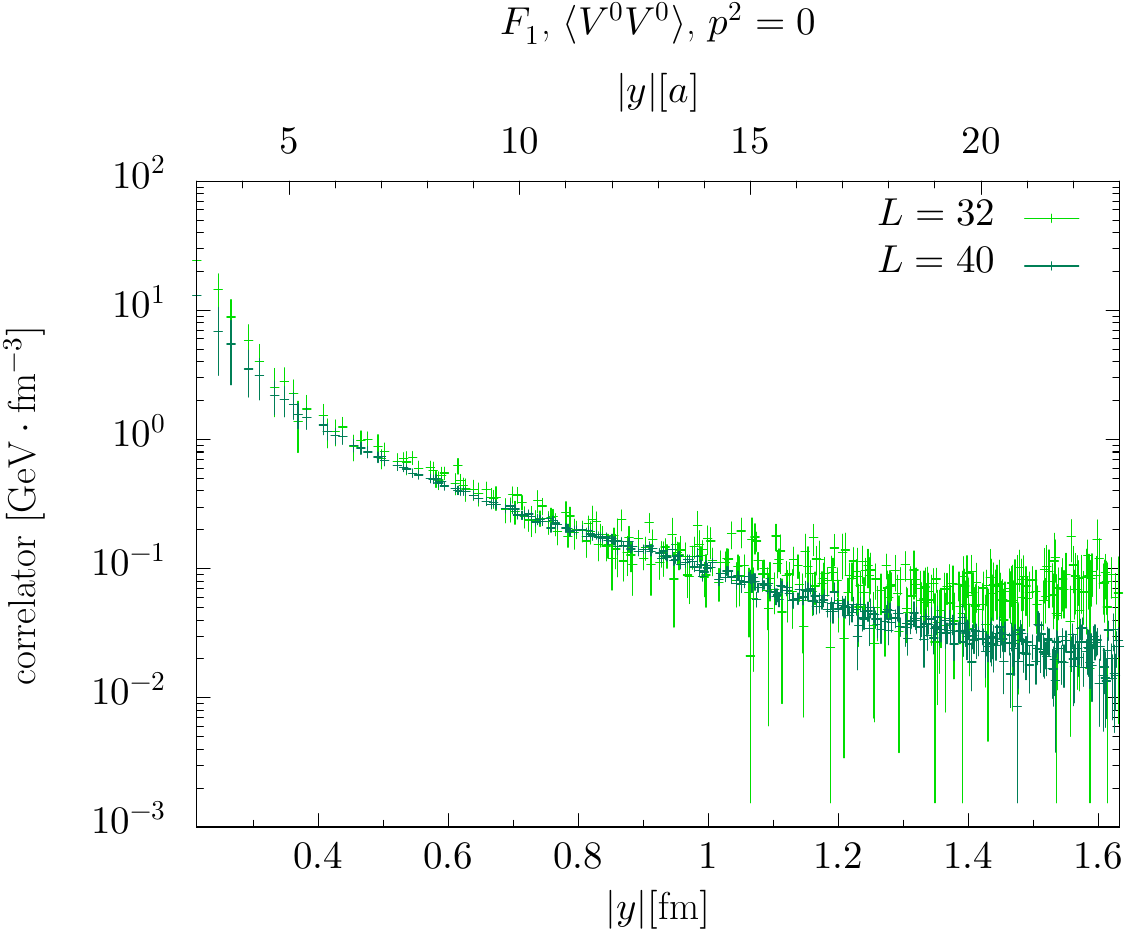}}
  \hfill
  \subfigure[$F_1, \oa, p=0$]{\includegraphics[width=0.45\textwidth,trim=0 0 0 17pt,clip]{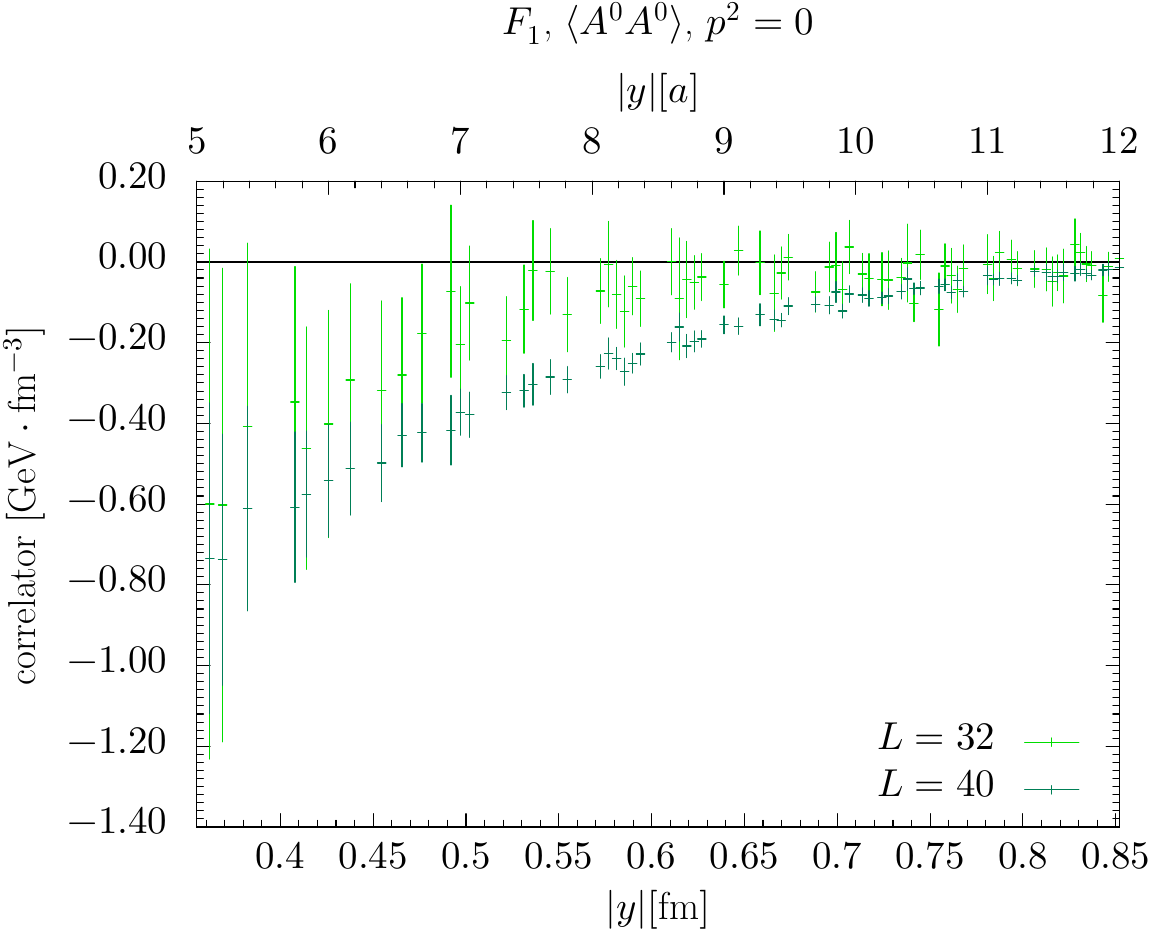}}
\\
  \subfigure[$F_0, \ov, p=0$]{\includegraphics[width=0.445\textwidth,trim=0 0 0 17pt,clip]{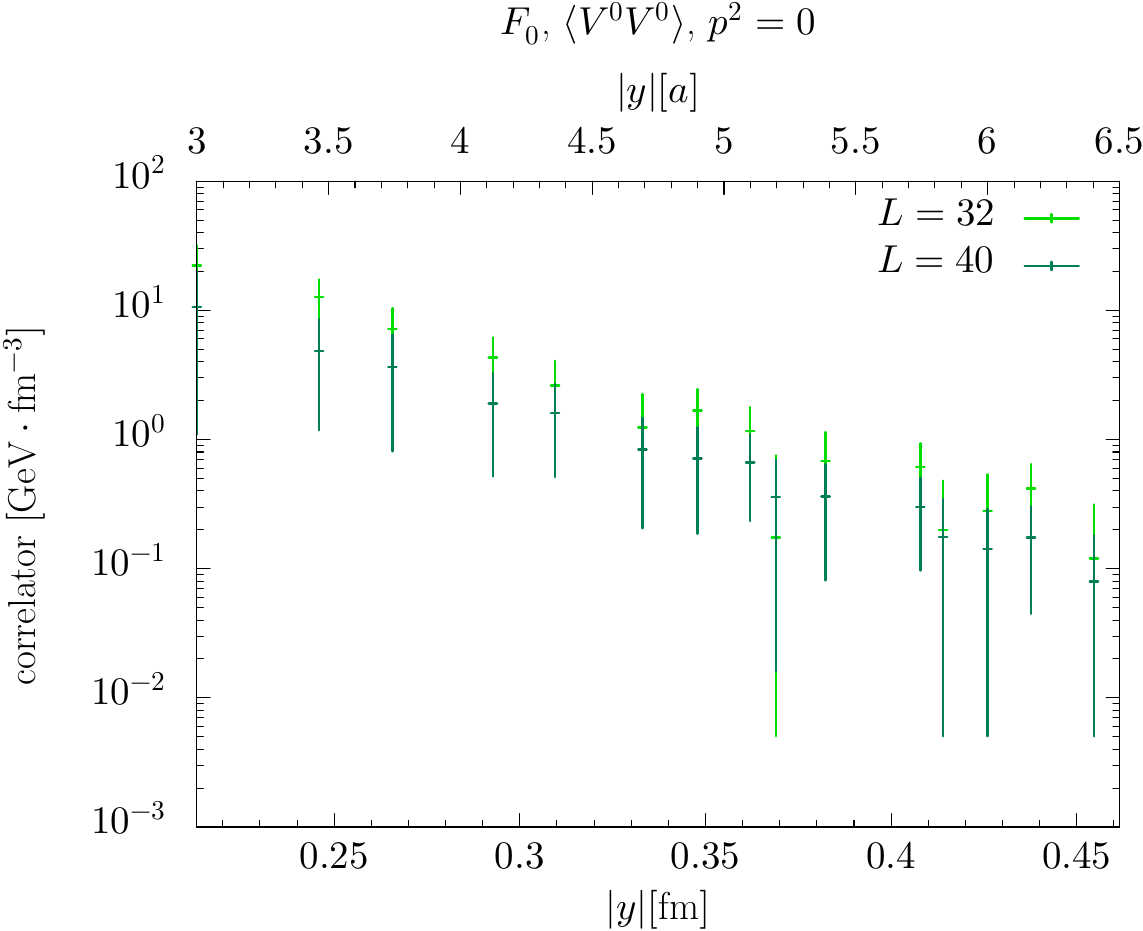}}
  \hfill
  \subfigure[$F_2, \ov, p=0$]{\includegraphics[width=0.445\textwidth,trim=0 0 0 17pt,clip]{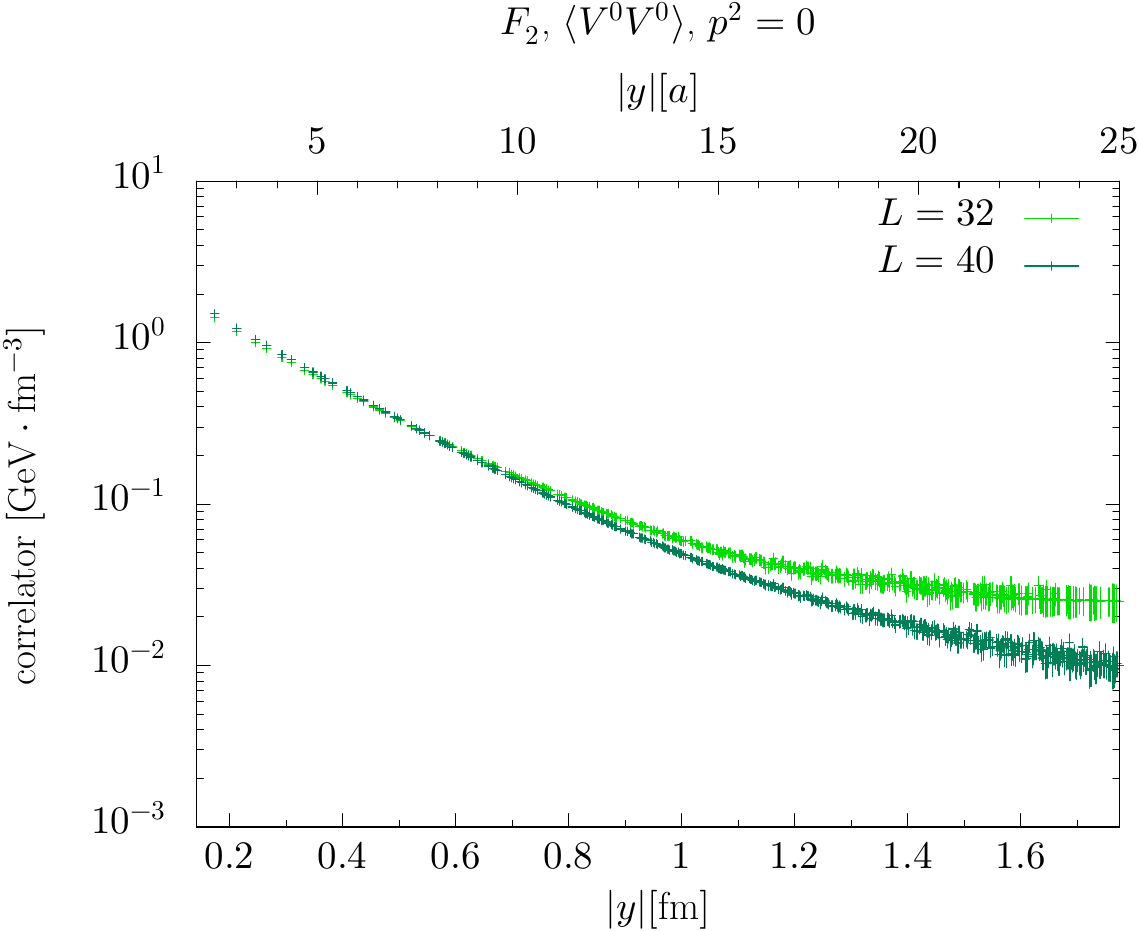}}
\\
  \subfigure[$F_2, \oa, p=0$]{\includegraphics[width=0.44\textwidth,trim=0 0 0 17pt,clip]{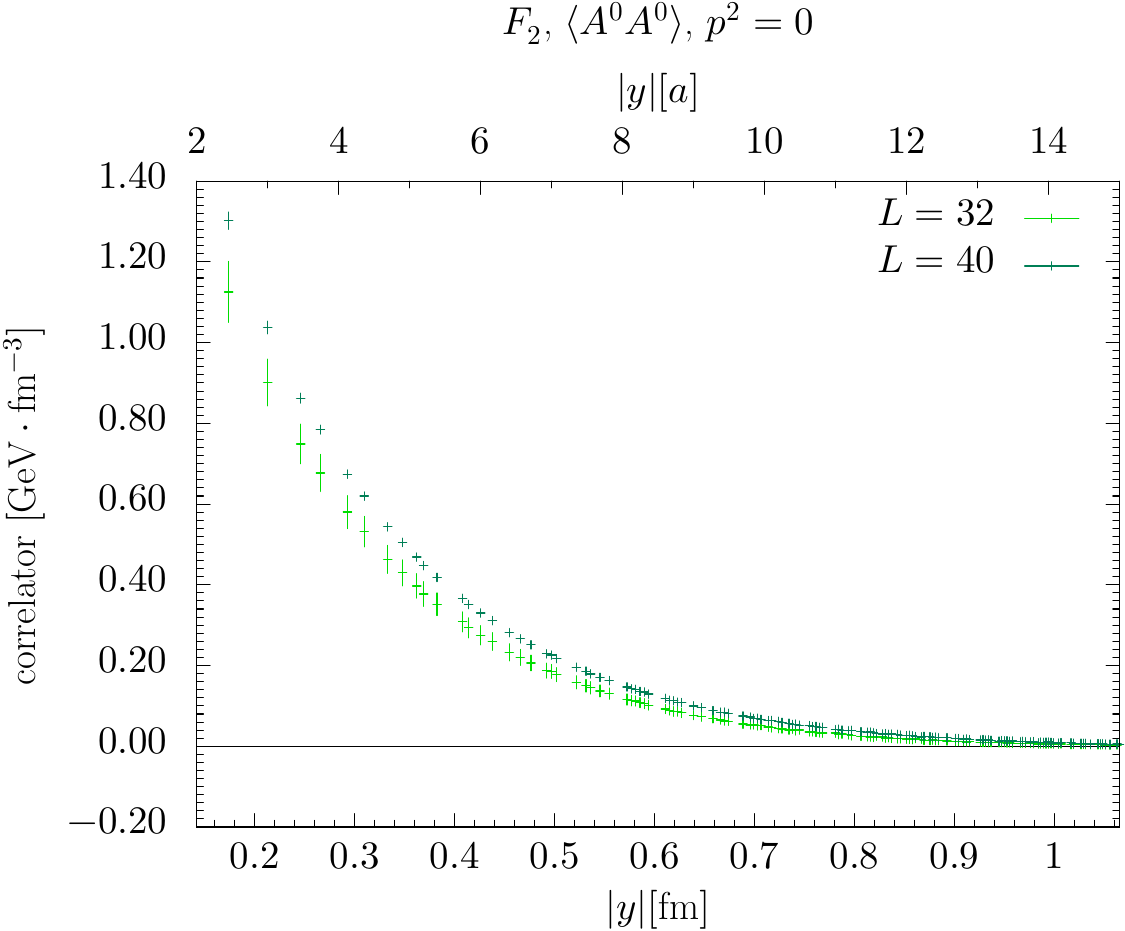}}
  \hfill
  \subfigure[$F_2, \oa, p=0$]{\includegraphics[width=0.45\textwidth,trim=0 0 0 17pt,clip]{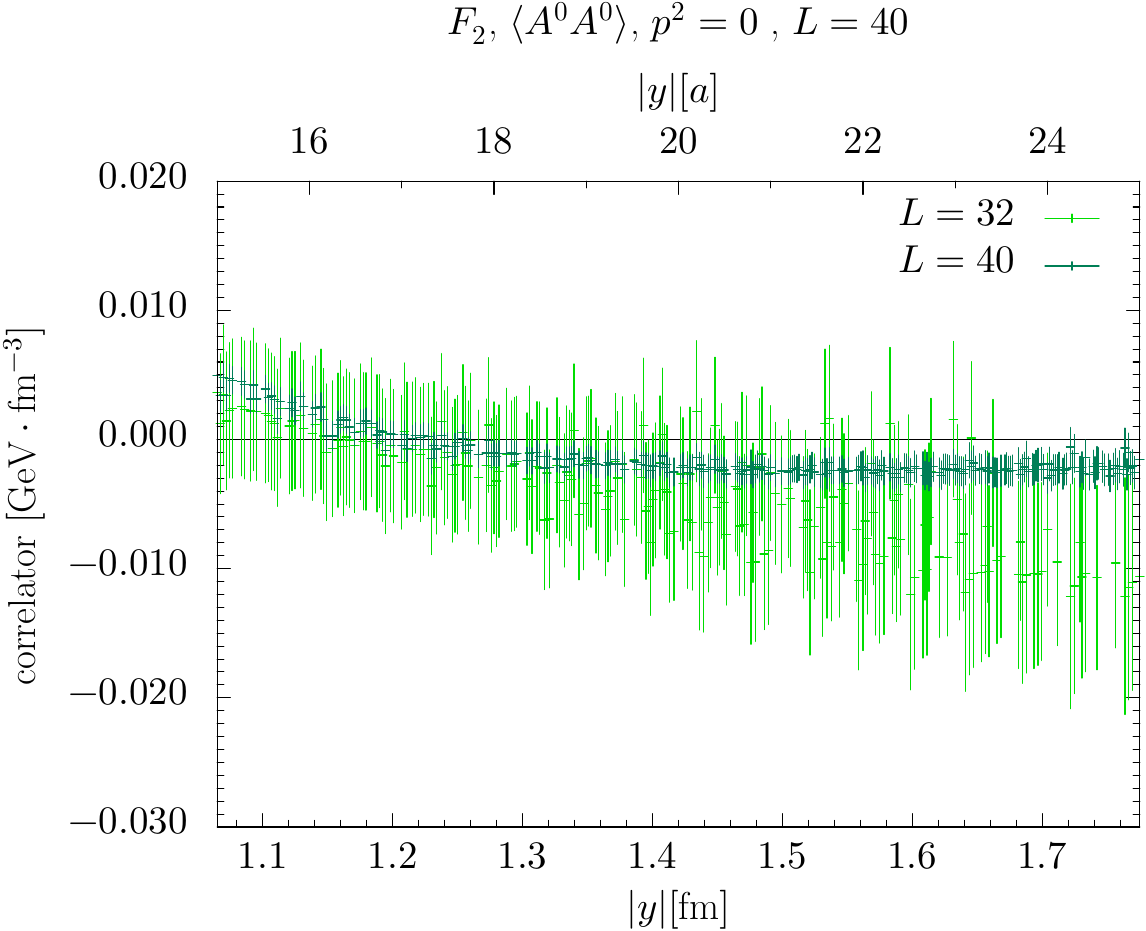}}
\\
  \caption{\label{fig:VA-vols} Lattice results for isospin amplitudes of $\ov$ and $\oa$.  The channel $\oa$ for $F_0$ has very large statistical errors and is not shown.}
\end{center}
\end{figure*}

\begin{figure*}
\begin{center}
  \subfigure[$F_0, \os, p=0$]{\includegraphics[width=0.44\textwidth,trim=0 0 0 17pt,clip]{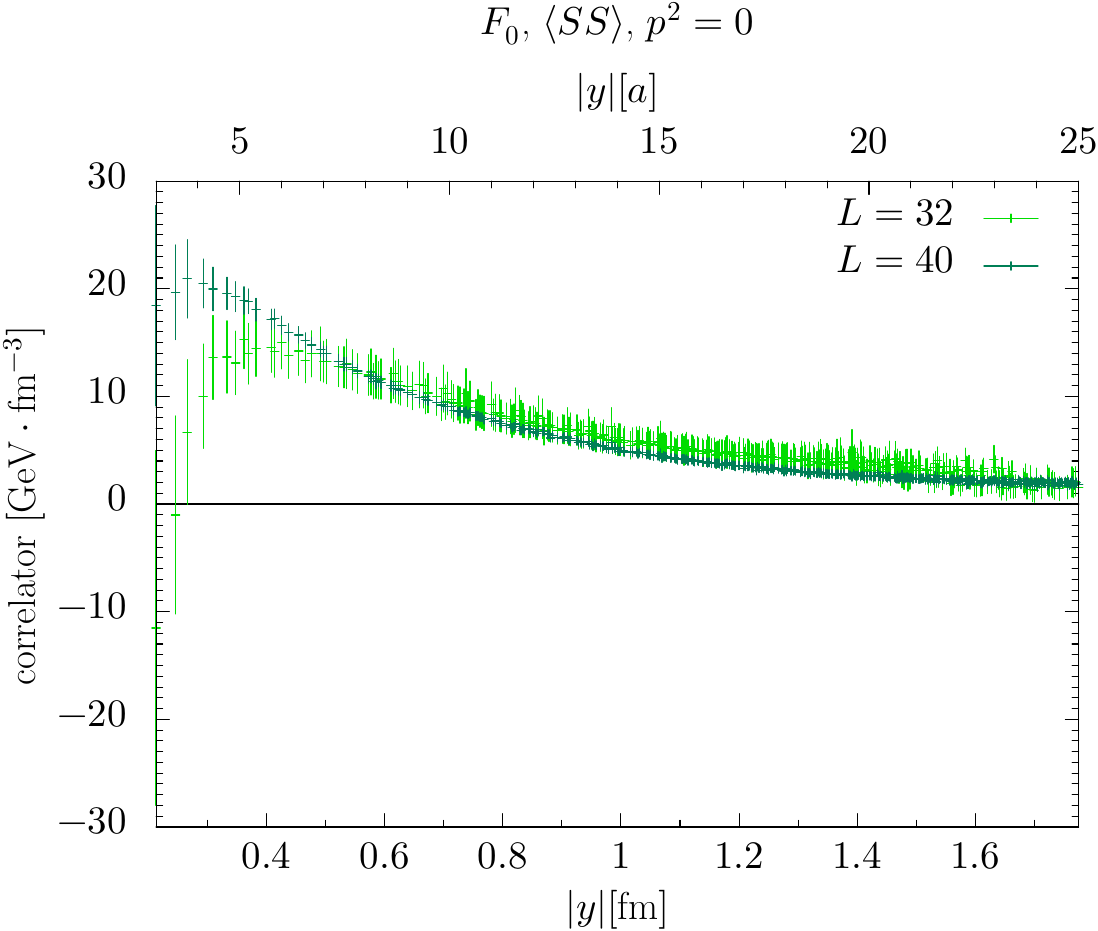}}
  \hfill
  \subfigure[$F_0, \op, p=0$]{\includegraphics[width=0.445\textwidth,trim=0 0 0 17pt,clip]{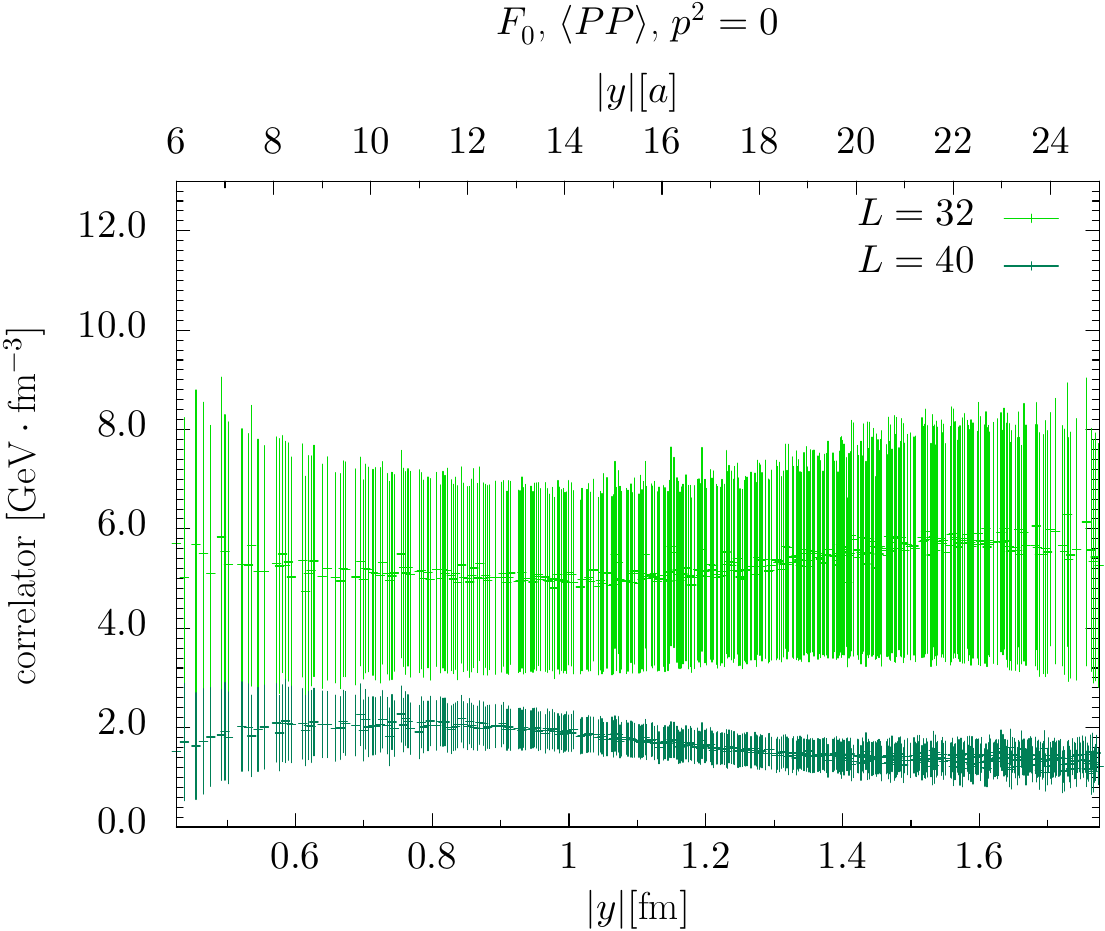}}
\\
  \subfigure[$F_1, \op, p=0$]{\includegraphics[width=0.44\textwidth,trim=0 0 0 17pt,clip]{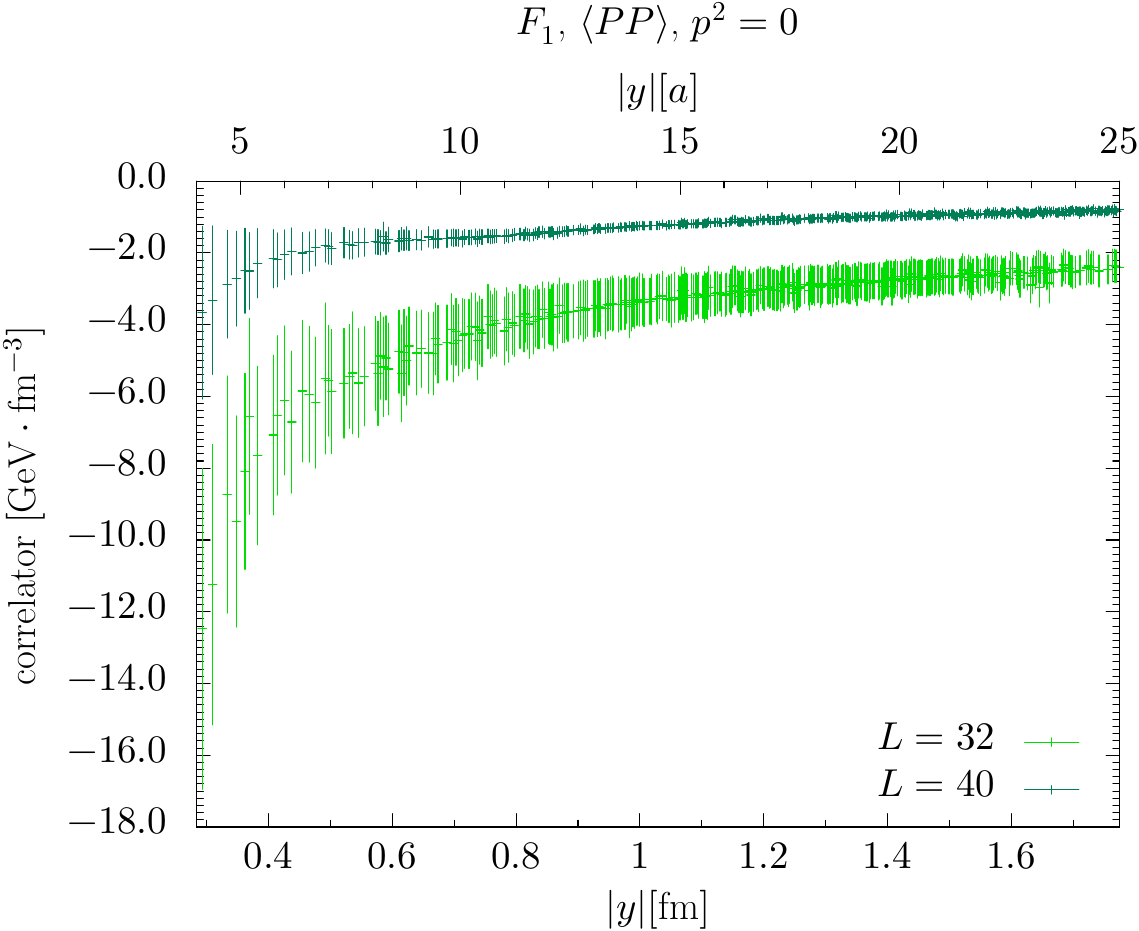}}
\\
  \subfigure[$F_2, \os, p=0$]{\includegraphics[width=0.44\textwidth,trim=0 0 0 17pt,clip]{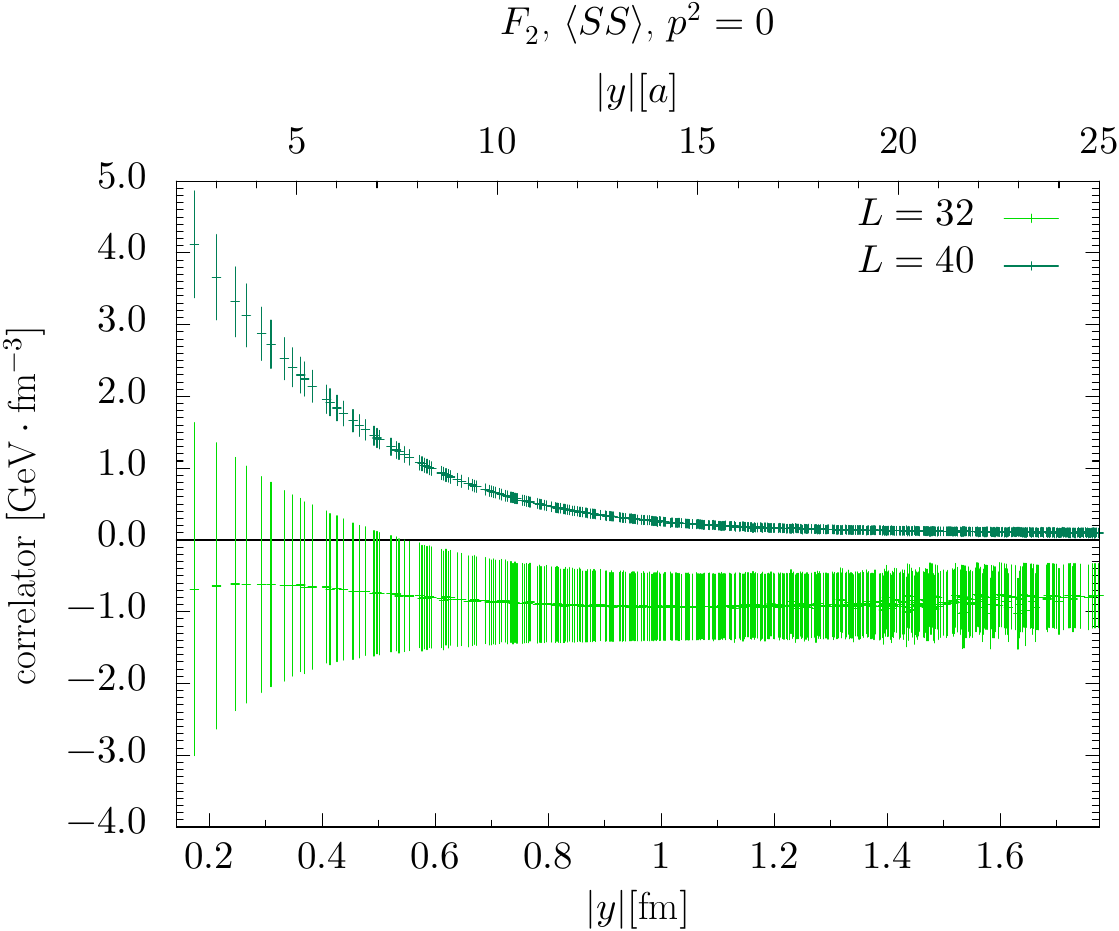}}
  \hfill
  \subfigure[$F_2, \op, p=0$]{\includegraphics[width=0.45\textwidth,trim=0 0 0 17pt,clip]{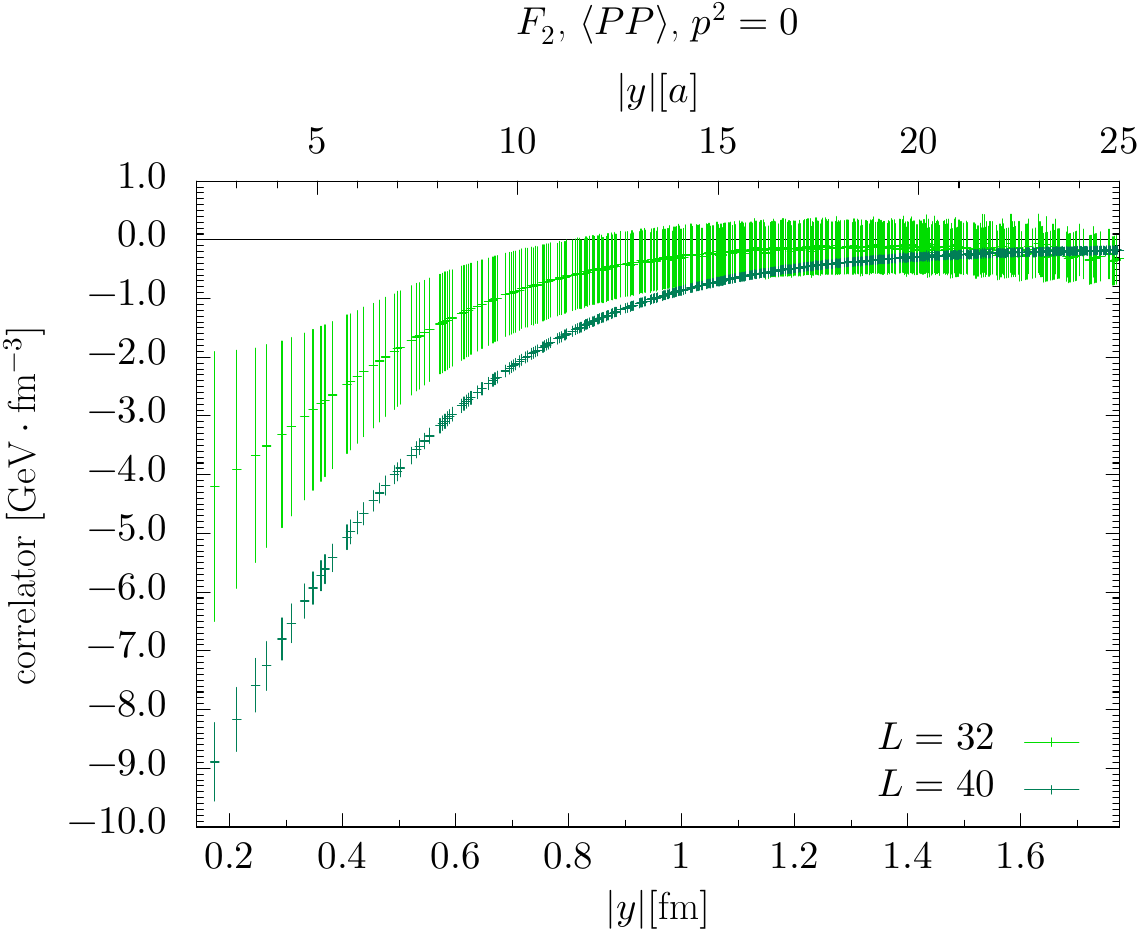}}
  \caption{\label{fig:SP-vols} Lattice results for isospin amplitudes of $\os$ and $\op$.  The channel $\os$ for $F_1$ has very large statistical errors and is not shown.}
\end{center}
\end{figure*}

In figures~\ref{fig:VA-vols} and \ref{fig:SP-vols}, we show our results for the isospin amplitudes in all channels except $\op$ for $F_0$ and $\os$ for $F_1$, where the statistical errors are too large to obtain a nonzero signal.  The $y$ range shown is selected such that regions where we see no signal have been omitted.

Comparison of the data for the $L=40$ and $L=32$ lattices shows a mixed situation regarding finite-volume effects.  These are significant for $\op$ in all three isospin combinations, for $\oa$ in $F_1$ and for $\os$ is $F_2$.  In the latter case, even the sign of the signal changes when going from $L=32$ to $L=40$.  For all other cases, volume effects are moderate, except for $\os$ in $F_0$ at small $y$ and for $\ov$ in $F_2$ at large $y$.

\begin{figure*}
\begin{center}
  \subfigure[$\os, p=0, L=40$]{\includegraphics[width=0.48\textwidth,trim=0 0 0 17pt,clip]{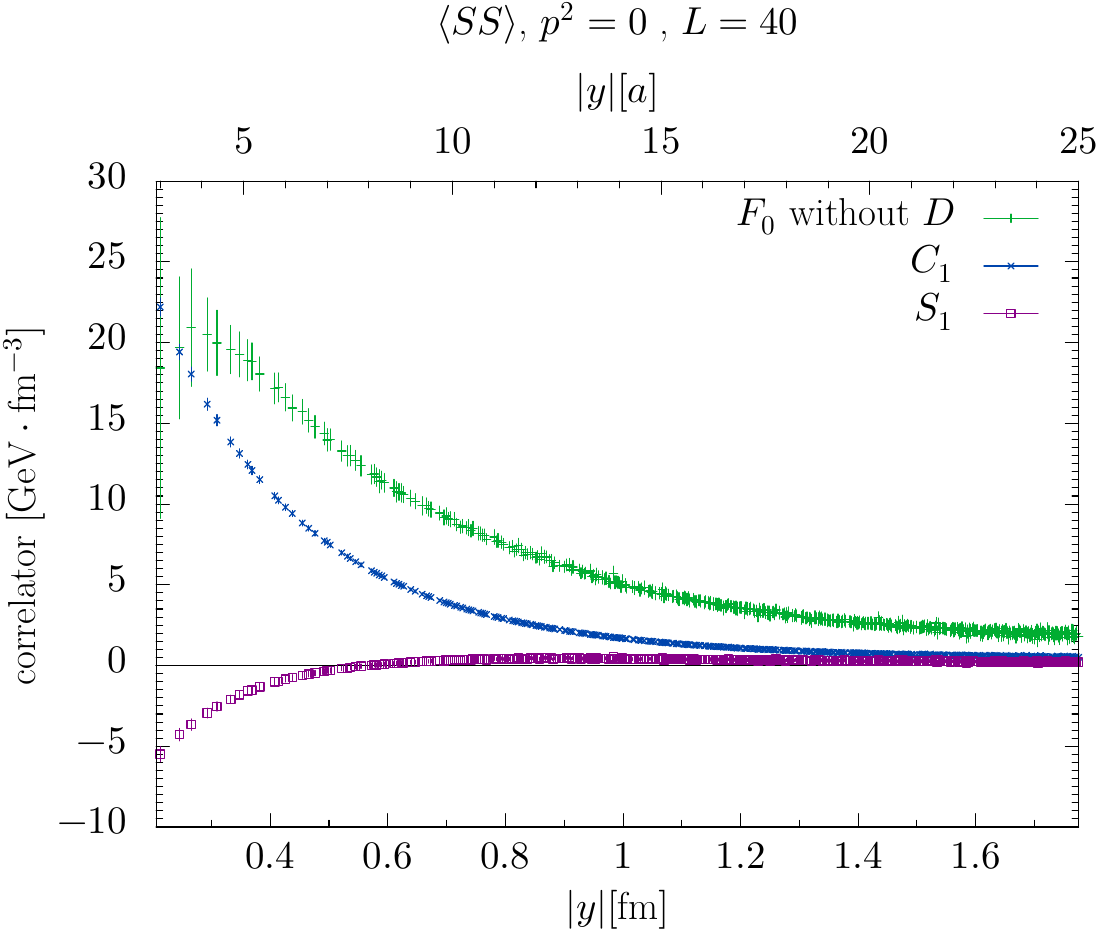}}
  \hfill
  \subfigure[$\op, p=0, L=40$]{\includegraphics[width=0.495\textwidth,trim=0 0 0 17pt,clip]{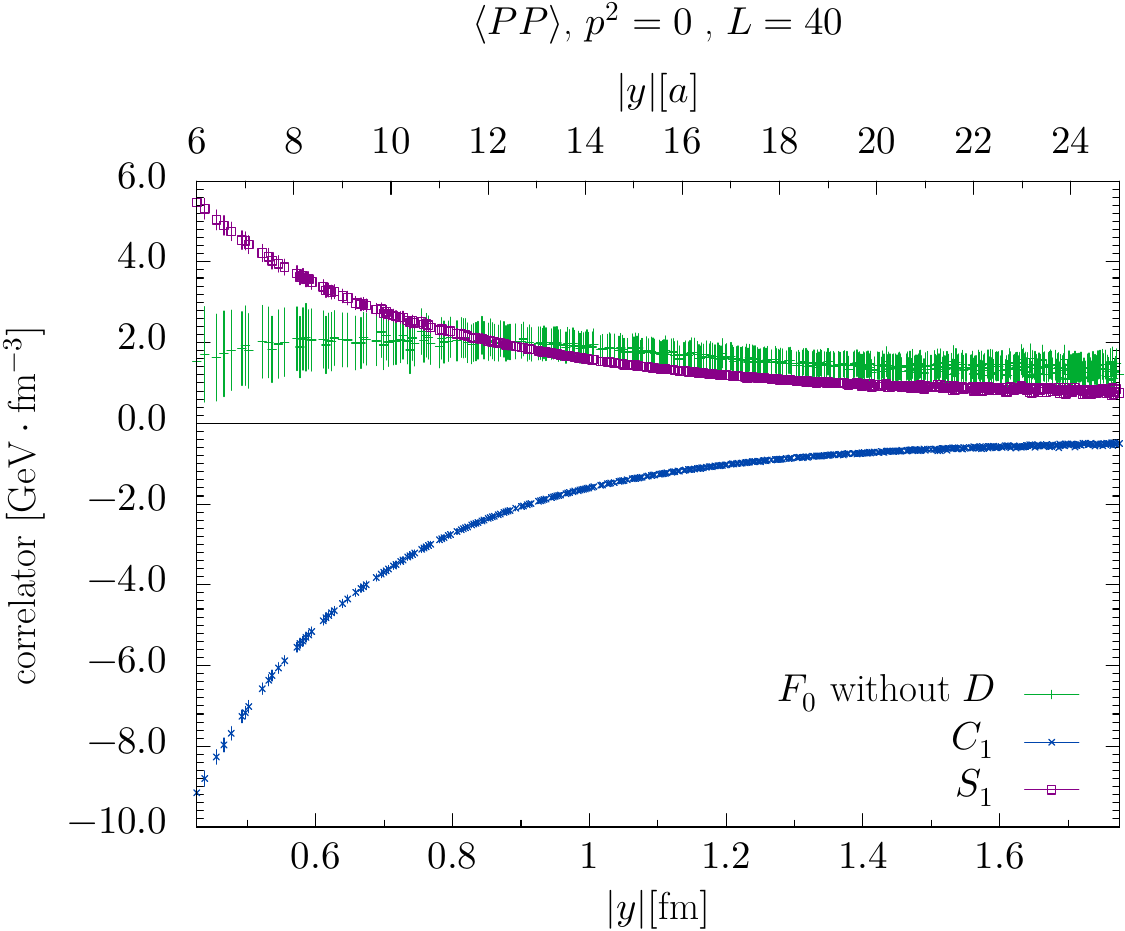}}
  \caption{\label{fig:F0-contractions} Comparison of $F_0$ with $C_1$ and $S_1$ for $\os$ and $\op$.  The contribution from $D$ to $F_0$ is omitted, as explained in the text.}
\end{center}
\end{figure*}

We must now return to the doubly disconnected graph $D$, which contributes to $F_0$ (but not to $F_1$ or $F_2$).  In all plots shown, the contribution from $D$ is omitted since the result would have too large errors to detect a signal.  We must therefore ask whether there is any reason to omit $D$ in $F_0 = C_1 + 2 S_1 + D$.  Given the structure of the corresponding graphs in figure~\ref{fig:contractions}, it seems plausible to assume that the ratio $D : S_1$ is similar in size to the ratio $S_1 : C_1$.  If one is willing to make this hypothesis, then one may justify the omission of $D$ in $F_0$ for cases in which the data shows that $|S_1| \ll |C_1|$.  In figure~\ref{fig:F0-contractions} we see that this condition is satisfied for $\os$ up to about $y \sim 1 \fm$, but not at all for $\op$.  For $\ov$ in the $y$ region shown in figure~\ref{fig:VA-vols}(c), we find a very small ratio $|S_1 /C_1| < 0.035$.  The correlator for $\oa$ is too noisy for $F_0$ even if we only consider the sum $C_1 + 2 S_1$.

\subsection{Comparison with chiral perturbation theory}
\label{sec:chiral-comp}

\begin{figure*}
\begin{center}
  \subfigure[$F_1, \ov, p=0, L=40$]{\includegraphics[width=0.45\textwidth,trim=0 0 0 17pt,clip]{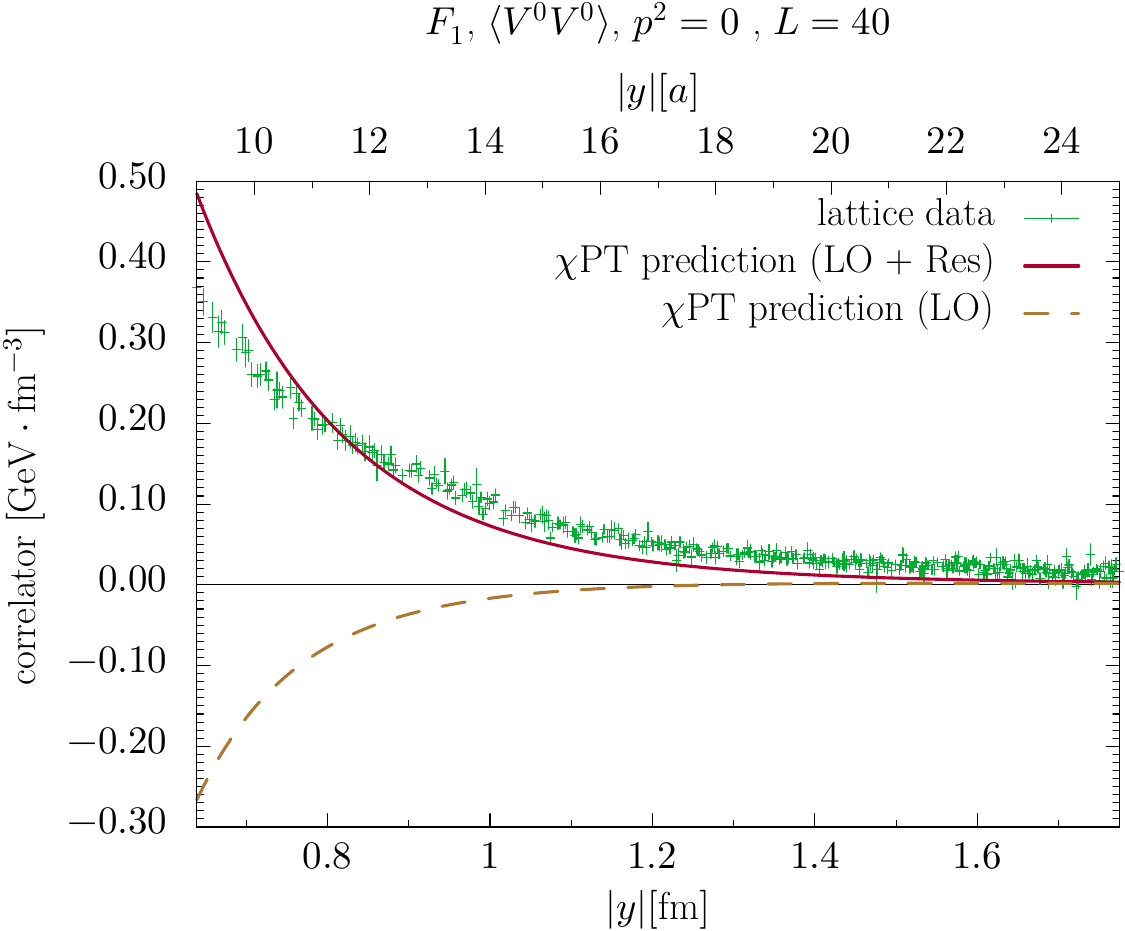}}
  \hfill
  \subfigure[$F_1, \oa, p=0, L=40$]{\includegraphics[width=0.45\textwidth,trim=0 0 0 17pt,clip]{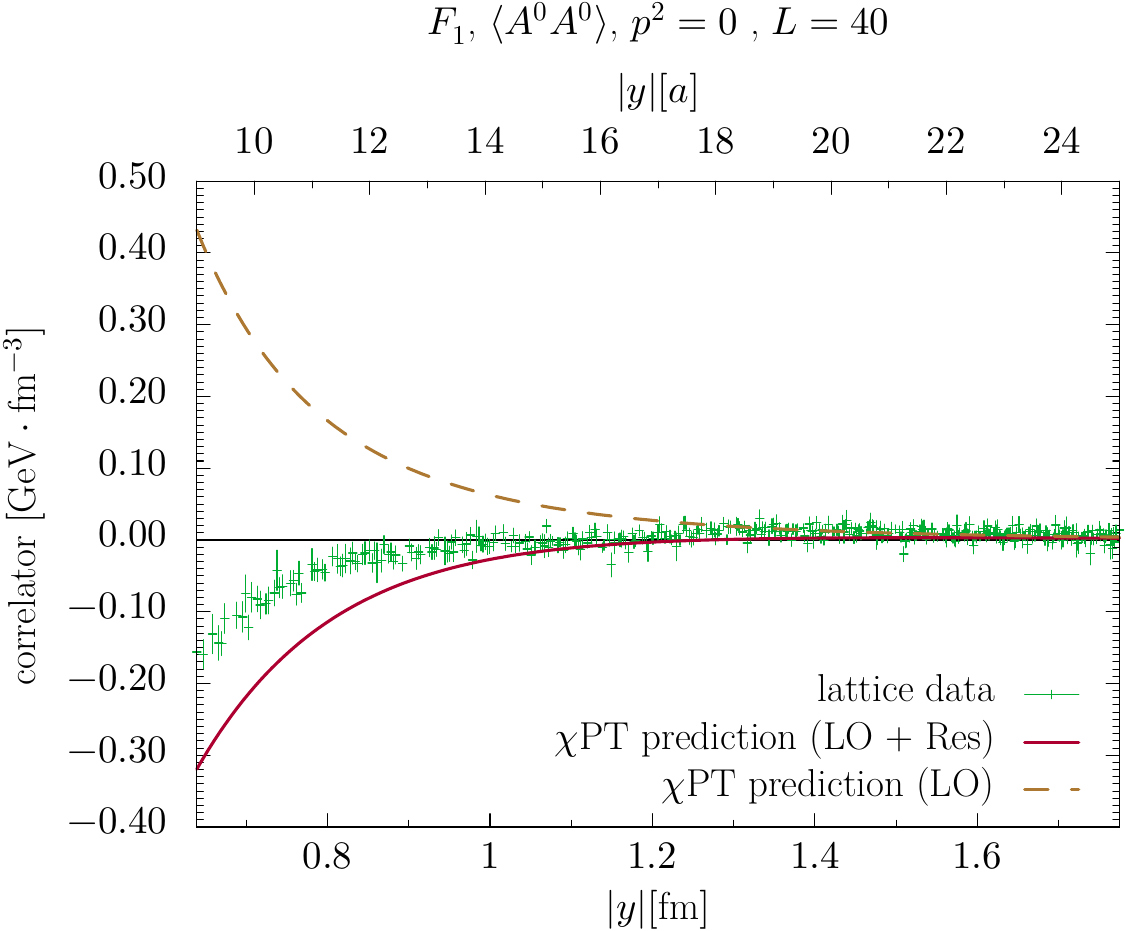}}
\\
  \subfigure[$F_2, \ov, p=0, L=40$]{\includegraphics[width=0.45\textwidth,trim=0 0 0 17pt,clip]{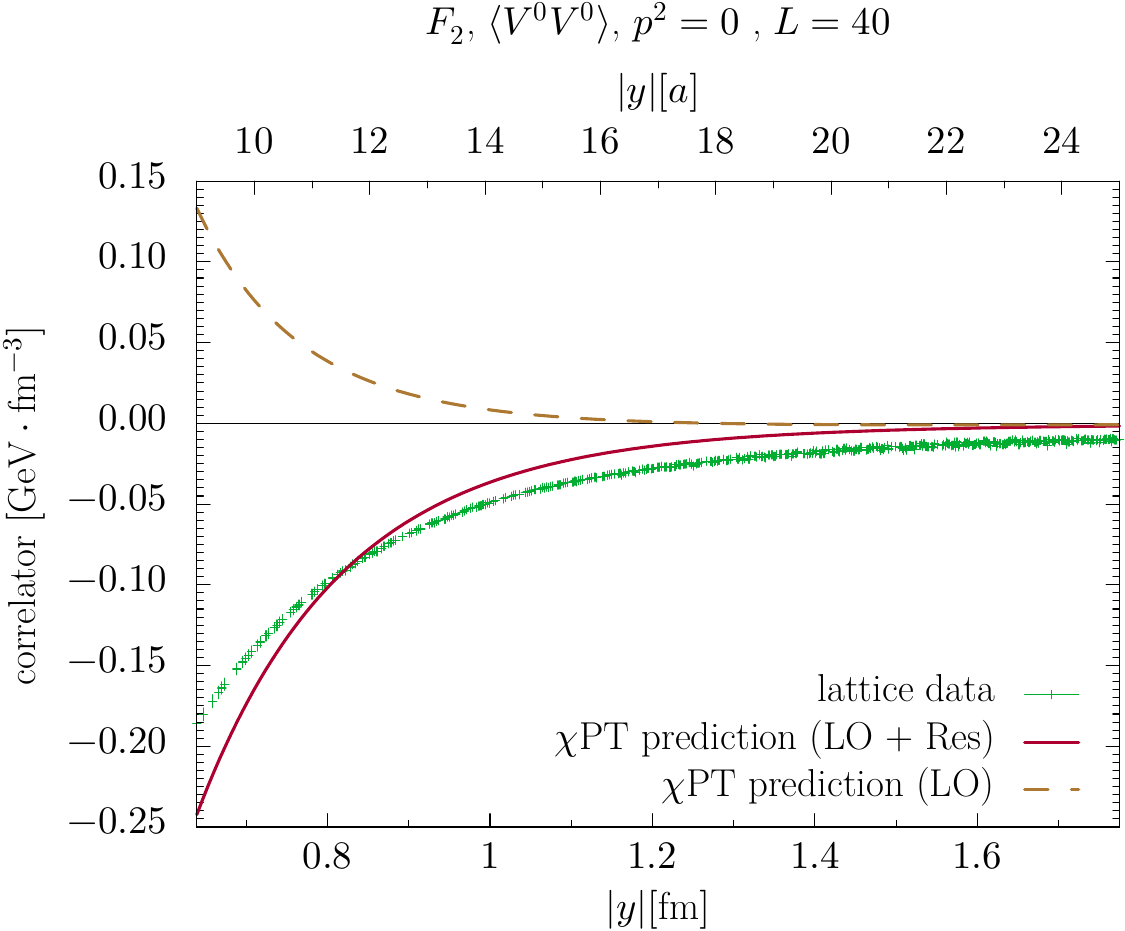}}
  \hfill
  \subfigure[$F_2, \oa, p=0, L=40$]{\includegraphics[width=0.45\textwidth,trim=0 0 0 17pt,clip]{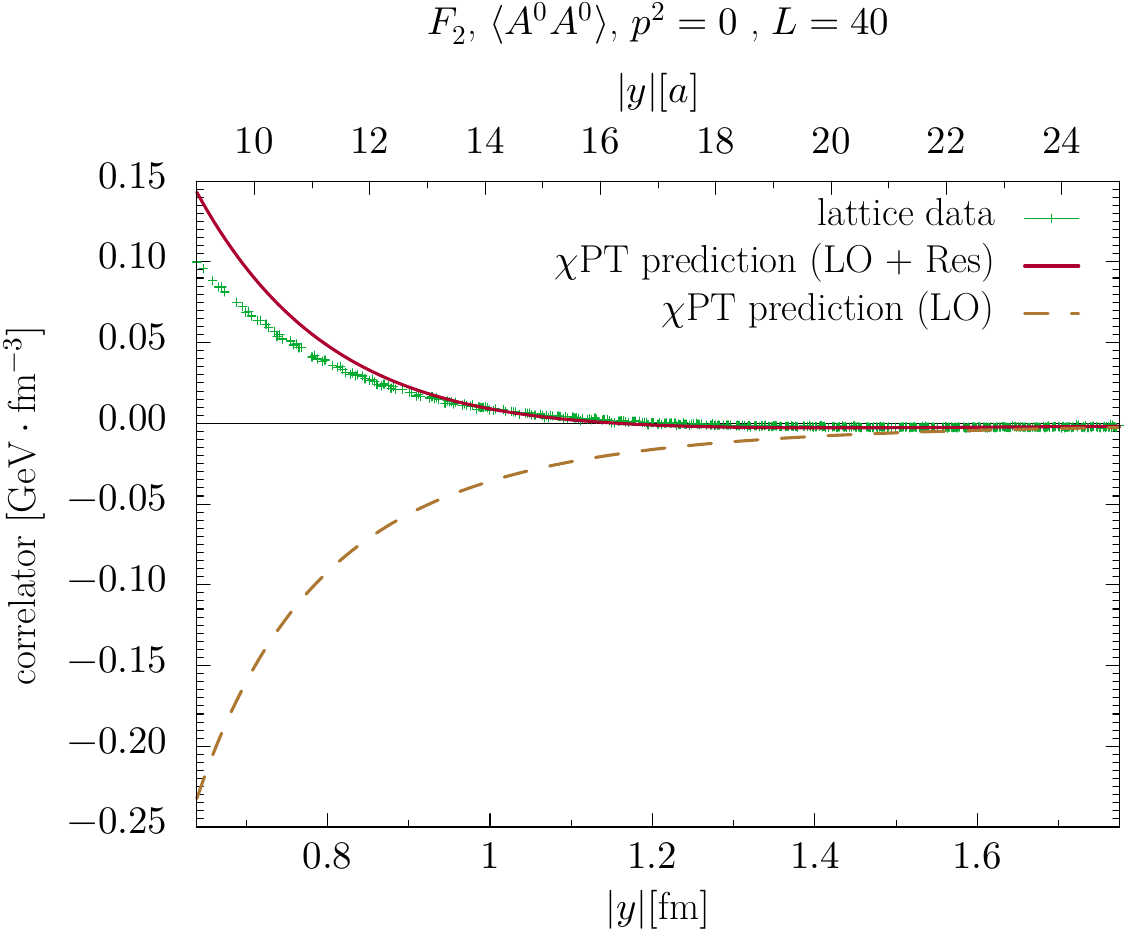}}
  \caption{\label{fig:chpt-VA} Comparison of our lattice results for $\ov$ and $\oa$ with chiral perturbation theory.  The results obtained from the leading-order chiral Lagrangian are denoted by ``LO'' and those including resonance exchange by ``LO $+$ Res''.}
\end{center}
\end{figure*}

\begin{figure*}
\begin{center}
  \subfigure[$F_0, \os, p=0, L=40$]{\includegraphics[width=0.433\textwidth,trim=0 0 0 17pt,clip]{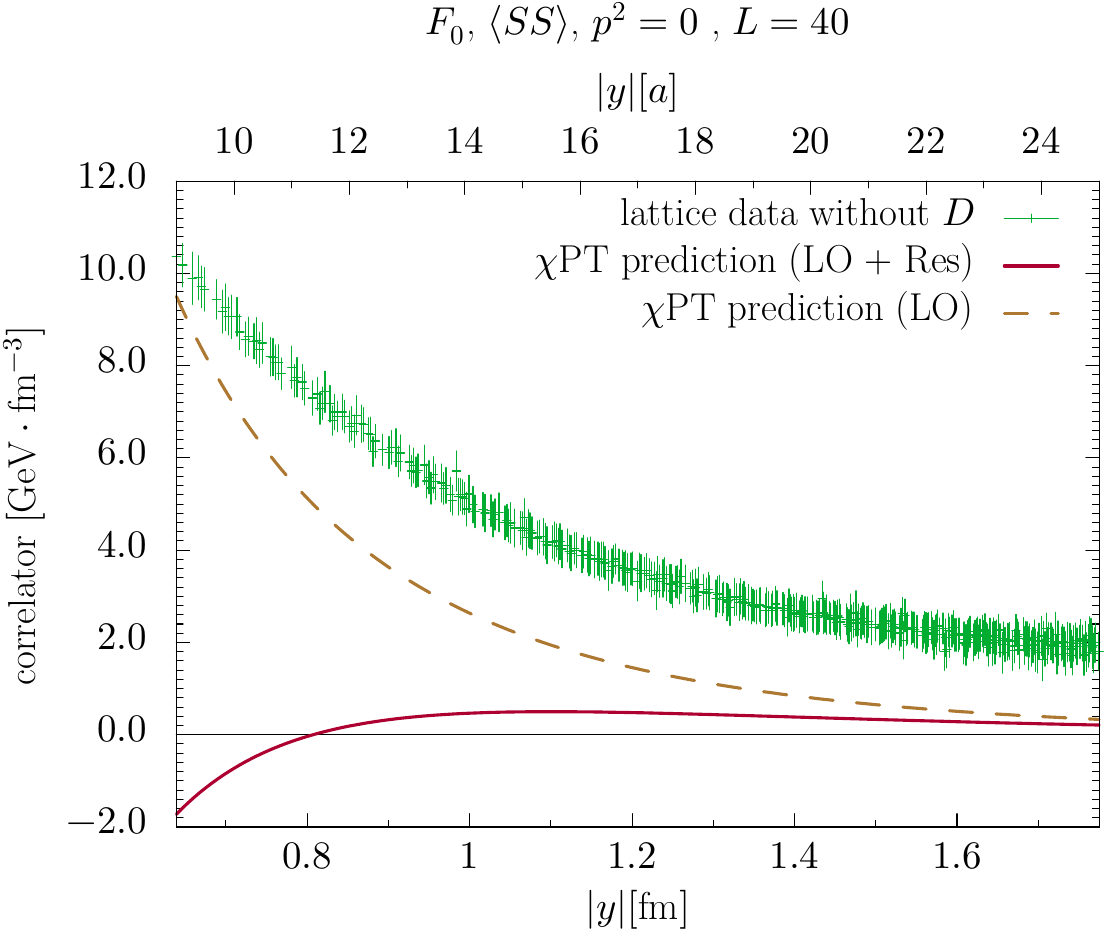}}
  \hfill
  \subfigure[$F_0, \op, p=0, L=40$]{\includegraphics[width=0.433\textwidth,trim=0 0 0 17pt,clip]{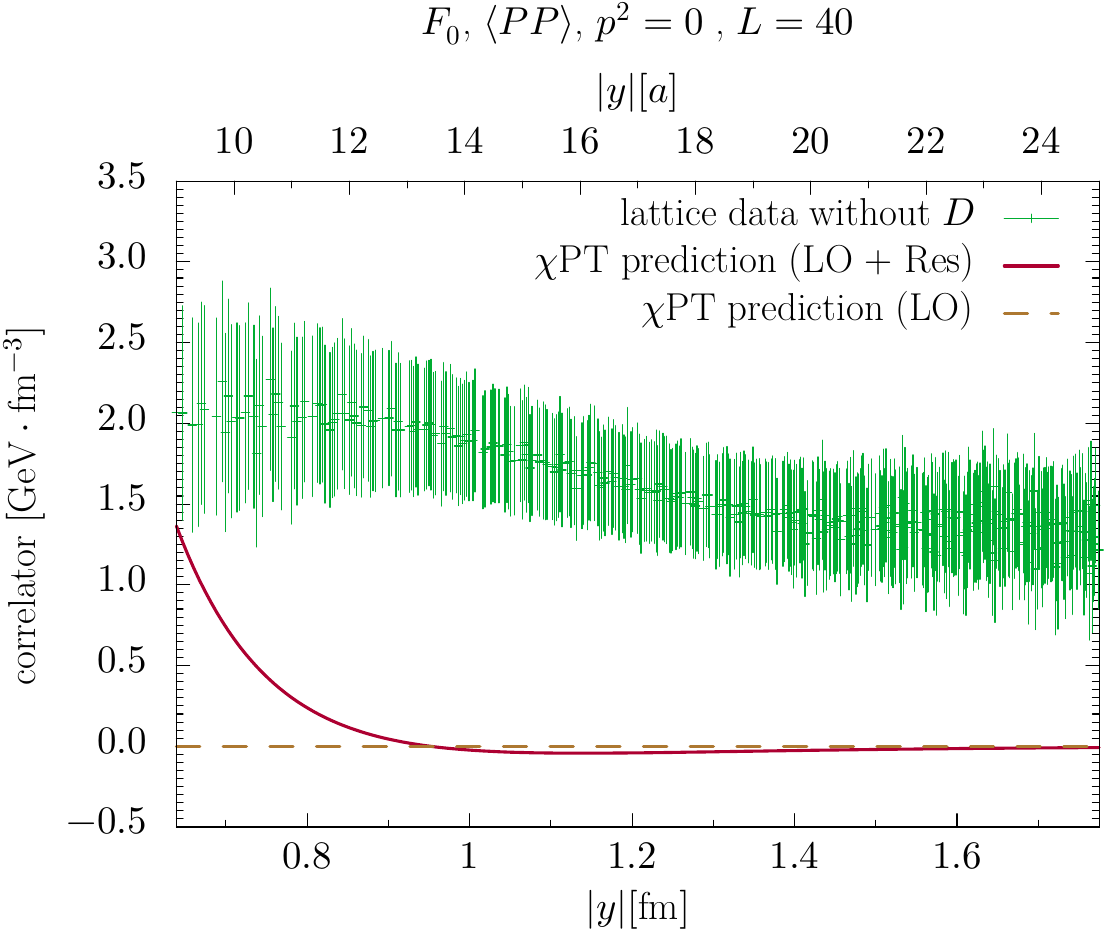}}
\\
  \subfigure[$F_1, \os, p=0, L=40$]{\includegraphics[width=0.433\textwidth,trim=0 0 0 17pt,clip]{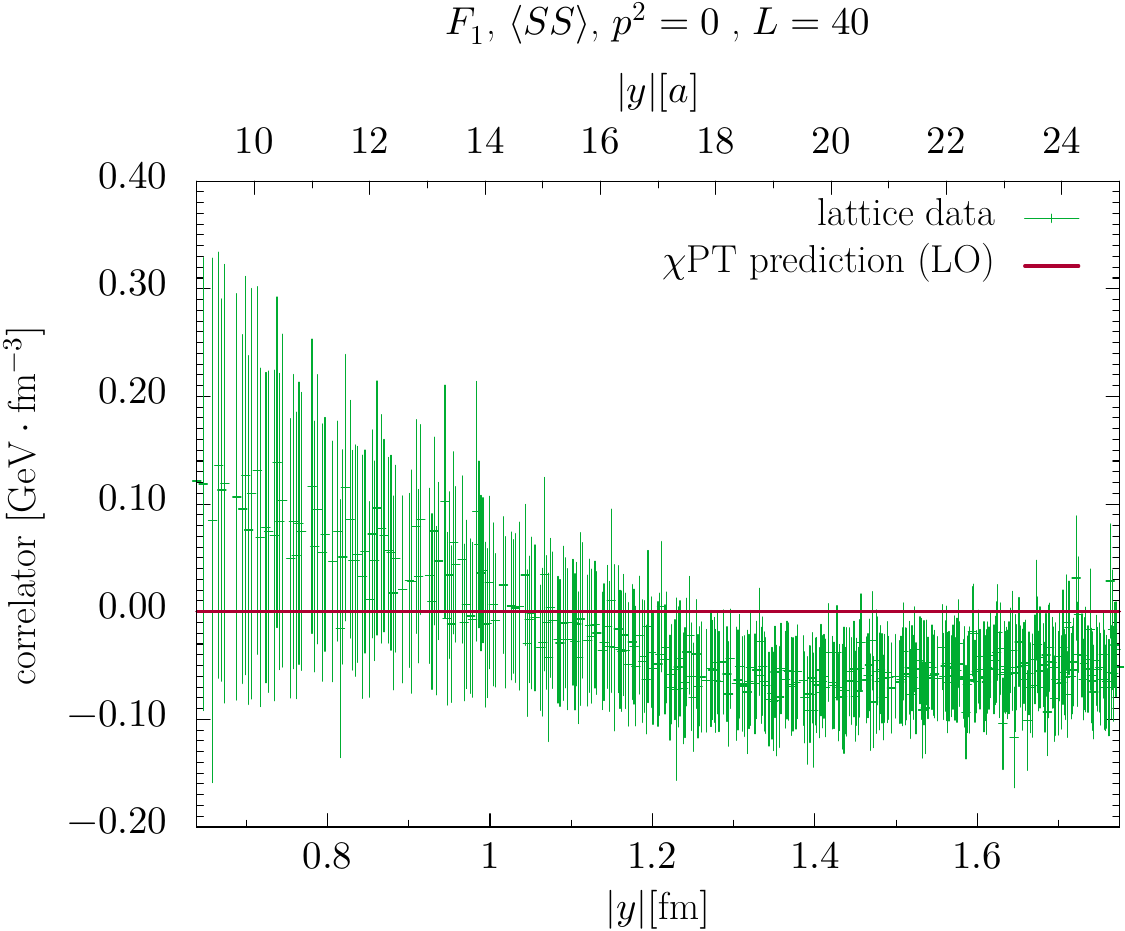}}
  \hfill
  \subfigure[$F_1, \op, p=0, L=40$]{\includegraphics[width=0.433\textwidth,trim=0 0 0 17pt,clip]{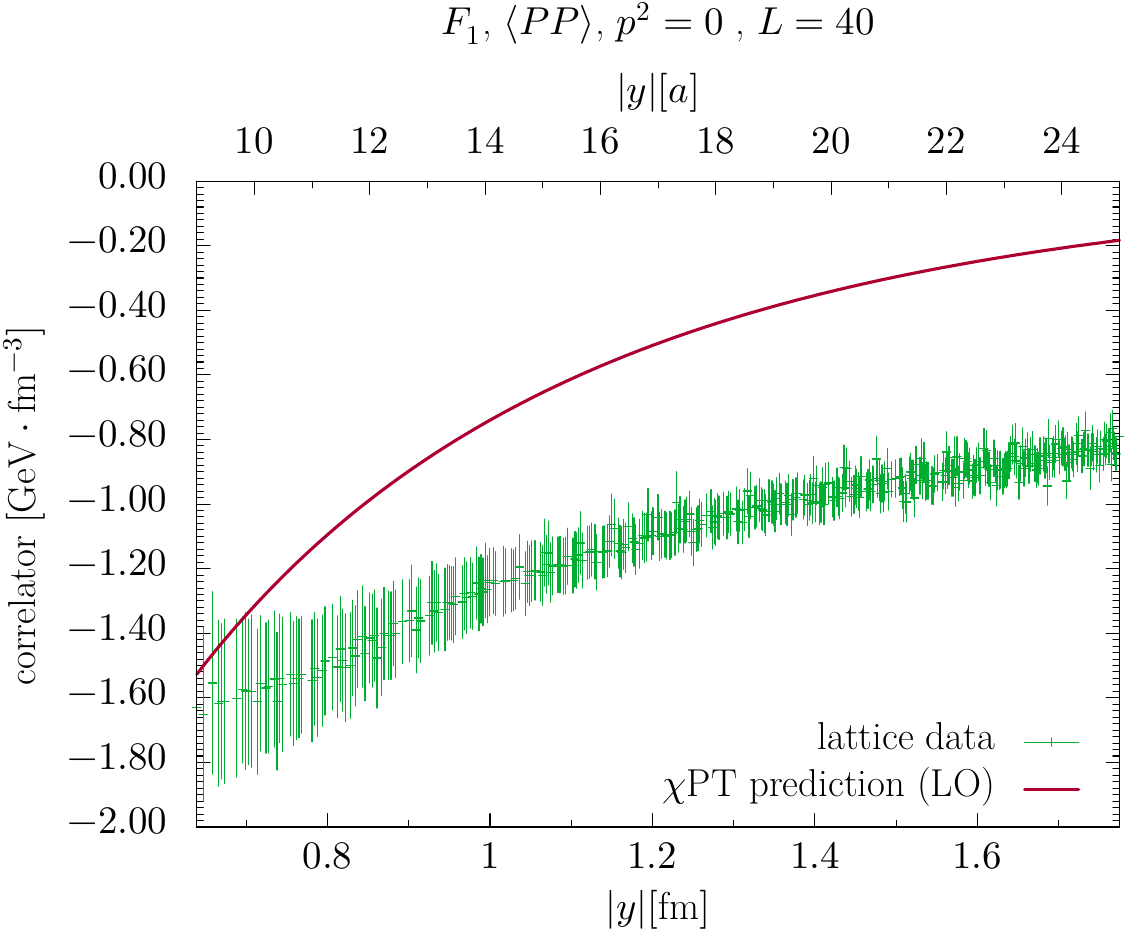}}
\\
  \subfigure[$F_2, \os, p=0, L=40$]{\includegraphics[width=0.435\textwidth,trim=0 0 0 17pt,clip]{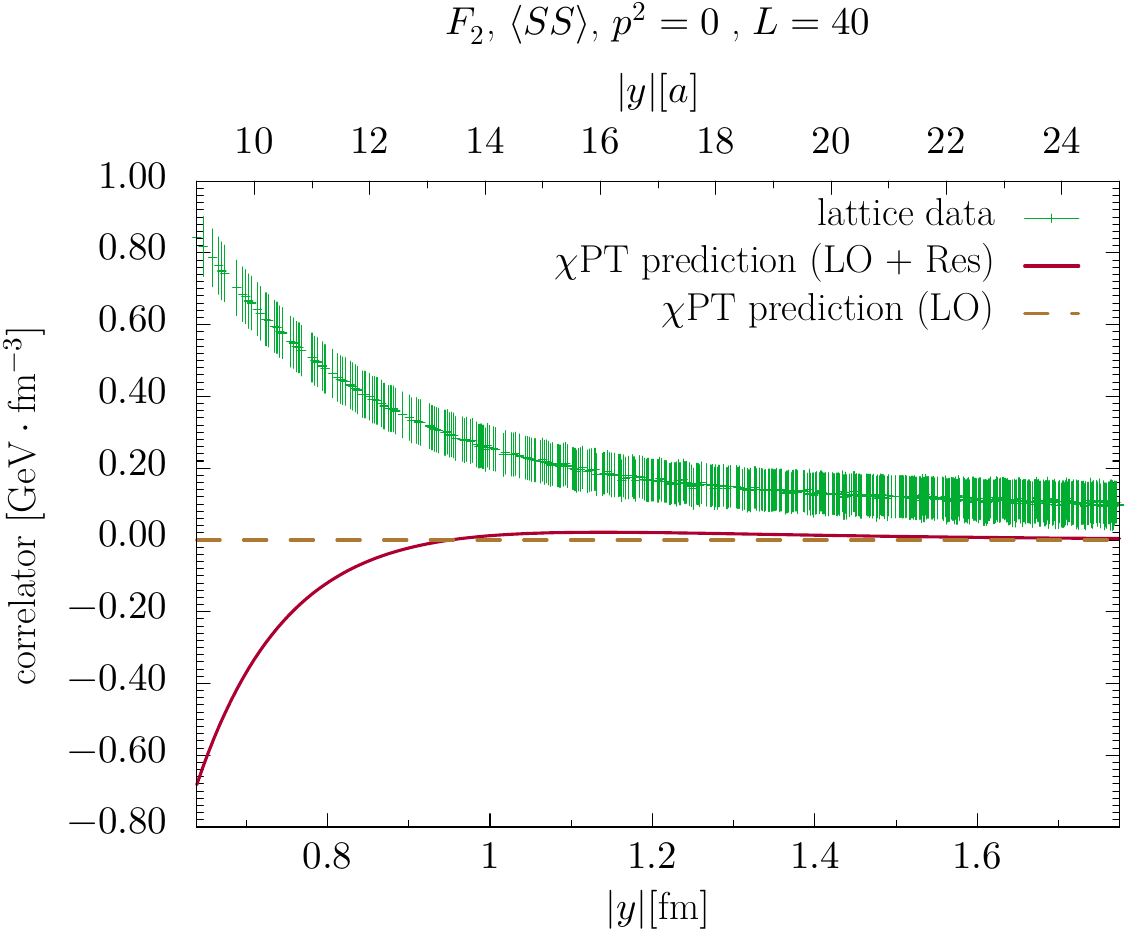}}
  \hfill
  \subfigure[$F_2, \op, p=0, L=40$]{\includegraphics[width=0.429\textwidth,trim=0 0 0 17pt,clip]{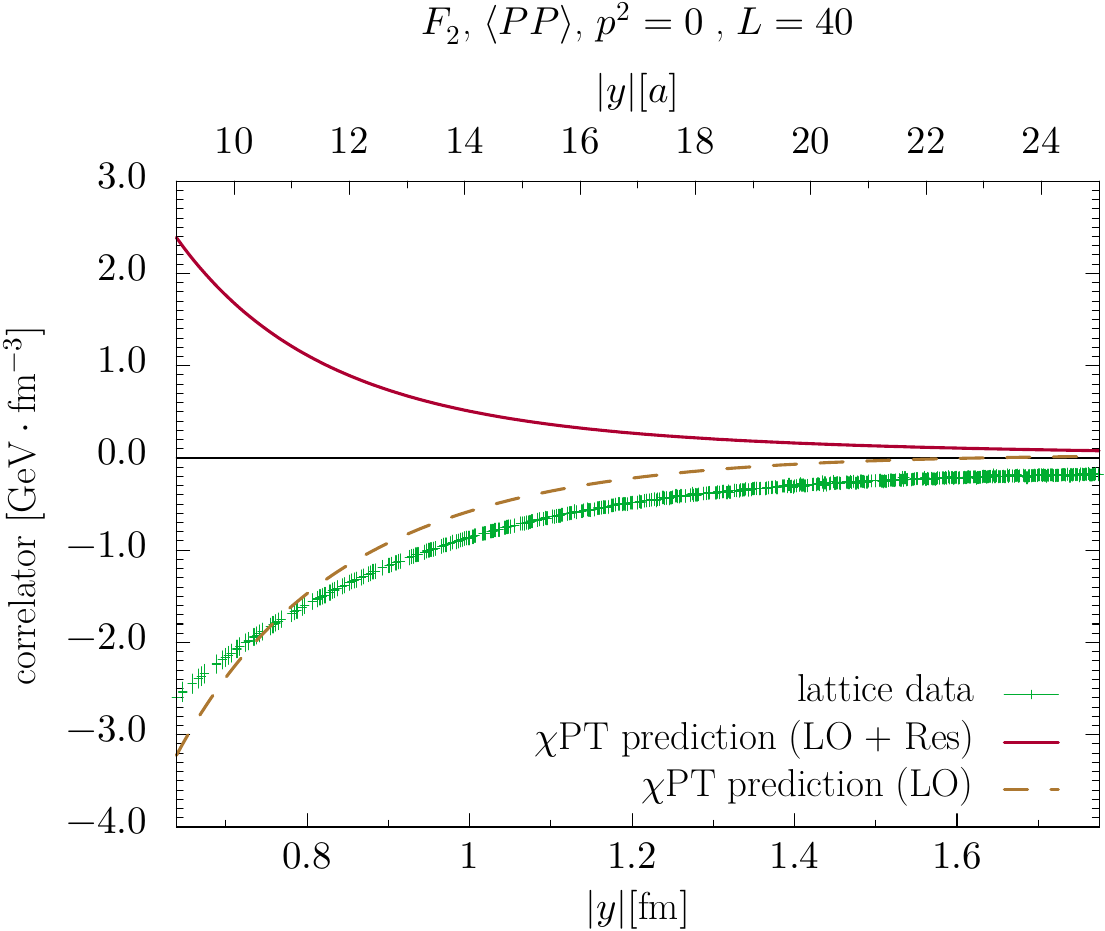}}
  \caption{\label{fig:chpt-SP} As figure~\protect\ref{fig:chpt-VA} but for the correlators $\os$ and $\op$.  The resonance exchange contribution is zero for $\os$ and $\op$ in $F_1$.}
\end{center}
\end{figure*}

In figures~\ref{fig:chpt-VA} and \ref{fig:chpt-SP}, we compare our lattice results with the predictions of chiral perturbation theory discussed in section~\ref{sec:chpt-pred}.  We remind the reader that in leading-order chiral perturbation theory, it is admissible to replace the pion decay constant $F$ in the chiral limit with its value $F_\pi$ at the considered pion mass, which is the value we have extracted in \eqref{chiral-params}.  Let us recall that predictions for the isospin amplitude $F_0$ are not available for the channels $\ov$ and $\oa$ in~\cite{Bruns:2015yto}.  Our lattice data in both channels is compatible with zero within errors for $y > 0.65 \fm$ and not shown here.

Since chiral perturbation theory requires large distances to be valid, we start our plots at $y \approx 0.65 \fm \approx 0.95 \ms \mpi^{-1} \approx 4.0 /(4 \pi F_\pi)$.  Comparing the leading-order chiral result with the resonance exchange contribution, we find three qualitatively different cases: for $\os$ and $\op$ in $F_1$ the resonance terms are zero, for $\op$ in $F_0$ and $\os$ in $F_2$ the leading-order terms vanish, whereas in all other channels both contributions are nonzero and there is a large difference between them in the $y$ range under study.  In the latter case, one may doubt whether the chiral expansion is stable, given that the resonance terms are formally subleading \rev{in the low-energy expansion}.  We also recall from the discussion in \cite{Bruns:2015yto} that the resonance terms \rev{are only meant to be an estimate of higher-order contributions}, and that the resonance parameters in the scalar and pseudoscalar channels are not well known.

With these caveats in mind, we now compare the chiral predictions with our lattice data.  As we see in figure~\ref{fig:chpt-VA}, the agreement between lattice data and theoretical results is quite good for $\ov$ in $F_1$ and $F_2$ and for $\oa$ in $F_2$.  These are the cases for which the volume effects seen in figure~\ref{fig:VA-vols} are moderate.  The agreement for $\oa$ in $F_1$ is still fair, and we recall that in this channel the volume dependence is somewhat larger.  For the vector and axial currents, we thus obtain a rather satisfactory picture.

The situation is quite different for the scalar and pseudoscalar currents.  For $\os$ in $F_0$, we find no large volume effects on the lattice and have argued that the neglect of graph $D$ should not have dramatic effects, at least in the lower $y$ range.  The comparison with the chiral prediction including resonance exchange is very bad in this case.
For the four channels shown in figure~\ref{fig:chpt-SP}(b), (d), (e) and (f), there is also a strong disagreement between lattice data and theory.  We recall from figure~\ref{fig:SP-vols} that in these cases the lattice results for $L=40$ and $L=32$ point to significant finite volume effects.
For $\os$ in $F_1$, our lattice signal is consistent with zero within errors, and the chiral prediction is zero as well at the order considered in \cite{Bruns:2015yto}.  In the sense that $\os$ in $F_1$ is small (compared for instance with $\op$ in the same isospin amplitude) we can state agreement between our lattice computation and chiral perturbation theory in this one case.

%% file: summary.tex
\section{Summary}
\label{sec:summary}

This paper contains a detailed study of two-current correlation functions in the pion in lattice QCD.  We use two gauge ensembles with a pion mass of $\mpi \approx 300 \mev$, a lattice spacing of $a \approx 0.07 \fm$ and spatial extensions of $L=32$ and $L=40$, respectively.  Additional results for heavier quarks (strange or charm) are obtained in a partially quenched setup.

We derive several general theoretical results concerning symmetry properties and the connection between lattice graphs and physical amplitudes.  In the remainder of our study, we focus on correlation functions of two currents $V^0$, $A^0$, $S$ or $P$ in a pion at rest.
In our lattice simulations, we make extensive use of stochastic sources, which allows us to obtain satisfactory (and in many cases excellent) signals for all lattice contractions except for the double disconnected graph $D$, where vacuum subtractions entail large cancellations and a loss of the signal in statistical noise.

We study a number of lattice artefacts in our data.  The comparison of results for two different time differences between the pion source and sink gives no indication for large contributions from excited states.  Comparing the Lorentz invariant correlation functions $\os$ and $\op$ for different pion momenta, we find that our lattice results are fully consistent with boost invariance.  We observe several types of anisotropic behaviour, indicating a loss of rotational invariance due to either the periodic boundary conditions or to discretisation effects.  We find that these effects are significantly reduced by selecting distances $\mvec{y}$ between the two currents that are close to the spatial diagonals of the lattice.  Finally, the comparison of data for two lattices shows moderate volume effects in several channels, but large ones in others.

In the connected graph $C_1$, the ``valence quark'' and ``valence antiquark'' in the pion are probed individually by the two currents.  One might naively expect this contraction to be dominant.  An important result of our study is that this is not true: we find that several lattice contractions are important for $\mpi \approx 300 \mev$.  This concerns especially the connected graph $C_2$ with two current insertions on the same quark line, but in the case of $\os$ and $\op$ also the disconnected graph $S_1$ and the annihilation contribution $A$.  For heavier quark masses, the importance of these graphs is reduced, although $C_2$ remains prominent in $\oa$ and $\op$ even for charm quarks.  Of course, not every contraction contributes to every physical matrix element, as shown in \eqref{phys-matels}.

For the connected graph $C_1$, we test the hypothesis that the two-current density can be computed in terms of the single-current density, assuming the absence of correlation effects between the quark and antiquark in the pion.  We find that this hypothesis clearly fails in a large range of current separations $\mvec{y}$, both for the vector and for the scalar current.

\rev{We also extract rms radii $r_{VV}$ and $r_{AA}$ of the correlator $C_1(\mvec{y})$ for the vector and for the axial current.  We find that a Fourier transform w.r.t.~$\mvec{y}$ mitigates finite-size effects.  The radius $r_{VV}$ shows a clear decrease with the quark mass.  Interestingly, it turns out that $r_{AA} < r_{VV}$ for light and strange quarks.}

Combining the different lattice graphs to physical amplitudes in an isospin basis, we can compare our lattice results with the computation in chiral perturbation theory performed in \cite{Bruns:2015yto}, which includes the contributions from the leading-order chiral Lagrangian and an estimate of higher-order contributions using resonance exchange graphs.  We find rather good agreement between the lattice data and the chiral calculation for vector and axial currents, whereas the comparison for scalar and pseudoscalar currents is poor in five out of six channels.  Since the exchanged resonances are $\rho$ and $a_1$ in the former case and $\sigma$, $a_0$, $\eta$ in the latter, our findings are consistent with the hypothesis that the exchange of the lowest-mass vector and axial vector mesons often gives a good estimate of higher orders in the chiral expansion, whereas the same does not necessarily hold for the exchange of spin-zero mesons.  This is in line with the conclusions drawn in \cite{Donoghue:1988ed,Ecker:1988te} and the  general success of models based on vector meson dominance.

The lattice methods presented in this work are suitable for the study of two-current correlators that can be related to double parton distributions in the pion, thus providing a connection with an active field of research in collider physics.  This will be presented in a forthcoming paper.